\documentclass[prd,nofootinbib,reprint,preprintnumbers,showkeys]{revtex4-1}
\usepackage{xcolor}
\usepackage{amsmath,amsfonts,bm}
\usepackage{graphicx, epstopdf,scrextend}
\usepackage{mathrsfs,mathtools}
\usepackage{enumitem}

\definecolor{darkgreen}{rgb}{0,0.4,0} 
\definecolor{darkblue}{rgb}{0,0,0.6} 
\usepackage[colorlinks=true,linkcolor=darkblue,citecolor=darkgreen]{hyperref}

\usepackage{pgfplots,pgfplotstable}
\usepgfplotslibrary{fillbetween}
\usetikzlibrary{intersections}
\pgfplotsset{compat=newest}

\newcommand{\as}{\alpha_{\mathrm{s}}}

\newcommand{\LA}{\mathrm{A}}

\newcommand{\LE}{\mathrm{E}}
\newcommand{\LF}{\mathrm{F}}

\newcommand{\LH}{\textsc{H}}
\newcommand{\scH}{\textsc{h}}

\newcommand{\LL}{\mathrm{L}}

\newcommand{\LR}{\mathrm{R}}
\newcommand{\scR}{\textsc{r}}

\newcommand{\LT}{\mathrm{T}}

\newcommand{\Lc}{\mathrm{c}}

\newcommand{\Lf}{\mathrm{f}}
\newcommand{\Lg}{\mathrm{g}}

\newcommand{\scV}{\textsc{v}}
\newcommand{\LZ}{\mathrm{Z}}

\newcommand{\mpone}{{m\!+\!1}}

\newcommand{\GeV}{\ \mathrm{GeV}}
\newcommand{\TeV}{\ \mathrm{TeV}}

\newcommand{\mug}{\mu^2}

\newcommand{\muf}{\mu_{\rm f}^2}
\newcommand{\muh}{\mu_{\scH}^2}

\newcommand{\cA}{\mathcal{A}}
\newcommand{\cB}{\mathcal{B}}
\newcommand{\cC}{\mathcal{C}}

\newcommand{\cI}{\mathcal{I}}

\newcommand{\cO}{\mathcal{O}}

\newcommand{\cR}{\mathcal{R}}
\newcommand{\cS}{\mathcal{S}}

\newcommand{\cU}{\mathcal{U}}

\newcommand{\cY}{\mathcal{Y}}

\definecolor{red}{rgb}{1,0,0}
\newcommand{\NOTE}[1]{\textcolor{red}{ \bf[NOTE: #1]}}

\def\P#1{{\color{red}\big[}{#1}{\color{red}\big]_\mathbb{P}}}
\def\omP#1{{\color{blue}\big[}{#1}{\color{blue}\big]_{1-\mathbb{P}}}}
\def\PL{{\color{red}\big[}}
\def\PR{{\color{red}\big]_\mathbb{P}}}
\def\iP#1{{\color{red}[}{#1}{\color{red}]_\mathbb{P}}}
\def\iomP#1{{\color{blue}[}{#1}{\color{blue}]_{1-\mathbb{P}}}}

\def\mi{{\mathrm i}}

\def\ket#1{\big|{#1}\big\rangle}
\def\bra#1{\big\langle{#1}\big|}
\def\brax#1{\big\langle{#1}}   
\def\<>#1{\big\langle{#1}\big\rangle}
\def\[]#1{\big[{#1}\big]}

\def\sket#1{\big|{#1}\big)}
\def\sbra#1{\big({#1}\big|}
\def\sbrax#1{\big({#1}}        

\def\isket#1{|{#1})}
\def\isbra#1{({#1}|}
\def\isbrax#1{({#1}}

\def\iket#1{|{#1}\rangle}
\def\ibra#1{\langle{#1}|}
\def\ibrax#1{\langle{#1}}   

\usepackage{pgfplots,pgfplotstable}

\newcommand{\errorband}[6][]{
  \pgfplotstableread{#3}\datatable
  \addplot[name path=pluserror,draw=none,no markers,forget plot]  
  table [x={#4},y expr=\thisrow{#5}+\thisrow{#6}]{\datatable};
  \addplot[name path=minuserror,draw=none,no markers,forget plot] 
  table [x={#4},y expr=\thisrow{#5}-\thisrow{#6}]{\datatable};
  \addplot[forget plot,#2] fill between[on layer={},of=pluserror and minuserror];
  \addplot [#1,no markers] table [x={#4},y={#5}]{\datatable};
}

\newcommand{\errorbandMOD}[6][]{
    \def\normalization{3.83333} 
  \pgfplotstableread{#3}\datatable
  \addplot[name path=pluserror,draw=none,no markers,forget plot]  
  table [x={#4},y expr=\normalization*\thisrow{#5}
  +\normalization*\thisrow{#6}]{\datatable};
  \addplot[name path=minuserror,draw=none,no markers,forget plot] 
  table [x={#4},y expr=\normalization*\thisrow{#5}
  -\normalization*\thisrow{#6}]{\datatable};
  \addplot[forget plot,#2] fill between[on layer={},of=pluserror and minuserror];
  \addplot [#1,no markers] table [x={#4},y expr = \normalization*\thisrow{#5}]{\datatable};
}

\pgfplotsset{every axis/.style={
    width=8.2cm,
    height=8.2cm,
    grid=both,
    scaled ticks=false,
    yticklabel style={/pgf/number format/.cd, fixed,precision=5}
  }
}

\newif\ifusefigs
\usefigstrue

\begin{document}

\title{Summations by parton showers of large logarithms in electron-positron annihilation}

\author{Zolt\'an Nagy}

\affiliation{
  DESY,
  Notkestrasse 85,
  22607 Hamburg, Germany
}

\email{Zoltan.Nagy@desy.de}

\author{Davison E.\ Soper}

\affiliation{
Institute for Fundamental Science,
University of Oregon,
Eugene, OR  97403-5203, USA
}

\email{soper@uoregon.edu}

\begin{abstract}

In a companion publication, we have explored how to examine the summation of large logarithms in a parton shower. Here, we apply this general program to the thrust distribution in electron-positron annihilation, using several shower algorithms. The method is to work with an appropriate integral transform of the distribution for the observable of interest. Then, we reformulate the parton shower calculation so as to obtain the transformed distribution as an exponential for which we can compute the terms in the perturbative expansion of the exponent.

\end{abstract}

\date{13 November 2020}

\preprint{DESY 20-182}

\maketitle


\section{Introduction}
\label{sec:introduction}

A parton shower event generator can provide a QCD based approximation for a cross section $\hat\sigma_J(\bm v)$ for an observable $J$ to take a value $\bm v$ in hadron-hadron, lepton-hadron, or electron-positron collisions. For example $J$ could denote the transverse momentum distribution in the Drell-Yan process and $\bm v$ could be $\bm v = \bm k_\LT$. Suppose that the observable $J$ is infrared safe with a scale $\hat Q^2_J(\bm v)$ substantially greater than $1 \GeV^2$. Then we can, at least in principle, omit a model for hadronization in the event generator. This leaves us with just an event generator based on a parton shower, which uses parton splitting functions based on the soft and collinear singularities of QCD.  Running the parton shower event generator gives us an approximation $\hat\sigma_J(\bm v; \mathrm{shower})$ for the cross section $\hat\sigma_J(\bm v)$.

The QCD perturbative expansion for $\hat\sigma_J(\bm v)$ will contain logarithms, $L =\log(\mu_\scH^2/\hat Q_J^2(\bm v))$, where $\mu_\scH^2$ is the scale of the hardest interaction in the event. Typically one finds perturbative contributions to $\hat\sigma_J(\bm v)$ proportional to $\as^n(\mu_\scH^2) L^{2n}$. If $\mu_\scH^2$ is close to $\hat Q_J^2(\bm v)$, then we do not need the parton shower at all. Rather, we can use just fixed order perturbation theory. However, if $1 \GeV^2 \ll \hat Q_J^2(\bm v) \ll \mu_\scH^2$, and $L^2 \gtrsim 1/\as(\mu_\scH^2)$, then fixed order perturbation theory is not adequate. One must try to sum the contributions at each order of perturbation theory that have the most powers of $L$. Since the splitting functions in a parton shower reflect the soft and collinear singularities of QCD and since it is these singularities that lead to the appearance of the logarithms $L$, we may hope that a parton shower provides an adequate approximation to the cross section $\hat\sigma_J(\bm v)$.

We caution the reader that we do not expect that a given parton shower algorithm correctly sums the logarithms for {\em all} infrared safe observables that generate large logarithms in perturbation theory. Thus we would not speak of a next-to-leading-log parton shower, without specifying just what logs are correctly summed.

For some observables $J$ one can derive an analytical approximation, $\hat\sigma_J(\bm v; \mathrm{analytical})$, to $\hat\sigma_J(\bm v)$ that sums the large logarithms in an appropriate sense. In some cases \cite{NSpT, NSdglap}, it is also possible to find an analytical formula that well approximates the shower result $\hat\sigma_J(\bm v; \mathrm{shower})$. Then one can tell whether $\hat\sigma_J(\bm v; \mathrm{shower})$ agrees with $\hat\sigma_J(\bm v; \mathrm{analytical})$ to the accuracy with which $\hat\sigma_J(\bm v; \mathrm{analytical})$ sums the large logarithms. However, this is usually difficult. 

Normally, the approximation $\hat\sigma_J(\bm v; \mathrm{shower})$ obtained with a parton shower is limited to a numerical result obtained by averaging over many generated events. In the limit of very large hard scattering scales $\mu_\scH^2$, $\hat\sigma_J(\bm v; \mathrm{shower})$ should match $\hat\sigma_J(\bm v; \mathrm{analytical})$. However, for $\mu_\scH^2$ in the kinematic range of experiments, $\hat\sigma_J(\bm v; \mathrm{shower})$ contains effects that are numerically important but are not included in $\hat\sigma_J(\bm v; \mathrm{analytical})$.  Thus it is difficult to tell whether $\hat\sigma_J(\bm v; \mathrm{shower})$ agrees with $\hat\sigma_J(\bm v; \mathrm{analytical})$.

One approach to comparing $\hat\sigma_J(\bm v; \mathrm{shower})$ to $\hat\sigma_J(\bm v; \mathrm{analytical})$ is to directly calculate $\hat\sigma_J(\bm v; \mathrm{shower})$ for a sequence of very large hard scattering scales $\mu_\scH^2$ that are far from the range of experiments. This approach can work \cite{DasguptaShowerSum}, and in fact we use it to a limited extent in this paper. However, it is difficult to maintain the required numerical accuracy at very large values of $\mu_\scH^2$ in a practical parton shower event generator.

In an analytical approach, one typically starts by taking an appropriate integral transform of $\hat\sigma_J(\bm v)$. Then one calculates a cross section $\sigma_J(\bm r)$ depending on a variable or variables $\bm r$. The cross section $\sigma_J(\bm r)$ contains logarithms $L(\bm r)$ that are large when $\bm r$ approaches a limit.  For instance, one might take the Fourier transform of the $\bm k_\LT$ distribution in the Drell-Yan process. Then $\bm r$ is the transverse position, usually called $\bm b$. The logarithm is $L = \log(\bm b^2 \muh)$, which is large when $\bm b^2 \to \infty$.

The aim of this paper is to redesign the calculation of the parton shower cross section so that it produces the same result for $\sigma_J(\bm r)$ as before but so that it produces a calculation of $\sigma_J(\bm r)$ and not a cross section for other observables. The redesigned calculation gives $\sigma_J(\bm r)$ as an exponential of a quantity that can be expanded in powers of the shower splitting operator. The leading order term in the exponent is simple and is the candidate for the summation of large logarithms produced by the shower. If the higher order contributions to the exponent are suitably small, then they to not interfere with the summation represented by the leading order terms.

We have presented a general formulation of this program in Ref.~\cite{NSlogsum}. The general formulation applies to hadron-hadron collisions, lepton-hadron collisions, and electron-positron collisions and to a general infrared-safe observable $J$. In this paper, we apply the general formalism to a simple example. We examine the thrust distribution in electron-positron annihilation. In this case, the original variable $\bm v$ is $\tau = 1-T$, where $T$ is the thrust. We take the Laplace transform of the $\tau$ distribution, so that the transformed variable $\bm r$ is the Laplace transform variable $\nu$.

There is much to be learned from this example. In particular, we learn that the shower result for $\sigma_J(\nu)$ depends on some details of the parton shower algorithm that one might have thought are not important.


\section{The parton shower framework}
\label{sec:deductor}

A parton shower can be described using operators on a vector space, the ``statistical space,'' that describes the momenta, flavors, colors, and spins for all of the partons created in a shower as the shower develops. We use this description in the parton shower event generator \textsc{Deductor} \cite{NSI, NSII, NSspin, NScolor, Deductor, ShowerTime, PartonDistFctns, ColorEffects}. The general theory includes parton spins but \textsc{Deductor} simply averages over spins, so our explanation here will leave out parton spins. With $m$ final state partons in electron-positron annihilation, the partons carry labels $1,2,\dots,m$. The partons have momenta $\{p\}_m = \{p_1,\dots,p_m\}$ and flavors $\{f\}_m$. We take the partons to be massless: $p_i^2 = 0$. For color, there are ket color basis states $\iket{\{c\}_m}$ and bra color basis states $\ibra{\{c'\}_m}$. We use the trace basis, as described in Ref.~\cite{NSI}. Color appears in the statistical space as the density matrix, with basis elements $\iket{\{c\}_m}\ibra{\{c'\}_m}$. Then the $m$-parton basis states for the statistical space are denoted by $\isket{\{p,f,c,c'\}_{m}}$.

\textsc{Deductor} uses specific choices with respect to shower kinematics, the shower ordering variable, the parton splitting functions, and the treatment of color. In this section, we outline some of these choices that play a role in the analysis of this paper.

An exact color treatment is used in the general formalism. However the code of \textsc{Deductor} mostly uses only an approximation, the leading-color-plus (LC+) approximation \cite{NScolor}. The LC+ approximation consists of simply dropping some terms in the splitting functions. In this paper, we generally use full color but sometimes use the LC+ approximation.

In \textsc{Deductor}, the default is to order splittings according to decreasing values of a hardness parameter $\Lambda^2$. The hardness parameter is based on virtuality. For massless final state partons in electron-positron collisions, the definition is\footnote{In hadron-hadron collisions, $Q$ in Eq.~(\ref{eq:Lambdadef}) is replaced by the momentum $Q_0$ of the final state partons at the start of the shower.}
\begin{equation}
\begin{split}
\label{eq:Lambdadef}
\Lambda^2 ={}& \frac{(\hat p_l + \hat p_{m+1})^2}{2 p_l\cdot Q}\, Q^2
\;.
\end{split}
\end{equation}
Here the mother parton in a final state splitting has momentum $p_l$ and the daughters have momenta $\hat p_l$ and $\hat p_{m+1}$. Here $Q$ is the total momentum $Q$ of all of the final state partons, which remains the same throughout the shower.  It proves convenient to use a dimensionless virtuality variable $y = \Lambda^2/Q^2$:
\begin{equation}
\label{eq:ydef}
y = \frac{(\hat p_l + \hat p_{m+1})^2}{2 p_l \cdot Q}
\;.
\end{equation}
Thus $y$ decreases from one shower splitting to the next.

One could use a hardness parameter other than $\Lambda$ to order the shower. We will consider also a shower ordered by the transverse momentum \cite{NSThreshold} in a splitting,
\begin{equation}
\label{eq:kTdef}
k_\LT^2 = z (1-z) (\hat p_l + \hat p_{m+1})^2
= z (1-z) y\, Q^2\!/a_l
\;,
\end{equation}
where $z$ is the momentum fraction in the splitting and
\begin{equation}
a_l = \frac{Q^2}{2p_l\cdot Q}
\;.
\end{equation}
We denote the hardness scale of a splitting by $\mu^2$. When we use the default ordering variable $\Lambda$ for the shower, then $\mu^2 = \Lambda^2$. If we use $k_\LT$ ordering, then $\mu^2 = k_\LT^2$.

To measure an infrared-safe observable $\cO_J$ in electron-positron annihilation, we can use the notation
\begin{equation}
\begin{split}
\label{eq:sigmaJ}
\sigma_J ={}& 
\sbra{1} \cO_J\, 
\cU(\mu_\Lf^2,Q^2)\,
\sket{\rho_\scH}
\;.
\end{split}
\end{equation}
Here $\isket{\rho_\scH}$ is the starting parton state for the hard scattering process. If we were to evaluate $\isket{\rho_\scH}$ beyond leading order, then it would contain appropriate subtractions to remove infrared singularities. In this paper, we evaluate $\isket{\rho_\scH}$ at lowest order so that it is simply a $q\bar q$ state. We associate a scale $\mu_\scH^2 = Q^2$ with the hard scattering, where $Q$ is the $q\bar q$ momentum. The operator $\cU(\mu_\Lf^2,Q^2)$ expresses the evolution of the system from the scale $Q^2$ to a scale $\mu_\Lf^2$ of order $1 \GeV^2$, at which the shower is turned off. After this evolution, we have a statistical state that can be expanded in the basis states $\isket{\{p,f,c,c'\}_{m}}$. This expansion is realized as an integral, which takes the form of a Monte Carlo integration that is obtained by generating many Monte Carlo events. We then apply an operator $\cO_J$ that embodies the desired measurement. We still have a sum and integral of basis states. We take the product with the statistical bra state $\isbra{1}$, which is defined by 
\begin{equation}
\label{eq:1def}
\sbrax{1} \sket{\{p,f,c,c'\}_{m}} = \brax{\{c'\}_{m}}\ket{\{c\}_{m}}
\;.
\end{equation}
This leaves us with the numerical result for $\sigma_J$.

The shower operator $\cU$ takes the form
\begin{equation}
\label{eq:jetUexponential}
\cU(\mu_2^2,\mu_1^2)
=\mathbb{T} \exp\!\left(
\int_{\mu_2^2}^{\mu_1^{2}}\!\frac{d\mu^2}{\mu^2}\,
\cS(\mu^2)
\right)
\;.
\end{equation}
There is an instruction $\mathbb{T}$ that indicates that if we expand the exponential, the operators $\cS(\mu^2)$ with the smallest values of $\mu^2$ belong on the left. This is simply a compact way of saying that $\cU(\mu_2^2,\mu_1^2)$ obeys the differential equation
\begin{equation}
\label{eq:cUdiffeqn}
\mu_1^2 \frac{\partial}{\partial \mu_1^2}\,\cU(\mu_2^2,\mu_1^2)
= \cU(\mu_2^2,\mu_1^2)\,\cS(\mu_1^2)
\;.
\end{equation}

In general, the generator $\cS(\mu^2)$ is a sum of terms with approximations to $n_\scR$ real emissions and $n_\scV$ virtual exchanges,
\begin{equation}
\label{eq:cSn}
\cS(\mu^2) = \sum_{\substack{n_\scR,n_\scV =0\\ n_\scR + n_\scV \ge 1}}^{\infty}
\cS^{[n_\scR,n_\scV]}(\mu^2)
\;.
\end{equation}
In existing parton shower event generators like \textsc{Deductor}, only the terms with $n_\scR + n_\scV = 1$ are implemented. This is also the case for other parton shower algorithms that we consider here. Thus in this paper we assume
\begin{equation}
\label{eq:cS1decomposition}
\cS(\mu^2) = \cS^{[1,0]}(\mu^2) + \cS^{[0,1]}(\mu^2)
\;.
\end{equation}
The operator $\cS^{[1,0]}(\mu^2)$ creates a splitting, changing an $m$ parton state to an $m+1$ parton state. The operator $\cS^{[0,1]}(\mu^2)$ leaves the number of partons and their momenta and flavors unchanged, although in a full color treatment it modifies the parton color state. The operator $\cS^{[0,1]}(\mu^2)$ is related to the inclusive sum over splitting variables in $\cS^{[1,0]}(\mu^2)$ by $\isbra{1}\cS^{[0,1]}(\mu^2) = -\isbra{1}\cS^{[1,0]}(\mu^2)$, so that
\begin{equation}
\label{eq:1cS}
\sbra{1}\cS(\mu^2) = 0
\;.
\end{equation}
If we had contributions to the shower generator with $n_\scR + n_\scV > 1$, we would still have $\isbra{1}\cS(\mu^2) = 0$ \cite{NSAllOrder}.

The operator $\cS^{[1,0]}(y Q^2)$ in \textsc{Deductor} \cite{NScolor,NSThreshold} is not simple.  However, in the cases for which we need an explicit expression in our analytical formulas here, we need only its form when $y \ll 1$ and $(1-z) \ll 1$. This is the limit in which $\cS^{[1,0]}(y Q^2)$ expresses the soft$\times$collinear double singularity of QCD. (However, our numerical results use the full  $\cS^{[1,0]}(y Q^2)$.) In this limit, we have 
\begin{equation}
\begin{split}
\label{eq:S10}
\cS^{[1,0]}&(y Q^2)\sket{\{p,f,c,c'\}_{m}}
\\
\approx{}& 
- \sum_{l = 1}^{m} \sum_{\substack{k = 1 \\ k \ne l}}^{m}\,
[T_{l} \otimes  T_{k}^\dagger
+ T_{k} \otimes  T_{l}^\dagger]\sket{\{c,c'\}_{m}}
\\&\times
\int\!\frac{d\phi}{2\pi}
\int\!\frac{dz}{1-z}\
\frac{\as(\lambda_\LR (1-z) y Q^2/a_l)}{2\pi}
\\&\times
\Theta\!\left(\frac{a_l y}{\vartheta(l,k)} < 1-z
< 1\right)
\\&\times
\sket{\{\hat p, \hat f\}_{m+1}}
\;.
\end{split}
\end{equation}
There is a sum over parton indices $l$ and $k$. We split parton $l$ with dipole partner parton $k$, creating a new parton $m+1$, which we consider to be a gluon. The momenta $\{\hat p\}_{m+1}$ of the partons after the splitting are functions of the momenta $\{p\}_{m}$ before the splitting and the splitting variables $y,z,\phi$, as specified in Eqs.~(\ref{eq:hatplpmp1}) and (\ref{eq:lorentztransformation}).

In Eq.~(\ref{eq:S10}), $[T_{l} \otimes  T_{k}^\dagger]$ and $[T_{k} \otimes T_{l}^\dagger]$ are operators on the parton color space. The notation $(C_\textrm{ket} \otimes C^\dagger_\textrm{bra})$ for color operators represents the following. A color basis vector $\isket{\{c,c'\}_m}$ in the statistical space represents the color density operator $\iket{\{c\}_m}\ibra{\{c'\}_m}$. Here $\iket{\{c\}_m}$ and $\iket{\{c'\}_m}$ are basis vectors for color amplitudes. Let $C_\textrm{ket}$ and $C_\textrm{bra}$ be operators on color amplitudes for $m$ partons that yield color amplitudes for $\hat m$ partons with $\hat m \ge m$. In the case of $\cS^{[1,0]}(y Q^2)$, $\hat m = m+1$. The statistical space vector $(C_\textrm{ket} \otimes C^\dagger_\textrm{bra})\isket{\{c,c'\}_m}$ then represents the color density operator $C_\textrm{ket}\iket{\{c\}_m}\ibra{\{c'\}_m} C^\dagger_\textrm{bra}$. In the case of $[T_{l} \otimes T_{k}^\dagger]$, the operator creates a new gluon with color index $a$ by inserting a color generator matrix $T^{a}$ on the color line for parton $l$ in the ket state and inserting $T^{a}$ on the color line for parton $k$ in the bra state. 

The argument of $\as$ in Eq.~(\ref{eq:S10}) contains the standard factor \cite{CMW}
\begin{equation}
\label{eq:CMW}
\lambda_\LR = \exp\left(- \frac{  C_\LA(67 - 3\pi^2)- 10\, n_\Lf}
{3\, (11 C_\LA - 2\,n_\Lf)}\right)
\;.
\end{equation}
The rest of the argument of $\as$ is $k_\LT^2$, Eq.~(\ref{eq:kTdef}), except that we drop the factor $z$ because we are interested only in small $1-z$. Although the operators $\cS^{[1,0]}(\mu^2)$ contain one power of $\as$, this $\as$ is evaluated at a scale that is not $\mug$. Thus if we expand $\cS(\mu^2)$ in powers of $\as(\mug)$, all powers will appear.

The parameter $\vartheta(l,k)$ is
\begin{equation}
\vartheta(l,k) = \frac12 [1 - \cos(\theta(l,k))]
\;,
\end{equation}
where $\theta(l,k)$ is the angle between partons $l$ and $k$ in $\isket{\{p, f, c, c'\}_m}$. With this definition, $\vartheta \approx \theta^2/4$ for small $\theta$. The angle $\hat\theta(l,\mpone)$ between partons $l$ and $m+1$ after a splitting is given by 
\begin{equation}
\begin{split}
[1-\cos\hat\theta(l,\mpone)]
={}&
\frac{Q^2}{\hat p_l\cdot Q\ \hat p_\mpone\cdot Q}\, \hat p_l\cdot \hat p_\mpone 
\\
\sim{}& \frac{2 a_l y}{z(1-z)} + \cO(y^2)
\;.
\end{split}
\end{equation}
For small $y$ and small $(1-z)$, this gives
\begin{equation}
\label{eq:splittinganglelimit}
\hat\vartheta(l,\mpone) \approx \frac{a_{l}y}{1-z}
\;.
\end{equation}
Thus the lower limit on $(1-z)$ is equivalent to an upper limit on the splitting angle, $\hat\vartheta(l,\mpone) < \vartheta(l,k)$. The splitting angle should be smaller than the angle between the two partons $l$ and $k$. The restriction $(1-z) < 1$ gives a lower limit on the splitting angle. The net range for the new splitting angle is
\begin{equation}
\label{eq:anglerange}
a_l y < \hat\vartheta(l,\mpone) < \vartheta(l,k)
\;.
\end{equation}
%


\section{Preview}
\label{sec:preview}

In this paper, we propose a way to gain more direct access to the summation of large logarithms in a parton shower than by simply running the shower and examining the result numerically. The analysis adapts the general formulation of this program in Ref.~\cite{NSlogsum} to the practical analysis of first order parton shower algorithms. Our example is the thrust distribution in electron-positron annihilation. Here is a brief preview.

\begin{itemize}[leftmargin = 0.3 cm]

\item We are interested in the thrust distribution $g(\tau)$ with $\tau = 1-T$, where $T$ is the thrust.

\item As in analytical approaches, we work with the Laplace transform $\tilde g(\nu)$ of $g(\tau)$.

\item $\tilde g(\nu)$ contains large logarithms, $\as(\muh)^n \log^{j}(\nu)$ with $j \le 2n$. 

\item We suppose that we know the proper summation of the $\log(\nu)$ factors in full QCD at a certain level of accuracy, but a leading order parton shower is not full QCD. We wish to know what result the parton shower gives.

\item The result of simply running the shower and examining the result numerically can be expressed as in Eq.~(\ref{eq:sigmaJ}),
\begin{equation}
\label{eq:sigmanustart}
\tilde g(\nu) =
\frac{1}{\sigma_\scH}
\sbra{1} \cO(\nu)\, 
\cU(\mu_\Lf^2,Q^2)\,
\sket{\rho_\scH}
\;.
\end{equation}
Here $\sigma_\scH$ is the total hard scattering cross section and $\isbra{1} \cdots \isket{\rho_\scH}$ indicates an ensemble average in the statistical state $\isket{\rho_\scH}$ representing the perturbative hard scattering. Then $\cU(\mu_\Lf^2,Q^2)$ represents the operator on the statistical space that generates the shower. This gives us states consisting of tens of partons. We could measure any operator $\cO_J$ that we like in this many-parton state. We apply a simple operator $\cO(\nu)$ that measures the Laplace transformed thrust distribution on this state.

\item In this paper, we rewrite $\tilde g(\nu)$ in the form
\begin{equation}
\label{eq:sigmanuresult1}
\tilde g(\nu) = 
\frac{1}{\sigma_\scH}
\sbra{1} 
\mathbb{T} \exp\!\big(\cI(\nu)\big)
\cO(\nu)
\sket{\rho_\scH}
\;.
\end{equation}
The notation $\mathbb{T}$ indicates an ordering instruction for the exponential, as in Eq.~(\ref{eq:jetUexponential}) and later in Eq.~(\ref{eq:cYexponential}). In the example used in this paper, the operator $\cO(\nu)$ applied to the hard state $\isket{\rho_\scH}$ simply gives an eigenvalue 1.

\item With this form, we have expressed $\tilde g(\nu)$ in terms of the exponential of an operator $\cI(\nu)$. This operator has an expansion\footnote{In Ref.~\cite{NSlogsum}, we expanded operators in powers of $\as(\mug)$ at a running scale $\mug$ appropriate to the operator. Here, we expand operators in powers of the splitting operator $\cS$ of the parton shower. This technique simplifies the analysis of a shower algorithm that is based on lowest order perturbation theory.}
\begin{equation}
\label{eq:cIk}
\cI(\nu) = \sum_{k=1}^\infty \cI^{[k]}(\nu)
\;,
\end{equation}
where each term in $\cI^{[k]}(\nu)$ contains $k$ factors of the splitting operator $\cS$.

\item We can further expand in powers of $\as$ evaluated at a fixed scale $Q^2/\nu$:
\begin{equation}
\label{eq:cIkn}
\cI^{[k]}(\nu) = \sum_{n=k}^\infty \left[\frac{\as(Q^2/\nu)}{2\pi}\right]^n
\cI^{[k]}_n(\nu)
\;.
\end{equation}

\item The most important feature of Eq.~(\ref{eq:sigmanuresult1}) is that the operators $\cI^{[k]}(\nu)$ can be computed using two fairly simple recursion relations.

\item The first order contribution, $\cI^{[1]}(\nu)$, is obtained rather trivially from one power of the shower splitting operator $\cS(\mu^2)$. This operator is then the obvious candidate for the exponentiation of $\tilde g(\nu)$ generated by the shower. If $\cS(\mu^2)$ is suitably defined, $\cI^{[1]}(\nu)$ matches the exponentiation in full QCD.

\item If $\cI^{[1]}(\nu)$ generates the desired exponentiation, then $\cI^{[k]}(\nu)$ for $k \ge 2$ should be small, so as not to destroy the desired exponentiation. 

\item For next-to-leading-log summation (NLL), this implies that $\cI^{[k]}_n(\nu)$ should not contain more than $n-1$ powers of $\log(\nu)$.

\item In one case examined in this paper, we can show analytically that $\cI^{[k]}_n(\nu)$ does not contain more than $n-1$ powers of $\log(\nu)$.

\item The operator $\cI^{[2]}_2(\nu)$ is of special interest. It should not contain more than one power of $\log(\nu)$.

\item In some cases, we can show analytically that $\cI^{[2]}_2(\nu)$ does not contain more than one power of $\log(\nu)$.

\item We can write the integral for $\cI^{[2]}_n(\nu)$ and evaluate it numerically to see if it contains more than $n-1$ powers of $\log(\nu)$.

\item For some shower algorithms examined here, $\cI^{[2]}_n(\nu)$ passes this test. For one algorithm examined, it fails.

\end{itemize}


\section{The thrust distribution and its Laplace transform}
\label{sec:laplace}

We will examine the distribution of thrust, $T$, defined for parton momenta $\{p\}_m$ by \cite{thrustdef1, thrustdef2}
\begin{equation}
\label{eq:thrustdef}
T = \max_{\vec n_\textsc{t}} \frac{\sum_i |\vec p_i \cdot \vec n_\textsc{t}|}{\sum_i |\vec p_i|}
= \frac{1}{\sqrt{Q^2}}\
\max_{\vec n_\textsc{t}} \sum_i |\vec p_i \cdot \vec n_\textsc{t}|
\;.
\end{equation}
The axis defined by the unit vector $\vec n_\textsc{t}$ that maximizes the sum is the {\em thrust axis}. We will be interested in the behavior of the thrust distribution for small values of 
\begin{equation}
\tau = 1 - T
\;.
\end{equation}

We can write $\tau$ in a useful form by defining sets $R$ and $L$ of partons by $\vec p_i \cdot \vec n_\textsc{t} > 0$ for $i \in R$ and $\vec p_i \cdot \vec n_\textsc{t} < 0$ for $i \in L$. Then, 
\begin{equation}
\label{eq:omT}
\tau 
= \frac{1}{\sqrt{Q^2}}
\left[ \sum_{i \in R} (E_i -\vec p_i \cdot \vec n_\textsc{t})
+ \sum_{i \in L} (E_i + \vec p_i \cdot \vec n_\textsc{t})
\right]
.
\end{equation}
Using the thrust axis, we define $\pm$ components of vectors by
\begin{equation}
\label{eq:plusminusdef}
p^\pm = \left[p_i^0 \pm \vec p \cdot \vec n_\textsc{t}\right]
/\sqrt{2}
\;.
\end{equation}
Then we can write
\begin{equation}
\label{eq:tauparts}
\tau = \tau_R + \tau_L
\;,
\end{equation}
where, using $Q^2 = 2 Q^+ Q^-$ with $Q^+ = Q^-$,
\begin{equation}
\begin{split}
\label{eq:tauLR}
\tau_R = \sum_{i \in R}\frac{p_i^-}{Q^-}
\;,
\quad\quad
\tau_L = \sum_{i \in L}\frac{p_i^+}{Q^+}
\;.
\end{split}
\end{equation}

In order to use a parton shower to analyze the thrust distribution, we begin with the cross section
\begin{equation}
\label{gdef}
g(\tau) = \frac{1}{\sigma_\scH}\frac{d\sigma}{d\tau}
\;,
\end{equation}
where $\sigma_\scH$ is the hard scattering cross section, equal to $d\sigma/d\tau$ integrated over $\tau$. We wish to analyze the small $\tau$ behavior of $g(\tau)$. For this purpose, it is standard to work with the Laplace transform of $g(\tau)$,
\begin{equation}
\label{eq:LaplaceTransform}
\tilde g(\nu) = \int_0^\infty\! d\tau\ e^{-\nu\tau} g(\tau)
\;.
\end{equation}
The coefficient of $\as^n$ in the perturbative expansion of $g(\tau)$ is not a normal function but is a distribution with $\log^{j-1}(\tau)/\tau$ singularities at $\tau = 0$. In order to work with normal functions, we define the integral of $g(\tau)$,
\begin{equation}
\label{eq:foftaudef}
f(\tau) = \int_0^\tau \! d\bar\tau\ g(\bar \tau)
\;.
\end{equation}
The coefficients in the perturbative expansion of $f(\tau)$ are functions with $\log^{j}(\tau)$ integrable singularities. The cross section $g(\bar \tau)$ vanishes for $\bar\tau > 1/2$, so $f(\tau) = 1$ for $\tau > 1/2$.

Consider the Laplace transform of $f(\tau)$:
\begin{equation}
\tilde f(\nu)=
\int_0^\infty\! d\tau\ e^{-\nu\tau} f(\tau)
\;.
\end{equation}
We have
\begin{equation}
\begin{split}
\tilde f(\nu)
={}& \int_0^\infty\! d\tau'\ e^{-\nu\tau'} 
\int_0^{\tau'}\! d\tau\ g(\tau)
\\
={}& \int_0^\infty\! d\tau\ g(\tau)
\int_{\tau}^\infty\! d\tau'\ e^{-\nu\tau'}
\\
={}& \frac{1}{\nu}
\int_0^\infty\! d\tau\ g(\tau)\,
e^{-\nu\tau}
\;.
\end{split}
\end{equation}
Thus
\begin{equation}
\tilde f(\nu)
=
\frac{\tilde g(\nu)}{\nu}
\;.
\end{equation}
The function $f(\tau)$ is given by the inverse Laplace transform of $\tilde f(\nu)$:
\begin{equation}
\label{eq:InverseLaplaceTransform}
f(\tau) = \frac{1}{2\pi\mi} \int_C\!  d\nu\ e^{\nu\tau}\,
\frac{\tilde g(\nu)}{\nu}
\;.
\end{equation}
The contour $C$ runs from $\nu_0 - \mi\infty$ to $\nu_0 + \mi\infty$ parallel to the imaginary $\nu$ axis, where $\nu_0 > 0$ so that the contour is to the right of the singularity of $\tilde g(\nu)/\nu$ at $\nu = 0$.

We expect the coefficient of $\as^n$ in the perturbative expansion of $f(\tau)$ to contain terms proportional to $\log^j(\tau)$ for $\tau \to 0$. To see how this translates to $\tilde g(\nu)$, we can start by noting that 
\begin{equation}
f(\tau) = \tau^A \implies 
\tilde g(\nu) = \Gamma(1+A)\, \nu^{- A}
\;.
\end{equation}
Thus
\begin{equation}
\begin{split}
f(\tau) ={}&
\sum_{j = 0}^\infty \frac{A^j}{j!}\,\log^{j}(\tau)
\implies
\\ 
\tilde g(\nu) ={}& \Gamma(1+A)
\sum_{j = 0}^\infty \frac{(-A)^j }{j!}\,\log^{j}(\nu)
\;.
\end{split}
\end{equation}
Matching powers of $A$, we learn that logarithms of $\tau$ for small $\tau$ translate into logarithms of $\nu$ for large $\nu$.

We wish to use the parton shower formalism to find an analytical formula that sums the logarithms of $\nu$ in $\tilde g(\nu)$. We can then compare what we find to the standard QCD formula that sums these logarithms. The final step needed to obtain something that can be compared to experiment would be to perform the inverse Laplace transform (\ref{eq:InverseLaplaceTransform}). This step is the same for the parton shower method or the normal analytical methods. We discuss this step only briefly in this paper.


\section{The measurement operator}
\label{sec:measurement}

If we want to measure the thrust distribution, then we define, following Eq.~(\ref{eq:sigmaJ}),
\begin{equation}
\begin{split}
\label{eq:sigmatau}
g(\tau) ={}& 
\frac{1}{\sigma_\scH}
\sbra{1} \delta(\tau - \tau_\mathrm{op})\, 
\cU(\mu_\Lf^2,\mu_\scH^2)\,
\sket{\rho_\scH}
\;,
\end{split}
\end{equation}
where $\tau$ is a real number times the unit operator on the statistical space and $\tau_\mathrm{op}$ is the operator defined by
\begin{equation}
\tau_\mathrm{op}\sket{\{p,f,c,c'\}_{m}}
=
\tau(\{p\}_{m})\sket{\{p,f,c,c'\}_{m}}
\;,
\end{equation}
where $\tau(\{p\}_{m})$ is $1-T$ for partons with momenta $\{p\}_{m}$, as defined in Eqs.~(\ref{eq:tauparts}) and (\ref{eq:tauLR}). Here $\sigma_\scH = \isbrax{1}\isket{\rho_\scH}$. This is the Born cross section for $e^+ e^ - \to q \bar q$ since, in this paper, we evaluate $\isket{\rho_\scH}$ at lowest order.

Rather than measuring $g(\tau)$, we wish to measure the Laplace transform $\tilde g(\nu)$. For this we have, using Eq.~(\ref{eq:LaplaceTransform}) in Eq.~(\ref{eq:sigmatau}),
\begin{equation}
\begin{split}
\label{eq:sigmanu}
\tilde g(\nu) ={}& 
\frac{1}{\sigma_\scH}
\sbra{1} \cO(\nu)\, 
\cU(\mu_\Lf^2,Q^2)\,
\sket{\rho_\scH}
\;,
\end{split}
\end{equation}
where
\begin{equation}
\label{cOnudef}
\cO(\nu) = e^{-\nu \tau_\mathrm{op}}
\;.
\end{equation}
We will analyze $\tilde g(\nu)$ in the subsequent sections. For this analysis, it is important that $\cO(\nu)$ has an inverse
\begin{equation}
\cO(\nu)^{-1} = e^{\nu \tau_\mathrm{op}}
\;.
\end{equation}
%
 

\section{Setting up the shower analysis}
\label{sec:showeranalysis}

Eq.~(\ref{eq:sigmanu}) allows us to calculate $\tilde g(\nu)$ numerically using the shower evolution operator $\cU(\mu_\Lf^2,\mu_\scH^2)$. We would now like to reformulate the shower result so that it takes an analytical form if, indeed, the shower generates $\tilde g(\nu)$ in a simple exponentiated form.


\subsection{The operators $\cY$ and $\cS_\cY$}
\label{sec:cY}

We begin with an operator $\cY(\mu^2;\nu)$, which is defined in Ref.~\cite{NSlogsum} using the all-order formalism of Ref.~\cite{NSAllOrder} for describing parton shower algorithms. The operator $\cY(\mu^2;\nu)$ is defined to have two properties. First, it does not change the number of partons or their momenta or flavors \cite{NSlogsum}. Second,
\begin{equation}
\label{eq:cYproperty}
\sbra{1}\cY(\mu^2;\nu)
= \sbra{1}\cO(\nu)\,\cU(\mu_\Lf^2,\mu^2)\,\cO^{-1}(\nu)
\;.
\end{equation}
These properties apply either for electron-positron annihilation or for hadron-hadron collisions. In this paper, we consider only electron-positron annihilation. Although $\cY(\mu^2;\nu)$ does not change the number of partons or their momenta or flavors \cite{NSlogsum}, it can change the parton colors. There is some freedom to define what $\cY(\mu^2;\nu)$ does to the parton color state. We will define the action of $\cY(\mu^2;\nu)$ on states in the statistical space in Eqs.~(\ref{eq:cScYk}) and (\ref{eq:cYk}) below.

The property Eq.~(\ref{eq:cYproperty}) can be written as
\begin{equation}
\label{eq:cYproperty2}
\sbra{1}\cO(\nu)\,\cU(\mu_\Lf^2,\mu^2)
=
\sbra{1}\cY(\mu^2;\nu)\cO(\nu)
\;.
\end{equation}
This result allows us to rewrite $\tilde g(\nu)$ as given by Eq.~(\ref{eq:sigmanu}) as
\begin{equation}
\begin{split}
\label{eq:sigmanuALT}
\tilde g(\nu) ={}& 
\frac{1}{\sigma_\scH}
\sbra{1}\cY(Q^2;\nu)\cO(\nu)
\sket{\rho_\scH}
\;.
\end{split}
\end{equation}
We see that instead of generating a complete parton shower as in Eq.~(\ref{eq:sigmanu}) and then measuring $\cO(\nu)$ for the resulting many parton state, we can measure $\cO(\nu)$ just on the hard state and then apply the operator $\cY(\mu^2;\nu)$ that depends on $\nu$ but leaves the number of partons unchanged.

How can one evaluate $\cY(\mu^2;\nu)$? We note first from the form of Eq.~(\ref{eq:cYproperty}),  that $\cY(\mu^2;\nu)$ has a perturbative expansion beginning with $\cY(\mu^2;\nu) = 1 + \cO(\as)$ and at $\mu^2 = \mu_\Lf^2$ it is exactly
\begin{equation}
\label{eq:cYendis1}
\cY(\mu_\Lf^2;\nu) = 1
\;.
\end{equation}

We define an infinitesimal generator $\cS_\cY(\mu^2;\nu)$ for $\cY(\mu^2;\nu)$ by
\begin{equation}
\label{eq:cScYdef}
\frac{1}{\mu^2 }\cS_\cY(\mu^2;\nu) =
\cY^{-1}(\mu^2;\nu)\,\frac{d}{d\mu^2 }\,\cY(\mu^2;\nu)
\;.
\end{equation}
Then $\cY(\mu^2;\nu)$ obeys the differential equation 
\begin{equation}
\label{eq:cYdiffeqn}
\mu^2 \frac{d}{d\mu^2 }\,\cY(\mu^2;\nu) = 
\cY(\mu^2;\nu)\,\cS_\cY(\mu^2;\nu)
\;,
\end{equation}
with boundary condition $\cY(\mu_\Lf^2;\nu) = 1$. We can use the notation
\begin{equation}
\label{eq:cYexponential}
\cY(\mu^2;\nu)
=\mathbb{T} \exp\!\left(
\int_{\mu_\Lf^2}^{\mu^{2}} \frac{d\bar \mu^2}{\bar\mu^2}\,
\cS_\cY(\bar \mu^2;\nu)
\right)
\end{equation}
to indicate the solution to Eq.~(\ref{eq:cYdiffeqn}). The instruction $\mathbb{T}$ indicates that the operators $\cS_\cY(\bar\mu^2; \nu)$ with the smallest values of $\bar\mu^2$ belong on the left. 

We will sometimes adopt the notation
\begin{equation}
\label{eq:Inudef}
\cI(\nu) = \int_{\mu_\Lf^2}^{Q^{2}}\!\frac{d\mu^2}{\mu^2}\,
\cS_\cY(\mu^2;\nu)
\end{equation}
when the upper integration limit is $Q^2$ and we do not need to explicitly display $\cS_\cY(\mu^2;\nu)$.\footnote{This is a useful definition even though $\cY(\mu^2;\nu)$ is not the exponential of $\cI(\nu)$ because of the $\mathbb{T}$ instruction in Eq.~(\ref{eq:cYexponential}).}


\subsection{Relation of $\cS_\cY$ to the shower generator $\cS$}
\label{sec:cScYandS}

We can relate $\cS_\cY(\mu^2,\nu)$ to $\cS(\mu^2)$. From Eq.~(\ref{eq:cUdiffeqn}), we have
\begin{equation}
\begin{split}
\label{eq:diffeqsUandY}
\mu^2 \frac{\partial}{\partial\mu^2 }\,\cU(\mu_\Lf^2,\mu^2) ={}& 
\cU(\mu_\Lf^2,\mu^2)\,\cS(\mu^2)
\;.
\end{split}
\end{equation}
Using Eqs.~(\ref{eq:diffeqsUandY}) and (\ref{eq:cYdiffeqn}) to differentiate Eq.~(\ref{eq:cYproperty}), we have
\begin{equation}
\begin{split}
\sbra{1}\cY(\mu^2;\nu)&\,\cS_\cY(\mu^2;\nu)
\\={}& 
\sbra{1}\cO(\nu)\,\cU(\mu_\Lf^2,\mu^2)\,\cS(\mu^2)\,\cO^{-1}(\nu)
\;.
\end{split}
\end{equation}
Using Eq.~(\ref{eq:cYproperty2}), this becomes
\begin{equation}
\begin{split}
\label{eq:cScYrecursively0}
\sbra{1}\cY(\mu^2;\nu)\,&\cS_\cY(\mu^2;\nu)
\\={}& 
\sbra{1}
\cY(\mu^2; \nu)\,
\cO(\nu)\,\cS(\mu^2)\,\cO^{-1}(\nu)
\;.
\end{split}
\end{equation}

We can also use Eq.~(\ref{eq:cYdiffeqn}), together with the boundary condition (\ref{eq:cYendis1}), to write an equation for $\cY(\mu^2;\nu)$,
\begin{equation}
\begin{split}
\label{eq:cYsoln}
\cY(\mu^2;\nu) ={}& 1
+ \int_{\mu_\Lf^2}^{\mu^2}\!\frac{d\bar \mu^2}{\bar \mu^2}\
\cY(\bar\mu^2;\nu)\,\cS_\cY(\bar\mu^2;\nu)
\;.
\end{split}
\end{equation}
%


\subsection{Operator mapping $\mathbb{P}$}
\label{sec:Pdef}

To use Eq.~(\ref{eq:cScYrecursively0}), we introduce some useful notation. Let $\cA$ be an operator that increases the number of partons or leaves the number of partons unchanged and changes momenta, flavors, and colors. Let $\cB$ be an operator on the statistical space that leaves the number $m$ of partons and their momenta and flavors $\{p,f\}_m$ unchanged, although it can change the parton color state.\footnote{In Ref.~\cite{NSlogsum}, $\cA$ is sometimes an operator that is defined in $d = 4 - 2 \epsilon$ dimensions that contains poles $1/\epsilon$ and singularities when the momenta of partons created by $\cA$ become soft or collinear. However, $\isbra{1}\cA$ is well defined in $d = 4$ dimensions. Then $\cB$ is well defined in 4 dimensions.}  Let $\cB$ be defined such that
\begin{equation}
\label{eq:Pdef1}
\sbra{1} \cB = \sbra{1} \cA
\;.
\end{equation}
We will define a linear relation $\cA \to \cB$ that realizes this relation. To represent this linear relation, we adopt the notation
\begin{equation}
\label{eq:Pdef2}
\cB = \P{\cA}
\;.
\end{equation}

The needed construction is straightforward. Suppose that $\cA$ maps states with $m$ partons into states with $\hat m$ partons, with $\hat m \ge m$. Let $\cA$ have the form
\begin{equation}
\label{eq:cBdef}
\cA = (C_\textrm{ket} \otimes C^\dagger_\textrm{bra})\,\cR
\;,
\end{equation}
where $\cR$ acts on the momentum and flavor factor of the statistical space and $(C_\textrm{ket} \otimes C^\dagger_\textrm{bra})$ acts on the color factor. Recall from Sec.~\ref{sec:deductor} the meaning of the color operators $(C_\textrm{ket} \otimes C^\dagger_\textrm{bra})$. Letting $\iket{\{c\}_m}$ and $\iket{\{c'\}_m}$ be basis vectors for color amplitudes, a color basis vector $\isket{\{c,c'\}_m}$ in the statistical space represents the color density operator $\iket{\{c\}_m}\ibra{\{c'\}_m}$.  Then $(C_\textrm{ket} \otimes C^\dagger_\textrm{bra})\isket{\{c,c'\}_m}$ represents the color density operator $C_\textrm{ket}\iket{\{c\}_m}\ibra{\{c'\}_m} C^\dagger_\textrm{bra}$.

Let us evaluate $\isbra{1}\cA \isket{\{p,f,c,c'\}_m}$ for an arbitrary $m$-parton basis state $\isket{\{p,f,c,c'\}_m}$. The inner product of $\isbra{1}$ with a statistical basis state is given in Eq.~(\ref{eq:1def}).  We insert a sum over the basis states \cite{NSI} with $\hat m$ partons,
\begin{equation}
\begin{split}
\sbra{1}\cA&\sket{\{p,f,c,c'\}_m}\\
 ={}&
\frac{1}{\hat m !}\int [d\{p,f\}_{\hat m}]\sum_{\{c,c'\}_{\hat m}}
\sbrax{1}\sket{\{\hat p,\hat f,\hat c,\hat c'\}_{\hat  m}}
\\ &\times
\sbra{\{\hat c,\hat c'\}_{\hat  m}}
C_\textrm{ket} \otimes C^\dagger_\textrm{bra}
\sket{\{c,c'\}_{m}}
\\&\times
\sbra{\{\hat p,\hat f\}_{\hat  m}}\cR\sket{\{p,f\}_{m}}
\;.
\end{split}
\end{equation}
For the color, this gives us the trace of the color density operator obtained by applying $C_\textrm{ket} \otimes C^\dagger_\textrm{bra}$ to $\isket{\{c,c'\}_{m}}$, namely the trace of $C_\textrm{ket}\iket{\{c\}_m}\ibra{\{c'\}_m} C^\dagger_\textrm{bra}$. The result is
\begin{equation}
\begin{split}
\sbra{1}\cA&\sket{\{p,f,c,c'\}_m}\\
 ={}&
\bra{\{c'\}_m}C^\dagger_\textrm{bra}C_\textrm{ket}\ket{\{c\}_m}
\\&\times
\frac{1}{\hat m !}\int [d\{p,f\}_{\hat m}]\
\sbra{\{\hat p,\hat f\}_{\hat  m}}\cR\sket{\{p,f\}_{m}}
\;.
\end{split}
\end{equation}

We now need to define $\cB = \iP{A}$ so that 
\begin{equation}
\sbra{1}\P{\cA} = \sbra{1} \cA
\;.
\end{equation}
We distinguish two cases. First, if $\hat m = m$ we leave the color operator in $\cA$ unchanged,
\begin{align}
\label{eq:Pcolorform1}
\P{(C_\textrm{ket} \otimes C^\dagger_\textrm{bra})\,\cR }
\sket{\{p,f,c,c'\}_{m}}
\hskip - 3.7 cm &
\\={}& 
(C_\textrm{ket} \otimes C^\dagger_\textrm{bra})
\sket{\{p,f,c,c'\}_{m}}
\notag
\\&\times
\frac{1}{m!}\int [d\{\hat p,\hat f\}_{m}]\
\sbra{\{\hat p,\hat f\}_{m}}\cR\sket{\{p,f\}_{m}}
\;.
\notag
\end{align}
Second, if $\hat m > m$ we define
\begin{align}
\label{eq:Pcolorform2}
\P{(C_\textrm{ket} \otimes C^\dagger_\textrm{bra})\,\cR }
\sket{\{p,f,c,c'\}_{m}}
\hskip - 3.7 cm &
\\={}& 
\frac{1}{2}
\left( C^\dagger_\textrm{bra} C_\textrm{ket}\otimes 1
+ 1 \otimes C^\dagger_\textrm{bra} C_\textrm{ket} \right)
\sket{\{p,f,c,c'\}_{m}}
\notag
\\&\times
\frac{1}{\hat m !}\int [d\{\hat p,\hat f\}_{\hat m}]\
\sbra{\{\hat p,\hat f\}_{\hat  m}}\cR\sket{\{p,f\}_{m}}
\;.
\notag
\end{align}
In either case, this satisfies $\isbra{1}\iP{\cA} = \isbra{1} \cA$.

There is a special case of some importance. Suppose that $\hat m =  m$ and, in addition, $\cA$ leaves the momenta and flavors of all partons unchanged. That is, $\isket{\{p,f\}_{m}}$ is an eigenvector of $\cR$:
\begin{equation}
\cR\sket{\{p,f\}_{m}} = r(\{p,f\}_{m}) \sket{\{p,f\}_{m}}
\;.
\end{equation}
Then $\cA$ applied to $\isket{\{p,f,c,c'\}_{m}}$ takes the form
\begin{equation}
\begin{split}
\cA \sket{\{p,f,c,c'\}_{m}} ={}& 
(C_\textrm{ket} \otimes C^\dagger_\textrm{bra})\,
r(\{p,f\}_{m})
\\&\times
\sket{\{p,f,c,c'\}_{m}}
\;.
\end{split}
\end{equation}
In this case, the definition (\ref{eq:Pcolorform1}) gives us
\begin{equation}
\label{eq:Bunchanged}
\P{\cA} = \cA
\;.
\end{equation}

There is some freedom available in fixing the color part of $\iP{(C_\textrm{ket} \otimes C^\dagger_\textrm{bra})\,\cR }$, as discussed in Sec.~VI D of Ref.~\cite{NSAllOrder}. We could add any operator $\cA'$ to $\iP{\cA}$ if $\cA'$ has the property that $\isbra{1}\cA' = 0$. The form in Eqs.~(\ref{eq:Pcolorform1}) and (\ref{eq:Pcolorform2}) is recommended by its simplicity, so we will use it in this paper.

This defines the operator $\iP{\cA}$ in general. However, when $\iP{\cA}$ acts on the $q \bar q$ initial hard scattering state in $e^+e^-$ annihilation, the action of $\iP{\cA}$ is simpler. The color space for $\isket{\{c,c'\}_{2}}$ has $\{c\}= \{c'\}$ and is one dimensional. Therefore, the color operator  in either of Eqs.~(\ref{eq:Pcolorform1}) and (\ref{eq:Pcolorform2}) acting on the $\isket{\{c,c'\}_{2}}$ state can only return an eigenvalue. That is, we have
\begin{equation}
\begin{split}
(C_\textrm{ket} \otimes C^\dagger_\textrm{bra})\sket{\{c,c\}_{2}}
={}&
\lambda_\mathrm{color}
\sket{\{c,c\}_{2}}
\end{split}
\end{equation}
or
\begin{equation}
\begin{split}
\frac{1}{2}
\Big( C^\dagger_\textrm{bra} C_\textrm{ket}\otimes 1
+  1 \otimes C^\dagger_\textrm{bra} C_\textrm{ket} \Big)&
\sket{\{c,c\}_{2}}
\\
={}&
\lambda_\mathrm{color}
\sket{\{c,c\}_{2}}
\;.
\end{split}
\end{equation}
This tells us that $\isket{\{p,f,c,c\}_{2}}$ is an eigenvector of $\iP{\cA}$:
\begin{equation}
\label{eq:cBonccbar}
\P{\cA}\sket{\{p,f,c,c\}_{2}}
= \lambda_\cA \sket{\{p,f,c,c\}_{2}}
\;,
\end{equation}
where
\begin{equation}
\lambda_\cA = 
\lambda_\mathrm{color}\,
\int [d\{\hat p,\hat f\}_{2}]\
\sbra{\{\hat p,\hat f\}_{2}}\cR\sket{\{p,f\}_{2}}
\;.
\end{equation}
Using $\isbrax{1}\isket{\{p,f,c,c\}_{2}} = \ibrax{\{c\}_2}\iket{\{c\}_2}$ and $\ibrax{\{c\}_2}\iket{\{c\}_2} = 1$ \cite{NSI}, we have a very simple result for the eigenvalue,
\begin{equation}
\label{eq:cBeigenvalue}
\lambda_\cA = \sbra{1}\cA\sket{\{p,f,c,c\}_{2}}
\;.
\end{equation}
%


\subsection{Recursive definition of $\cS_\cY$}
\label{sec:cScYrecursion}

We can now define $\cS_\cY(\mu^2;\nu)$ so that it satisfies Eq.~(\ref{eq:cScYrecursively0}). Recall that $\cY(\mu^2;\nu) = 1 + \cO(\as)$. Because of this, it is possible to isolate $\cS_\cY(\mu^2;\nu)$ on the left hand side of Eq.~(\ref{eq:cScYrecursively0}):
\begin{equation}
\begin{split}
\label{eq:cScYrecursively}
\sbra{1}\cS_\cY(\mu^2;\nu)
={}& 
\sbra{1}
\Big\{
\cY(\mu^2; \nu)\,
\cO(\nu)\,\cS(\mu^2)\,\cO^{-1}(\nu)
\\ &
+ \left[1-\cY(\mu^2;\nu)\right]\cS_\cY(\mu^2;\nu)\Big\}
\;.
\end{split}
\end{equation}
Using the operator mapping $\P{\cdots}$, this is
\begin{equation}
\begin{split}
\label{eq:cScYrecursion}
\cS_\cY(\mu^2;\nu)
={}& 
\P{
\cY(\mu^2; \nu)\,
\cO(\nu)\,\cS(\mu^2)\,\cO^{-1}(\nu)
}
\\ &
+ \P{\left[1-\cY(\mu^2;\nu)\right]\cS_\cY(\mu^2;\nu)}
\;.
\end{split}
\end{equation}
Note that the operators $\cY(\mu^2;\nu)$ and $\cS_\cY(\mu^2;\nu)$ in the second line of Eq.~(\ref{eq:cScYrecursion}) leave the number of partons, their momenta, and their flavors unchanged. Thus Eq.~(\ref{eq:Bunchanged}) applies and the $\iP{\cdots}$ operation has no effect. 

Equation~(\ref{eq:cScYrecursion}) can be used to define $\cS_\cY(\mu^2;\nu)$ and $\cY(\mu^2;\nu)$ recursively. We can write $\cS_\cY(\mu^2;\nu)$, $\cY(\mu^2;\nu)$, and $\cI(\nu)$ as expansions in powers of the shower evolution operator $\cS$:
\begin{equation}
\begin{split}
\label{eq:cIkexpansion}
\cS_\cY(\mu^2;\nu) ={}& \sum_{k=1}^\infty \cS_\cY^{[k]}(\mu^2;\nu)
\;,
\\
\cY(\mu^2;\nu) ={}& 1 + \sum_{k=1}^\infty \cY^{[k]}(\mu^2;\nu)
\;,
\\
\cI(\nu) ={}& \sum_{k=1}^\infty \cI^{[k]}(\nu)
\;,
\end{split}
\end{equation}
where each of $\cS_\cY^{[k]}(\mu^2;\nu)$, $\cY^{[k]}(\mu^2;\nu)$, and $\cI^{[k]}(\nu)$ contain $k$ factors of $\cS$. Then we can write Eq.~(\ref{eq:cScYrecursion}) as
\begin{equation}
\begin{split}
\label{eq:cScYk}
\cS_\cY^{[k]}(\mug; \nu) 
= {}&
\P{
\cY^{[k-1]}(\mug; \nu)\,
\cO(\nu)
\cS(\mug)
\cO^{-1}(\nu)}
\\&-
\sum_{j=1}^{k-1}
\cY^{[k-j]}(\mug; \nu)\,
\cS_\cY^{[j]}(\mug; \nu) 
\;.
\end{split}
\end{equation}
Similarly, we can write Eq.~(\ref{eq:cYsoln}) as
\begin{equation}
\label{eq:cYk}
\cY^{[k]}(\mug; \nu)
=  
\sum_{j=1}^{k}
\int_{\muf}^{\mug}\!\frac{d\bar\mu^2}{\bar\mu^2}\,
\cY^{[k-j]}(\bar\mu^2; \nu)\,
\cS_\cY^{[j]}(\bar\mu^2; \nu)
.
\end{equation}
These equations apply for $k = 1,2,\dots$ with $\cY^{[0]}(\mug; \nu) = 1$. 

We now illustrate this for the first two orders. At order 1, Eq.~(\ref{eq:cScYk}) gives us
\begin{equation}
\begin{split}
\label{eq:cScY1soln}
\cS_\cY^{[1]}(\mu^2;\nu)
={}& 
\P{\cO(\nu)\,\cS(\mu^2)\,\cO^{-1}(\nu)}
\;.
\end{split}
\end{equation}
At order $\as^2$, we have
\begin{equation}
\begin{split}
\label{eq:cScY2soln1}
\cS_\cY^{[2]}(\mu^2;\nu)
={}& 
\P{
\cY^{[1]}(\mu^2;\nu)\,
\cO(\nu)\,\cS(\mu^2)\,\cO^{-1}(\nu)
}
\\& - 
\cY^{[1]}(\mu^2;\nu)\,
\cS_\cY^{[1]}(\mu^2;\nu)
\;.
\end{split}
\end{equation}
From Eq.~(\ref{eq:cYsoln}) at first order, we have
\begin{equation}
\label{eq:cY1}
\cY^{[1]}(\mu^2;\nu) =
\int_{\mu_\Lf^2}^{\mu^2}\!\frac{d\bar\mu^2}{\bar\mu^2}\
\cS_\cY^{[1]}(\bar\mu^2;\nu)
\;.
\end{equation}
For $\cS_\cY^{[1]}(\mu^2;\nu)$ we can use Eq.~(\ref{eq:cScY1soln}). This gives us
\begin{equation}
\begin{split}
\label{eq:cScY2soln}
\cS_\cY^{[2]}(\mu^2;\nu)
={}& 
\int_{\mu_\Lf^2}^{\mu^2}\!\frac{d\bar\mu^2}{\bar\mu^2}\
\PL \P{\cO(\nu)\,\cS(\bar\mu^2)\,\cO^{-1}(\nu)}
\\&\quad\quad\times
\omP{\cO(\nu)\,\cS(\mu^2)\,\cO^{-1}(\nu)}
\PR
\;.
\end{split}
\end{equation}
Here we use the abbreviation
\begin{equation}
\label{eq:omPdef}
\omP{\cA} = \cA - \P{\cA}
\;.
\end{equation}

The operator $\cS_\cY(\mu^2;\nu)$ is a complicated operator in general. However, it is significant that, because of Eqs.~(\ref{eq:cBonccbar}) and (\ref{eq:cBeigenvalue}), the initial $q\bar q$ state is an eigenvector of $\cS_\cY(\mu^2;\nu)$:
\begin{equation}
\label{eq:cScYeigenvector}
\cS_\cY(\mu^2;\nu)\sket{\{p,f,c,c\}_{2}}
= \lambda_{\cS_\cY}
\sket{\{p,f,c,c\}_{2}}
\;,
\end{equation}
where
\begin{equation}
\label{eq:cScYeigenvalue}
\lambda_{\cS_\cY} = \sbra{1}\cS_\cY(\mu^2;\nu)\sket{\{p,f,c,c\}_{2}}
\;.
\end{equation}
%


\section{Evaluation of $\cS_\cY^{[1]}(\mu^2;\nu)$}
\label{sec:cScY1analysis}

Let us see what we can say about $\cS_\cY^{[1]}(\mu^2; \nu)$ as given in Eq.~(\ref{eq:cScY1soln}). In a first order shower, like $\textsc{Deductor}$, we divide $S(\mu^2)$ into its real emission and virtual parts as in Eq.~(\ref{eq:cS1decomposition}). Then Eq.~(\ref{eq:cScY1soln}) gives us
\begin{equation}
\begin{split}
\label{eq:cVS1parts1}
\cS_\cY^{[1]}(\mu^2; \nu) ={}&
\PL \cO(\nu)\,
\cS^{[1,0]}(\mu^2)\,
\cO^{-1}(\nu)
\\&
+ \cO(\nu)\,
\cS^{[0,1]}(\mu^2)\,
\cO^{-1}(\nu) \PR
\;.
\end{split}
\end{equation}
The virtual operator $\cS^{[0,1]}(\mu^2; 0)$ leaves the momentum and flavor state unchanged, so this is
\begin{equation}
\begin{split}
\label{eq:cVS1parts2}
\cS_\cY^{[1]}(\mu^2;& \nu) 
\\={}&
\PL \cO(\nu)\,
\cS^{[1,0]}(\mu^2)\,
\cO^{-1}(\nu)
+ \cS^{[0,1]}(\mu^2)\,
\PR
\;.
\end{split}
\end{equation}
Recall from  Eq.~(\ref{eq:1cS}) that $\isbra{1}\cS^{[0,1]}(\mu^2) = - \isbra{1}\cS^{[1,0]}(\mu^2)$. This tells us that
\begin{equation}
\label{eq:cS1realvirtual}
\P{\cS^{[0,1]}(\mu^2)}
= - \P{\cS^{[1,0]}(\mu^2)}
\;.
\end{equation}
Using Eq.~(\ref{eq:cS1realvirtual}), Eq.~(\ref{eq:cVS1parts2}) becomes
\begin{equation}
\begin{split}
\label{eq:cVS1parts3}
\cS_\cY^{[1]}(\mu^2;& \nu) \\={}&
\PL \cO(\nu)\,
\cS^{[1,0]}(\mu^2)\,
\cO^{-1}(\nu)
- \cS^{[1,0]}(\mu^2)\,
\PR
\;.
\end{split}
\end{equation}
This is a convenient form for calculations.


\section{Change in $\tau$ induced by a splitting}
\label{sec:changeintau}

The operator $\cO(\nu)\,\cS^{[1,0]}(\mu^2)\,\cO^{-1}(\nu)$ appears in Eq.~(\ref{eq:cVS1parts3}) for $\cS_\cY^{[1]}(\mu^2;\nu)$. This operator is
\begin{equation}
\begin{split}
\cO(\nu)\,
\cS^{[1,0]}(\mu^2)\,
\cO^{-1}(\nu) ={}& 
e^{-\nu \tau_\textrm{op}}
\cS^{[1,0]}(\mu^2)
e^{+\nu \tau_\textrm{op}}
\;.
\end{split}
\end{equation}

The operator $\cS^{[1,0]}(\mu^2)$ is a sum of operators,
\begin{equation}
\label{eq:cSlsum}
\cS^{[1,0]}(\mu^2) = \sum_{l=1}^\infty \cS_l^{[1,0]}(\mu^2)
\;,
\end{equation}
where $l$ is the label of the parton that splits. When we apply $\cS_l^{[1,0]}(\mu^2)$ to a state $\isket{\{p,f,c,c'\}_m}$, the splitting operator creates a new state $\isket{\{\hat p,\hat f,\hat c,\hat c'\}_{m+1}}$ as long as $l \le m$. For $l>m$, $\cS_l$ just gives zero. The operators $\tau_\textrm{op}$ measure the values of $\tau$ before and after the splitting. Thus
\begin{equation}
\begin{split}
\sbra{\{\hat p,\hat f,\hat c,\hat c'\}_{m+1}}
\cO(\nu)\, 
\cS_l^{[1,0]}(\mu^2)\,
\cO^{-1}(\nu)\sket{\{p,f,c,c'\}_m} 
\hskip -7.2 cm &
\\={}& 
e^{-\nu(\hat \tau - \tau)}
\sbra{\{\hat p,\hat f,\hat c,\hat c'\}_{m+1}}
\cS_l^{[1,0]}(\mu^2)\sket{\{p,f,c,c'\}_m}
\;,
\end{split}
\end{equation}
where $\tau = \tau(\{p\}_m)$ and $\hat\tau = \tau(\{\hat p\}_{m+1})$. Thus we need to know how $\tau$ changes in a splitting. We are looking for the leading contributions to logarithms of $\nu$, so we can use the approximations that $\tau$ is small and that the splitting is nearly soft or collinear.

We start with momenta $\{p\}_m$ and suppose that the parton that splits is in the right thrust hemisphere, $l \in R$. The splitting produces a new parton $l$ and a parton $m+1$. After the splitting, we have partons with momenta $\{\hat p\}_{m+1}$. 

The emission of a parton changes the thrust axis. However, in the case of a nearly soft or collinear splitting of a parton in a state with small $\tau$, the thrust axis changes by very little. For this reason, we calculate $\tau(\{\hat p\}_{m+1})$ for the new parton state using the thrust axis of the old parton state $\{p\}_{m}$. We also assume that after the splitting partons $l$ and $m+1$ are still in the right thrust hemisphere.

Now we turn to the calculation of $\hat\tau - \tau $. We use the definition, Eqs.~(\ref{eq:tauparts}) and  (\ref{eq:tauLR}), to write
\begin{equation}
\begin{split}
\label{eq:deltatau1}
\hat \tau - \tau ={}&
\frac{\hat p_l^- + \hat p_{m+1}^- - p_l^-}{Q^-}
\\&
+ \sum_{\substack{i \in R \\ i \ne \{l, m+1\}}}\frac{\hat p_i^- - p_i^-}{Q^-}
+ \sum_{i \in L}\frac{\hat p_i^+ - p_i^+}{Q^+}
\;.
\end{split}
\end{equation}

Now we need to evaluate $({\hat p_l^- + \hat p_{m+1}^- - p_i^-})/{Q^-}$ and $(\hat p_i^\pm - p_i^\pm)/Q^\pm$. Following the notation of Appendix B of Ref.~\cite{NSThreshold}, we define 
\begin{equation}
\begin{split}
\label{eq:splittingparameters}
h_\pm ={}& (1 + y \pm \lambda)/2
\;,
\\
\lambda ={}& \sqrt{(1+y)^2 - 4 a_l y}
\;,
\\
a_l ={}& \frac{Q^2}{2 p_l\cdot Q}
\;,
\end{split}
\end{equation}
where $y$ was defined in Eq.~(\ref{eq:ydef}). We suppose that $y \ll 1$. We define a lightlike vector $n_l$ by
\begin{equation}
\label{eq:nldef}
n_l  = \frac{2 p_l\cdot Q}{Q^2}\,Q - p_l
\;.
\end{equation}
Note that $n_l$ is independent of the normalization of $Q$. 

We write the momentum vectors for partons $l$ and $m+1$ after the splitting as
\begin{equation}
\begin{split}
\label{eq:hatplpmp1}
\hat p_l ={}& h_+ z\, p_l + h_- (1-z) n_l + k_\perp
\;,
\\
\hat p_{m+1} ={}& h_+ (1-z) p_l + h_- z\, n_l - k_\perp
\;,
\end{split}
\end{equation}
where $k_\perp \cdot p_l = k_\perp \cdot n_l = 0$. The splitting is specified by $y$, the momentum fraction $z$ in Eq.~(\ref{eq:hatplpmp1}), and the azimuthal angle $\phi$ of $k_\perp$. The magnitude of $k_\perp$ is determined by the condition $\hat p_l^2 = 0$ or $\hat p_{m+1}^2 = 0$:
\begin{equation}
\label{eq:ktfromyz}
- k_\perp^2 = z (1-z) y\, 2 p_l\cdot Q
\;.
\end{equation}

Define
\begin{equation}
P_l = \hat p_l + \hat p_{m+1} 
= h_+ p_l + h_- n_l
\;.
\end{equation}
This gives us $P_l^2 = 2 p_l \cdot  Q\, y$. Using these results we obtain 
\begin{equation}
\label{eq:massconservation}
(Q - p_l)^2 = (Q - P_l)^2
\;.
\end{equation}
We require that momentum be conserved in the splitting, so that
\begin{equation}
Q - p_l = \sum_{\substack{i=1 \\ i \ne l}}^m p_i \;,\qquad
Q - P_l = \sum_{\substack{i=1 \\ i \ne l}}^m \hat p_i \;.
\end{equation}
The relation (\ref{eq:massconservation}) allows the $\hat p_i$ for $i \notin \{l,m+1\}$ to be obtained from the $p_i$ by a Lorentz transformation,
\begin{equation}
\label{eq:lorentztransformation}
\hat p_i^\mu = \Lambda^\mu_\nu\, p_i^\nu \;,
\hskip 1 cm 
i \notin \{l,\mpone\}
\;.
\end{equation}
The needed Lorentz transformation can be a small boost in the $p_l$-$Q$ plane. Let
\begin{equation}
p_i = \alpha_i p_l + \beta_i n_l + p_{i,\perp}
\;,
\end{equation}
where  $p_{i,\perp}\cdot p_l = p_{i,\perp}\cdot n_l = 0$. Then define $\hat p_i$ for $i \notin \{l,m+1\}$ by
\begin{equation}
\hat p_i = e^\omega \alpha_i p_l + e^{-\omega} \beta_i n_l + p_{i,\perp}
\;.
\end{equation}
The needed boost angle is small:
\begin{equation}
\label{eq:omegafromy}
\omega = y + \cO(y^2)
\;.
\end{equation}

Using Eq.~(\ref{eq:lorentztransformation}) in Eq.~(\ref{eq:deltatau1}), we have
\begin{equation}
\begin{split}
\label{eq:deltatau2}
\hat \tau& - \tau =
\frac{\hat p_l^- + \hat p_{m+1}^- - p_l^-}{Q^-}
\\&
+ \sum_{\substack{i \in R \\ i \ne l}}
(\Lambda^-_\nu - \delta^-_\nu) \frac{p_i^\nu}{Q^-}
+ \sum_{i \in L}
(\Lambda^+_\nu - \delta^+_\nu) \frac{p_i^\nu}{Q^+}
\;.
\end{split}
\end{equation}
We will see momentarily that $(\hat p_l^- + \hat p_{m+1}^- - p_l^-)/Q^-$ is small, of order $y$. This allows $\hat \tau - \tau$ to be of order $y$. 

In the third term, for $i \in \LR$, $(\Lambda^\mu_\nu - \delta^\mu_\nu)$ is of order $y$. The thrust axis defines the $\pm$ components of vectors in Eq.~(\ref{eq:deltatau2}). If $p_l$ were exactly aligned with the thrust axis, then the only nonvanishing index choice for $\Lambda^-_\nu$ would be $\nu = -$. But $p_i^-/Q^- \ll 1$ for $i\in \LR$, since this quantity is of order $\tau$ and we suppose that $\tau \ll 1$. This restriction on the index choices is not exact. However, for $i\in \LR$, the components $p_i^\nu/Q^-$ for $\nu \in \{1,2\}$ are of order $p_i^\nu/Q^- \sim [{p_i^+ p_i^-}]^{1/2}/Q^-$, which is at most of order $\sqrt{\tau}$. The component $p_i^\nu/Q^-$ for $\nu = +$ can be of order 1. However, $\Lambda^-_+ = \Lambda^{--}$ is at most of order $y^2$ since $\Lambda = \exp(\omega w)$ where $\omega$ is given by Eq.~(\ref{eq:omegafromy}) and the first order contribution to $\Lambda^{--}$ vanishes because the generator matrix $w^{\mu\nu}$ is antisymmetric. Thus the second term in Eq.~(\ref{eq:deltatau2}) is of order $y$ times a small factor, either $\tau$, $\sqrt{\tau}$, or $y$. The same reasoning applies to the third term.

We conclude that the only surviving term in Eq.~(\ref{eq:deltatau2}) is the first:
\begin{equation}
\begin{split}
\label{eq:deltatau3}
\hat \tau - \tau \approx{}&
\frac{\hat p_l^- + \hat p_{m+1}^- - p_l^-}{Q^-}
\;.
\end{split}
\end{equation}
We have
\begin{equation}
\begin{split}
\frac{\hat p_l^- + \hat p_{m+1}^- - p_l^-}{Q^-} \approx{}& 
\frac{(1-a_l)y p_l^- + a_l y  n_l^-}{Q^-} 
\;.
\end{split}
\end{equation}
With our kinematic conventions,
\begin{equation}
\begin{split}
\frac{p_l^- }{Q^-} ={}& \frac{1 - \cos\theta(l,\vec n_\LT)}{2 a_l}
\;,
\\
\frac{n_l^- }{Q^-} ={}& \frac{1 + \cos\theta(l,\vec n_\LT)}{2 a_l}
\;,
\end{split}
\end{equation}
where 
\begin{equation}
\cos\theta(l,\vec n_\LT) = \frac{|\vec p_l\cdot \vec n_\textsc{t}|}{|\vec p_l|}
\;.
\end{equation}
This gives us
\begin{equation}
\label{eq:taulresult}
\frac{\hat p_l + \hat p_{m+1}^- - p_l^-}{Q^-} \approx 
\xi_l\, y
\;,
\end{equation}
where
\begin{equation}
\label{eq:cldef}
\xi_l 
= 1 - \left(1 - \frac{1}{2 a_l}\right)[1 - \cos(\theta(l,\vec n_\LT))]
\;.
\end{equation}
That is
\begin{equation}
\label{eq:tau2result}
\hat\tau - \tau \approx 
\xi_l\, y 
\;.
\end{equation}
The same result holds for $l \in L$ if we change $1 - \cos(\theta(l,\vec n_\LT))$ to $1 + \cos(\theta(l,\vec n_\LT))$.

If we are splitting the quark or the antiquark in the two parton state created initially in $e^+e^-$ annihilation, then $a_l = 1$ and $\theta(l,\vec n_\LT) = 0$. Then $\xi_l = 1$. 

In the general case, $0 < 1 - \cos(\theta(l,\vec n_\LT)) < 1$ for $l \in R$ and $1/2 < (2 a_l - 1)/(2 a_l) < 1$, so 
\begin{equation}
 0 < \xi_l < 1
 \;.
\end{equation}
We get $\xi_l \to 0$ only when $\theta_l \to \pi/2$ and parton $l$ is very soft, $1/a_l \to 0$. Notice that there is no singularity for $\theta_l \to \pi/2$, so there is no singularity for $\xi_l \to 0$. There is a singularity for $\theta(l,\vec n_\LT) \to 0$ for all partons $l$. This corresponds to $\xi_l \to 1$. Thus in the general case we can treat $\xi_l$ as being close to 1. We will argue in Appendix \ref{sec:NLLproof} that for the purpose of finding next-to-leading logarithms of $\nu$ we can simply set $\xi_l$ to 1.

We conclude that the effect of the operators $\cO(\nu)$ in a splitting of parton $l$ can be approximated by
\begin{equation}
\begin{split}
\label{eq:cSlapproximation}
\cO(\nu)\,
\cS^{[1,0]}_l(\mu^2)&\,
\cO^{-1}(\nu)
\sket{\{p,f,c,c'\}_m}
\\
\approx{}&
\cS^{[1,0]}_l(\mu^2)\,
e^{-\nu \xi_l^\mathrm{op} y}
\sket{\{p,f,c,c'\}_m}
\;,
\end{split}
\end{equation}
where $\xi_l^\mathrm{op}$ is an operator that, acting on a state $\isket{\{p,f,c,c'\}_m}$, has eigenvalue $\xi_l$ as defined in Eq.~(\ref{eq:cldef}) as long as $l \le m$. For $l > m$, we can simply define $\xi_l$ to have eigenvalue 1. We recall that $\xi_l$ is generally of order 1 and equals 1 exactly in the case of a splitting of one of the partons in a two parton state. Using this in Eq.~(\ref{eq:cVS1parts3}) gives us
\begin{equation}
\begin{split}
\label{eq:cVS1parts4}
\cS_\cY^{[1]}&(\mu^2; \nu)
\approx
-\sum_l
\P{\cS_l^{[1,0]}(\mu^2)}
(1 - e^{-\nu \xi_l^\mathrm{op} y})
\;.
\end{split}
\end{equation}
%


\section{$\cS_\cY^{[1]}$ for a quark-antiquark state}
\label{sec:cScY1}

For the $q\bar q$ state created initially in electron-positron annihilation, Eq.~(\ref{eq:cVS1parts4}) simplifies considerably. First, the index $l$ denoting the parton that splits can take only the values $l=1$ (for the quark) and $l=2$ (for the antiquark). Each choice gives the same result, so we can take $l=1$ and multiply by two. Also, the color factors are trivial. In $\iP{\cS_l^{[1,0]}(\mu^2)}$ we encounter color operators $\bm T_1\cdot \bm T_1$, $\bm T_2\cdot \bm T_2$, and $\bm T_1\cdot \bm T_2$, where $\bm T_i\cdot \bm T_j = \sum_a T_i^a T_j^a$ and $T_i^a$ inserts a color matrix $T^a$ on parton line $i$.  The operators $\bm T_1\cdot \bm T_1$ and $\bm T_2\cdot \bm T_2$ simply give an eigenvalue $C_\LF$ times the unit color operator, while $\bm T_1\cdot \bm T_2$ gives $-C_\LF$. This gives us a result of the form
\begin{equation}
\begin{split}
\label{eq:cVS1m2}
\cS_\cY^{[1]}(\mu^2;& \nu)\sket{\{p,f,c,c'\}_2} 
\\\approx{}&
-(1 - e^{- \nu y})\,
\lambda(y)
\sket{\{p,f,c,c'\}_2}
\;.
\end{split}
\end{equation}
The eigenvalue $\lambda(y)$ is obtained in a straightforward calculation from the $q \to q + \Lg$ splitting functions used in \textsc{Deductor} \cite{NScolor}. There is an integral over the splitting variables $z$ and $\phi$. The $\phi$ integral is trivial and gives simply a factor $2\pi$. The integration over the momentum fraction $z$ remains,
\begin{equation}
\begin{split}
\label{eq:lambday}
\lambda(y) ={}& 2 C_\LF \int_0^1\!dz\ 
\Big\{
\frac{\alpha_s(\lambda_\LR (1-z) y Q^2)}{2\pi}\
f_\mathrm{sing}(z,y)
\\&\quad
+
\frac{\alpha_s(\lambda_\LR y Q^2)}{2\pi}\
f_\mathrm{reg}(z,y)
\Big\}
\;.
\end{split}
\end{equation}

The argument of $\as$ contains the standard factor $\lambda_\LR$, Eq.~(\ref{eq:CMW}), and, in the first term, a factor $(1-z)$, as in Eq.~(\ref{eq:S10}) with $a_l = 1$. The functions $f_\mathrm{sing}(z,y)$ and $f_\mathrm{reg}(z,y)$ are taken directly from \textsc{Deductor} and are quite complicated. However, they are simple in the relevant limits, $y \to 0$ with fixed $z$ and $y \to 0$ with $1-z \propto y$. In these limits, they are
\begin{equation}
\begin{split}
\label{eq:fzylimits}
f_\mathrm{sing}(z,y) \approx{}& \frac{2}{1 - z + y} -2
\;,
\\ 
f_\mathrm{reg}(z,y) \approx{}&  1-z
\;.
\end{split}
\end{equation}
Note that $f_\mathrm{sing}(z,0) + f_\mathrm{reg}(z) = (1+z^2)/(1-z)$ is just the DGLAP splitting kernel for $q \to q + \Lg$. However in $f_\mathrm{sing}(z,y)$ the singularity at $(1-z) \to 0$ is regulated by adding $y$ in the denominator.

We have written these results in the form used in \textsc{Deductor}. In $f_\mathrm{sing}(z,y)$, we could recognize that the second term could have been transferred to $f_\mathrm{reg}(z,y)$.

We would now like to compare this to the standard results for the summation of logs of $\tau$ in Ref.~\cite{thrustsum}. We begin by inserting Eq.~(\ref{eq:fzylimits}) into Eq.~(\ref{eq:lambday}): 

\begin{equation}
\begin{split}
\label{eq:lambday1}
\lambda(y) \approx{}& 2 C_\LF \int_0^1\!dz\ 
\Big\{
\frac{\alpha_s(\lambda_\LR (1-z) y Q^2)}{2\pi}\
\frac{2}{1-z+y}
\\ &\quad
-\frac{\alpha_s(\lambda_\LR (1-z) y Q^2)}{2\pi}\
2
\\&\quad
+
\frac{\alpha_s(\lambda_\LR y Q^2)}{2\pi}\
(1-z)
\Big\}
\;.
\end{split}
\end{equation}
We will want to evaluate this approximately for small $y$ in such a way that if we expand the result in powers of $\alpha_s(y Q^2)$ we retain all terms proportional to $\alpha_s^n(y Q^2) \log^{n}(y)$ and $\alpha_s^n(y Q^2) \log^{n-1}(y)$. After integrating over $\bar \mu^2 = y Q^2$ as in Eq.~(\ref{eq:cYexponential}), this will give contributions $\alpha_s^n(Q^2) \log^{n+1}(\nu)$ and $\alpha_s^n(Q^2) \log^{n}(\nu)$. These are the leading log (LL) and next-to-leading log (NLL) terms. In $\lambda(y)$, we neglect contributions proportional to fewer powers of $\log(y)$ or to powers of $y$. 

In order to carry out this approximate evaluation, we note first that we can use 
\begin{equation}
\label{eq:asscalechange}
\as(A \mu^2) = \as(\mu^2) - \beta_0 \log(A)\, \as^2(\mu^2) + \cO(\as^3)
\;,
\end{equation}
where $\beta_0 = (11 C_\LA - 2 n_\Lf)/(12 \pi)$. Then we can omit the $\lambda_\LR(1-z)$ factor in the argument of $\as$ in the second term in Eq.~(\ref{eq:lambday1}) and the $\lambda_\LR$ in the third term, since these terms do not have $1/(1-z+y)$ singularities that could produce $\log(y)$ factors after integration.  In the first term, there is a $1/(1-z+y)$ singularity. For this term, we need to keep the $\as^2$ contribution in Eq.~(\ref{eq:asscalechange}). After performing the $z$ integration in the last two terms, this gives us
\begin{equation}
\begin{split}
\label{eq:lambday2}
\lambda(y) \approx{}& 4 C_\LF \int_0^1\!d(1-z)\ 
\frac{1}{1-z+y}
\\&\times
\frac{\alpha_s((1-z) y Q^2) 
-\beta_0 \log(\lambda_\LR)\, \alpha_s^2((1-z) y Q^2)}{2\pi}\
\\&\quad
- 3 C_\LF\,
\frac{\alpha_s(y Q^2)}{2\pi}\
\;.
\end{split}
\end{equation}
Now we note that the $y$ in the denominator in the first term of Eq.~(\ref{eq:lambday2}) places an effective lower cutoff on $(1-z)$ at about $(1-z) = y$. This observation suggests that the integration over $(1-z)$ can be written in a simpler form:
\begin{align}
\label{eq:lambday3}
\notag
\lambda(y) \approx{}& 4 C_\LF \int_y^1\!
\frac{d(1-z)}{1-z}
\\&\times \notag
\frac{\alpha_s((1-z) y Q^2) 
-\beta_0 \log(\lambda_\LR)\, \alpha_s^2((1-z) y Q^2)}{2\pi}\
\\&\quad
- 3 C_\LF\,
\frac{\alpha_s(y Q^2)}{2\pi}\
\;.
\end{align}
In fact, this correctly reproduces the $\as^n \log^n(y)$ terms and the $\as^n \log^{n-1}(y)$ terms in the expansion of the integral. To see this, one can approximately solve the renormalization group equation for $\as$ in the form \cite{SurguladzeSoper}
\begin{equation}
\begin{split}
\label{eq:asrunning}
\frac{1}{\as(A \mug)}
={}& 
\frac{1 + \beta_0 \log(A)\,\as(\mug) }{\as(\mug)}
\\ &
+ \frac{\beta_1}{\beta_0}\,\log\big(1 + \beta_0 \log(A)\,\as(\mug)\big)
\\&
+ \cdots \;,
\end{split}
\end{equation}
with $\mug = yQ^2$ and $A = 1-z$. Here $\beta_1 = (17 C_\LA^2 - 5 C_\LA n_\Lf - 3 C_\LF n_\Lf)/(24 \pi^2)$. This yields $\as(A\mug)$ as a series
\begin{equation}
\begin{split}
\as(A\mug) ={}& \as(\mug)\Bigg\{ 1 + 
\sum_{n=2}^\infty \as^n(\mug) 
\big[c_n\log^n(A) 
\\ & 
+ d_n \log^{n-1}(A) 
+ \cdots \big]
\Bigg\}
\;.
\end{split}
\end{equation}
Then one can check that the integral  (\ref{eq:lambday2}) agrees with the integral (\ref{eq:lambday3}) at the NLL level.

The current code in \textsc{Deductor} does not include the $\beta_1$ contributions in evaluating the $z$ dependence of $\as((1-z)yQ^2)$. This appears to be not particularly significant numerically, but it is significant in principle because it means that some of the NLL contributions to $\cS_\cY^{[1]}(\mu^2;\nu)$ are absent.

We can now compare to Ref.~\cite{thrustsum} by changing the integration variable to $q^2 = (1-z)yQ^2$:
\begin{equation}
\begin{split}
\label{eq:lambday4}
\lambda(y) \approx{}& 4 C_\LF \int_{y^2Q^2}^{yQ^2}\!
\frac{d q^2}{q^2}
\frac{\alpha_s(q^2) 
-\beta_0 \log(\lambda_\LR) \alpha_s^2(q^2)}{2\pi}\
\\&
- 3 C_\LF\,
\frac{\alpha_s(y Q^2)}{2\pi}\
\;.
\end{split}
\end{equation}
This agrees with the result in Eq.~(64) of Ref.~\cite{thrustsum} for the LL and NLL contributions to $\lambda(y)$.

We have been seeking a formula for the summation of logarithms of $\nu$ in the Laplace transform $\tilde g(\nu)$ of the thrust distribution. We use Eq.~(\ref{eq:sigmanuALT}) for $\tilde g(\nu)$, choosing for $\isket{\rho_\scH}$ the state with a quark and an antiquark with opposite momenta. The operator $\cO(\nu)$ acting on this state is just 1. Then
\begin{equation}
\begin{split}
\label{eq:tildeg3}
\tilde g(\nu) ={}&
\frac{1}{\sigma_\scH}
\sbra{1}
\cY(Q^2; \nu) 
\sket{\rho_\LH}
\;.
\end{split}
\end{equation}
We approximate $\cY(Q^2; \nu)$, using Eq.~(\ref{eq:cYexponential}), as the exponential of the integral of the first order generator $\cS_\cY^{[1]}$, which we take from Eq.~(\ref{eq:cVS1m2}). This gives 
\begin{equation}
\begin{split}
\label{eq:cYexponential1}
\tilde g(\nu)
\approx{}& 
\exp\!\left(
-\int_{\mu_\Lf^2/Q^2}^{1}\!\frac{dy}{y}\,
(1 - e^{- \nu y})\,
\lambda(y)
\right)
\;.
\end{split}
\end{equation}
Here $\lambda(y)$ can be either the exact function from \textsc{Deductor}, as in Eq.~(\ref{eq:lambday}), or else the approximate function given in Eq.~(\ref{eq:lambday3}). The factor $(1 - e^{- \nu y})$ puts an effective lower cutoff on the $y$ integration at $y = 1/\nu$. Then a factor $\log^n(y)$ in $\lambda(y)$ produces a factor $\log^{n+1}(\nu)$ in the exponent of Eq.~(\ref{eq:cYexponential1}).

We have seen that one can start with Eq.~(\ref{eq:sigmanu}) for $\tilde g(\nu)$ as given by a parton shower and rearrange the operators to express $\tilde g(\nu)$ in the form Eq.~(\ref{eq:tildeg3}). Then approximating $\cS_\cY(\mug;\nu)$ by $\cS_\cY^{[1]}(\mug;\nu)$ in $\cY$ gives us a candidate result (\ref{eq:cYexponential1}) for the summation of logarithms of $\nu$ in $\tilde g(\nu)$. We do note that the shower splitting functions contain ingredients related to the argument of $\as$ in the parton splitting function. These ingredients are somewhat {\em ad hoc} from the perspective of just representing the soft and collinear singularities of a single splitting. Their purpose was to build into the first order splitting functions some approximation to splitting functions beyond leading order so as to improve the effectiveness of a parton shower in summing large logarithms. We have seen the effect of these ingredients in giving us the standard summation of thrust logarithms at the NLL level.

Our analysis uses primarily the Laplace transform $\tilde g(\nu)$ of the thrust distribution. One can take the inverse Laplace transform of $\tilde g(\nu)$ to obtain the thrust distribution $g(\tau)$, Eq.~(\ref{gdef}), itself. The function $g(\tau)$ is the derivative of $f(\tau)$, Eq.~(\ref{eq:foftaudef}):
\begin{equation}
\label{eq:gfromf}
g(\tau) = \frac{d f(\tau)}{d\tau}
\;.
\end{equation}
We can follow Ref.~\cite{thrustsum} to evaluate $f(\tau)$ at NLL accuracy: 
\begin{equation}
\begin{split}
\label{eq:Laplacef}
f(\tau) ={}& 
\exp\!\left(
-\frac{C_\LF}{\pi \beta_0^2}\left\{
\frac{f_1(\lambda)}{\as(Q^2)}
+ f_2(\lambda)
\right\}
\right)
\\&\times
\frac{1}{\Gamma(1-\gamma(\lambda))}\,
\;.
\end{split}
\end{equation}
Here
\begin{equation}
\lambda = \beta_0 \as(Q^2) \log(1/\tau)
\;,
\end{equation}
the LL function $f_1(\lambda)$ is
\begin{equation}
\label{eq:f1lambda}
f_1(\lambda) = (1-2\lambda)\log(1-2\lambda) 
- 2(1-\lambda)\log(1-\lambda) 
\;,
\end{equation}
the NLL function $f_2(\lambda)$ is
\begin{equation}
\begin{split}
\label{eq:f2lambda}
f_2(\lambda) ={}& 
- \frac{\beta_1}{2\beta_0}\,2 \log^2(1-\lambda)
+ \frac{\beta_1}{2\beta_0}  \log^2(1-2\lambda)
\\&
+ 2\beta_0 \gamma_\LE\log\left(
\frac{1-\lambda}{1-2\lambda}\right)
\\&
-\left(
\frac{\beta_1}{\beta_0} + \beta_0\log(\lambda_\LR)\right)
\log\!\left(\frac{(1-\lambda)^2}{1-2\lambda}\right)
\\&
+
\frac{3\beta_0}{2}\log\!\left(1-\lambda\right)
\;,
\end{split}
\end{equation}
where $\gamma_\LE$ is Euler's constant, and the function $\gamma(\lambda)$ is
\begin{equation}
\gamma(\lambda) 
= -\frac{2C_\LF}{\pi \beta_0}\,
\log\!\left(\frac{1-\lambda}{1-2\lambda}\right)
\;.
\end{equation}
The logarithm of $f(\tau)$ contains LL contributions proportional to $\as(Q^2)^n \log^{n+1}(1/\tau)$ and NLL contributions proportional to $\as(Q^2)^n \log^{n}(1/\tau)$, but contributions proportional to $\as(Q^2)^n \log^{j}(1/\tau)$ with $j < n$ are dropped. Of course, a parton shower does not drop terms beyond NLL.


\section{Result from the parton shower}
\label{sec:showerresult}

We have manipulated the operators used in a parton shower to produce a candidate formula (\ref{eq:cYexponential1}) for the summation of logarithms for the thrust distribution. We have seen that this formula reproduces the known result \cite{thrustsum} for $\tilde g(\nu)$ in QCD at the NLL level. We now ask what the result for $\tilde g(\nu)$ is in a first order parton shower that uses the \textsc{Deductor} algorithm or another algorithm of interest. That is, what do we get from Eqs.~(\ref{eq:sigmanu}) and (\ref{eq:jetUexponential}),
\begin{equation}
\begin{split}
\label{eq:sigmanuencore}
\tilde g(\nu) ={}& 
\frac{1}{\sigma_\scH}
\sbra{1} \cO(\nu) 
\mathbb{T} \exp\!\left(
\int_{\mu_\Lf^2}^{Q^2}\!\frac{d\mu^2}{\mu^2}
\cS(\mu^2)\!
\right)\!
\sket{\rho_\scH}
,
\end{split}
\end{equation}
when the shower generator $\cS(\mu^2)$ represents a first order shower? This must be the same as the result of using Eq.~(\ref{eq:cYexponential}) in Eq.~(\ref{eq:sigmanuALT}),
\begin{equation}
\begin{split}
\label{eq:sigmanu1}
\tilde g(\nu) ={}& 
\frac{1}{\sigma_\scH}
\sbra{1}
\mathbb{T} \exp\!\left(
\int_{\mu_\Lf^2}^{Q^2}\!\frac{d\mu^2}{\mu^2}
\cS_\cY(\mu^2;\nu)
\right)
\cO(\nu)
\sket{\rho_\scH}
\;.
\end{split}
\end{equation}
Here we take $\isket{\rho_\scH}$ to be the initial $q\bar q$ state in $e^+e^-$ annihilation (with massless quarks). Then there is some simplification because $\cO(\nu)\isket{\rho_\scH} = \isket{\rho_\scH}$. There is a more significant simplification because $\isket{\rho_\scH}$ is an eigenvector of $\cS_\cY(\mu^2; \nu)$. We use Eq.~(\ref{eq:cScYeigenvector}), Eq.~(\ref{eq:cScYeigenvalue}), and $\isbrax{1}\isket{\rho_\scH} = \sigma_\scH$ to give
\begin{equation}
\begin{split}
\label{eq:sigmanualt1}
\tilde g(\nu) ={}& 
\exp\!\left(
\int_{\mu_\Lf^2}^{Q^2}\!\frac{d\mu^2}{\mu^2}
\sbra{1}\cS_\cY(\mu^2;\nu)\sket{\{p,f,c,c\}_{2}}
\right)
\;.
\end{split}
\end{equation}
Here $\isket{\{p,f,c,c\}_{2}}$ is a color singlet $q \bar q$ basis state with $p_1 + p_2 = Q$. The results are independent of the direction of $\vec p_1 = - \vec p_2$ and independent of the quark flavor $f_1 = - f_2$. There is only one possible color state. The basis state is normalized to $\isbrax{1}\isket{\{p,f,c,c\}_{2}} = 1$ \cite{NSI}.

We use the operator $\cI(\nu)$ defined in Eq.~(\ref{eq:Inudef}),
\begin{equation}
\label{eq:Inu}
\cI(\nu) = \int_{\mu_\Lf^2}^{Q^2}\!\frac{d\mu^2}{\mu^2}\ 
\cS_\cY(\mu^2;\nu)
\;,
\end{equation}
to write Eq.~(\ref{eq:sigmanualt1}) as
\begin{equation}
\begin{split}
\label{eq:sigmanualt}
\tilde g(\nu) ={}& 
\exp\!\left[
\sbra{1}\cI(\nu)\sket{\{p,f,c,c\}_{2}}
\right]
\;.
\end{split}
\end{equation}
In Eq.~(\ref{eq:sigmanualt}), $\cI(\nu)$ is obtained from just $\cS(\mu^2)$, not from any higher order splitting functions that might be present in a higher order shower algorithm. The result for $\tilde g(\nu)$ in Eq.~(\ref{eq:sigmanualt}) could be very different from $\tilde g(\nu)$ as given by Eq.~(\ref{eq:cYexponential1}) because a first order parton shower is not the same as full QCD. 

Using Eq.~(\ref{eq:cIkexpansion}), we expand $\cI(\nu)$ as a series of terms $\cI^{[k]}(\nu)$, where  $\cI^{[k]}(\nu)$ contains $k$ powers of the shower splitting operator $\cS(\mug)$. Thus $\cI^{[k]}(\nu)$ contains $k$ powers of $\as$ evaluated at a running scale inside the integrations that give $\cI^{[k]}(\nu)$. We can expand  $\cI^{[k]}(\nu)$ in powers of $\as$ evaluated at a fixed scale. A convenient choice\footnote{In Sec.~\ref{sec:cScY1}, we used $\mu_\mathrm{fixed}^2 = Q^2$. Using Eq.~(\ref{eq:asrunning}), one can transform between expansions $\sum c(n,j)\as^n(Q^2/\nu) \log^j(\nu)$ and $\sum c'(n,j)\as^n(Q^2) \log^j(\nu)$ with $j \le n$  in each case, so both choices of $\mu_\mathrm{fixed}^2$ work equally well in an analytical treatment. In a numerical evaluation, $\mu_\mathrm{fixed}^2 = Q^2/\nu$ has the advantage that this scale is closer to the running scale at which $\as$ is evaluated inside the integrals for $\cI^{[k]}(\nu)$.} is $\mu_\mathrm{fixed}^2 = Q^2/\nu$. Thus we write
\begin{equation}
\cI^{[k]}(\nu) = \sum_{n=k}^\infty 
\left[\frac{\as(Q^2/\nu)}{2\pi}\right]^n
\cI^{[k]}_n(\nu)
\;.
\end{equation}

In $\cI^{[k]}_k(\nu)$ there are $k$ integrations over scale variables $y$ and $k$ integrations over momentum fractions $z$, so $\cI^{[k]}_k(\nu)$ could contain $2k$ factors of $\log(\nu)$. Changing the scale in $\as$ can produce one more factor $\log(\nu)$ for each factor $\as$, so that $\cI^{[k]}_n(\nu)$ could contain $n+k$ factors of $\log(\nu)$. However the exponent in $\tilde g(\nu)$ in Eq.~(\ref{eq:cYexponential1}) contains only contributions proportional to $\as^n(Q^2/\nu) \log^{j}(\nu)$ with $j \le n+1$. Thus a minimal expectation for the parton shower is that $\cI^{[k]}_n(\nu)$ contains only $j$ factors of $\log(\nu)$ with $j \le n+1$. If this is the case, we can say that the $\log(\nu)$ factors {\em exponentiate}. 

If we expand the QCD result for the exponent in $\tilde g(\nu)$ as given by Eq.~(\ref{eq:cYexponential1}) in powers of $\as(Q^2/\nu)$, the coefficients of $\as^n(Q^2/\nu) \log^{n+1}(\nu)$ and $\as^n(Q^2/\nu) \log^{n}(\nu)$ take particular values. These values are generated by $\cI^{[1]}(\nu)$ using $\as$ with its argument suitably specified by the shower algorithm. Thus for $k \ge 2$, $\cI^{[k]}_n(\nu)$ must not contain a factor $\log^{n+1}(\nu)$ if we are to maintain the logarithmic summation at LL level and additionally must not contain a factor $\log^{n}(\nu)$ if we are to maintain the logarithmic summation at NLL level.

We investigate how many powers of $\log(\nu)$ are contained in $\cI^{[k]}_n(\nu)$ in the following two sections.


\section{Parton shower at leading log}
\label{sec:showeratLL}

In this section we examine the operators $\cS_\cY^{[k]}(\mu^2; \nu)$ with the aim of discovering the behavior of $\tilde g(\nu)$ as given by a leading order parton shower using the $\Lambda$-ordered \textsc{Deductor} algorithm. The Laplace transform of $\tilde g(\nu)$ can be represented according to Eq.~(\ref{eq:sigmanualt}) in terms of the integral $\cI(\nu)$ of $\cS_\cY(\mu^2; \nu)$ defined in Eq.~(\ref{eq:Inudef}). We write the definition in the form
\begin{equation}
\label{eq:Inux}
\cI(\nu) = \int_0^\nu\!\frac{dx}{x}\ 
\cS_\cY(x Q^2/\nu;\nu)
\;.
\end{equation}
Here we have defined a standard scale $Q^2/\nu$ and a scale variable $x$ that gives the ratio of $\mu^2$ to this standard scale: $\mu^2 = x Q^2/\nu$. If we expand the exponential (not just the exponent) in Eq.~(\ref{eq:sigmanualt}) in powers of $\as(Q^2/\nu)$, we will find terms proportional to $\as^n(Q^2/\nu) \log^j(\nu)$ with $j \le 2n$. 

The simplest expectation would be that $\cI(\nu)$ also has an expansion with terms $\as^n(Q^2/\nu) \log^j(\nu)$ with $j \le 2n$. Such a representation would not be very useful, even if we knew all of the coefficients for $j = 2n$. It is much more useful if there are nonzero contributions $\as^n(Q^2/\nu) \log^j(\nu)$ only for $j \le n+1$ and we knew the coefficients for terms with $j = n+1$. We then call the $j = n+1$ terms the leading log, LL, terms.

In the notation of this paper, the operator $\cI^{[1]}(\nu)$ is proportional to one power of the shower splitting operator and thus to one power of a running $\as$ rather than the fixed $\as(Q^2/\nu)$. As we have seen, this operator generates a whole LL series $\as^n(Q^2/\nu) \log^j(\nu)$ with $j = n+1$. We may hope that this is all that survives at the LL level. That is, we may hope that $\cI^{[k]}(\nu)$ for $k \ge 2$ generates only terms $\as^n(Q^2/\nu) \log^j(\nu)$ with $j \le n$. If so, we will say that $\tilde g(\nu)$ as given by the leading order parton shower exponentiates at the LL level.

In this section, we demonstrate that $\tilde g(\nu)$ does exponentiate at the LL level in this sense. In the following section, we will turn our attention to the NLL level.

We will need a small preliminary analysis. We see from Eq.~(\ref{eq:cScY2soln}) that for $\cS_\cY^{[2]}(x Q^2/\nu; \nu)$  we will need $\iP{\cO(\nu)\,\cS(\bar\mu^2)\,\cO^{-1}(\nu)}$ and $\iomP{\cO(\nu)\,\cS(\bar\mu^2)\,\cO^{-1}(\nu)}$. 

For $\iP{\cO(\nu)\,\cS(\bar\mu^2)\,\cO^{-1}(\nu)}$, we briefly repeat the derivation that gave us Eq.~(\ref{eq:cVS1parts4}). We use
Eq.~(\ref{eq:cS1decomposition}), then Eq.~(\ref{eq:cS1realvirtual}), then Eqs.~(\ref{eq:cSlsum}) and (\ref{eq:cSlapproximation}):
\begin{align}
\label{eq:cS1P}
\P{\cO(\nu)\,\cS(x Q^2/\nu)\,\cO^{-1}(\nu)}
\hskip - 3 cm &
\notag\\={}& 
\P{\cO(\nu)\,\cS^{[1,0]}(x Q^2/\nu)\,\cO^{-1}(\nu) + 
\cS^{[0,1]}(x Q^2/\nu)}
\notag\\={}& 
\P{\cO(\nu)\,\cS^{[1,0]}(x Q^2/\nu)\,\cO^{-1}(\nu) - 
\cS^{[1,0]}(x Q^2/\nu)}
\notag\\={}& 
\sum_l
\P{\cS_l^{[1,0]}(x Q^2/\nu)e^{- \xi_l^\mathrm{op} x} - 
\cS_l^{[1,0]}(x Q^2/\nu)}
\notag\\={}& 
- \sum_l
\P{\cS_l^{[1,0]}(x Q^2/\nu)}
(1 - e^{- \xi_l^\mathrm{op} x})
\;.
\end{align}

For $\omP{\cO(\nu)\,\cS(\bar\mu^2)\,\cO^{-1}(\nu)}$, we need a somewhat different argument. We use Eqs.~(\ref{eq:omPdef}) and (\ref{eq:cS1decomposition}). Then we note that $\iP{\cS^{[0,1]}(x Q^2/\nu)} = \cS^{[0,1]}(x Q^2/\nu)$ according to Eq.~(\ref{eq:Bunchanged}) because $\cS^{[0,1]}(x Q^2/\nu)$ leaves the parton momenta and flavors unchanged. Then we use Eqs.~(\ref{eq:cSlsum}) and (\ref{eq:cSlapproximation}). Finally, we use the definition (\ref{eq:omPdef}) again. This gives
\begin{align}
\label{eq:cS1omP}
\omP{\cO(\nu)\,\cS(x Q^2/\nu)\,\cO^{-1}(\nu)}
\hskip - 3.8 cm &
\notag\\={}& 
\Big\{\cO(\nu)\,\cS^{[1,0]}(x Q^2/\nu)\,\cO^{-1}(\nu) + 
\cS^{[0,1]}(x Q^2/\nu)
\notag\\&
-\P{\cO(\nu)\,\cS^{[1,0]}(x Q^2/\nu)\,\cO^{-1}(\nu) + 
\cS^{[0,1]}(x Q^2/\nu)}
\Big\}
\notag\\={}& 
\Big\{
\cO(\nu)\,\cS^{[1,0]}(x Q^2/\nu)\,\cO^{-1}(\nu)
\notag\\&
-\P{\cO(\nu)\,\cS^{[1,0]}(x Q^2/\nu)\,\cO^{-1}(\nu)}
\Big\}
\notag\\={}& 
\Big\{
\sum_l \cS_l^{[1,0]}(x Q^2/\nu) e^{- \xi_l^\mathrm{op} x}
\notag\\& 
- \sum_l \P{\cS_l^{[1,0]}(x Q^2/\nu)}e^{- \xi_l^\mathrm{op} x}
\Big\}
\notag\\={}& 
\sum_l  
\omP{\cS_l^{[1,0]}(x Q^2/\nu)}e^{- \xi_l^\mathrm{op} x}
\;.
\end{align}

Now we can start with $\cS_\cY(x Q^2/\nu; \nu)$ at first order. Eq.~(\ref{eq:cVS1parts4}) gives us the result on the right hand side of Eq.~(\ref{eq:cS1P}):
\begin{equation}
\begin{split}
\label{eq:cVS1parts4encore}
\cS_\cY^{[1]}(x &Q^2/\nu; \nu) \\
\approx{}&
-\sum_l
\P{\cS_l^{[1,0]}(x Q^2/\nu)}\,
(1 - e^{- \xi_l^\mathrm{op} x})
\;.
\end{split}
\end{equation}
Recall that the eigenvalue $\xi_l$ of $\xi_l^\mathrm{op}$, given by Eq.~(\ref{eq:cldef}), is of order 1. We will also need $\cY^{[1]}(x Q^2/\nu;\nu)$. When we substitute Eq.~(\ref{eq:cVS1parts4encore}) into Eq.~(\ref{eq:cY1}), we obtain 
\begin{equation}
\begin{split}
\label{eq:cY1result}
\cY^{[1]}&(x Q^2/\nu;\nu)
\\={}&
-\int_{0}^{x}\!\frac{d\bar x}{\bar x}\
\sum_l
\P{\cS_l^{[1,0]}(\bar x Q^2/\nu)}\,
(1 - e^{- \xi_l^\mathrm{op} \bar x})
\;.
\end{split}
\end{equation}
Here, and in the remainder of this section, we set the infrared cutoff $\mu_\Lf^2$ to zero. We notice that the factor $(1 - e^{- \xi_l \bar x})$ is small for $\bar x \ll 1$ and approaches zero like $\bar x$ when $\bar x \to 0$. This provides an infrared cutoff for the $\bar x$ integration.

Now look at $\cS_\cY(x Q^2/\nu; \nu)$ at second order. We use Eq.~(\ref{eq:cScY2soln}):
\begin{equation}
\begin{split}
\label{eq:cScY2solnbis}
\cS_\cY^{[2]}(&x Q^2/\nu;\nu)
\\={}&  
\int_{0}^{x}\! \frac{d\bar x}{\bar x}
\PL \P{\cO(\nu)\,\cS(\bar x Q^2/\nu)\,\cO^{-1}(\nu)}
\\&\quad\quad\times
\omP{\cO(\nu)\,\cS(x Q^2/\nu)\,\cO^{-1}(\nu)}
\PR
\;.
\end{split}
\end{equation}
With the results (\ref{eq:cS1P}) and (\ref{eq:cS1omP}), we obtain
\begin{equation}
\begin{split}
\label{eq:cScY2result1}
\cS_\cY^{[2]}(&x Q^2/\nu;\nu)
\\
={}&  
- \sum_{\bar l,l}
\int_{0}^{x}\! \frac{d\bar x}{\bar x}
\PL \P{\cS_{\bar l}^{[1,0]}(\bar x Q^2/\nu)}
(1 - e^{- \xi_{\bar l}^\mathrm{op} \bar x})
\\&\times
\omP{\cS_{l}^{[1,0]}(x Q^2/\nu)}
e^{- \xi_l^\mathrm{op} x}
\PR
\;.
\end{split}
\end{equation}

We integrate this to form the contribution to $\cI$, Eq.~(\ref{eq:Inudef}), with two powers of $\cS$:
\begin{equation}
\label{eq:cI2}
\cI^{[2]}(\nu) = \int_0^\nu\!\frac{dx}{x}\ 
\cS_\cY^{[2]}(x Q^2/\nu;\nu)
\;.
\end{equation}
There are potentially two $\log(\nu)$ factors from the $z$ integrations inside the two factors of $\cS_{l}^{[1,0]}$. After expanding the running couplings in $\cS_{l}^{[1,0]}$, at order $\as^n(Q^2/\nu)$ there could be a total of $n$ factors of $\log(\nu)$. Then we integrate over $x$ and $\bar x$. This could produce two more factors of $\log(\nu)$, giving $\log^{n+2}(\nu)$ at order $\as^n(Q^2/\nu)$. But what happens in the $x$ and $\bar x$ integrations that we find based on Eq.~(\ref{eq:cScY2result1})? If $\bar x \ll 1$, the factor $(1 - e^{- \xi_{\bar l} \bar x})$ is small, so that the $\bar x$ integration is effectively limited to the range $1 \lesssim \bar x$. If $1 \ll x$, the factor $e^{- \xi_{\bar l} x}$ is small, so that the $x$ integration is effectively limited to the range $x \lesssim 1$. We also have $\bar x < x$. Thus the net effective integration range is $1 \lesssim \bar x < x \lesssim 1$. This leaves only $\bar x \sim x \sim 1$. There are no $\log(\nu)$ factors from the $\bar x$ and $x$ integrations.

A contribution to $\cI^{[2]}$ proportional to $\as^n(Q^2/\nu) \log^{n+1}(\nu)$ can be designated leading log. The result (\ref{eq:cScY2result1}) shows that there is no LL contribution to $\cI^{[2]}$. Rather, the LL contributions to the integral $\cI$ of $\cS_\cY$ come from $\cS_\cY^{[1]}$ after we account for the argument of the strong coupling $\as$ in $\cS_\cY^{[1]}$, Eq.~(\ref{eq:lambday4}). This leaves the possibility of a NLL, $\as^n(Q^2/\nu) \log^{n}(\nu)$, contribution to  $\cI^{[2]}$. We will investigate the NLL contribution in the following section by looking at the $z$ integrations in $\cS_\cY^{[2]}$.

We will also need some qualitative information about the behavior of $\cY^{[2]}$. From Eq.~(\ref{eq:cYk}) we have
\begin{align}
\label{eq:cYs2}
\cY^{[2]}(x Q^2/\nu;\nu) ={}& 
\int_{0}^{x}\!\frac{d\bar x}{\bar x}\
\Big\{\cS_\cY^{[2]}(\bar x Q^2/\nu;\nu)
\\&
+
\cY^{[1]}(\bar x Q^2/\nu;\nu)\,
\cS_\cY^{[1]}(\bar x Q^2/\nu;\nu)
\Big\}
\;. \notag
\end{align}
Using Eqs.~(\ref{eq:cScY2result1}), (\ref{eq:cVS1parts4encore}), and (\ref{eq:cY1result}),
\begin{equation}
\begin{split}
\label{eq:cY2}
\cY^{[2]}(x &Q^2/\nu;\nu) 
\\={}& \sum_{l_1,l_2}
\int_{0}^{x}\!\frac{dx_1}{x_1}
\int_{0}^{x_1}\!\frac{dx_2}{x_2}
\\&\times
\Big\{
\PL \P{\cS_{l_2}^{[1,0]}(x_2 Q^2/\nu)}\,
(1 - e^{- \xi_{l_2}^\mathrm{op} x_2})
\\&\quad\times
\omP{\cS_{l_1}^{[1,0]}(x_1 Q^2/\nu)}\,
e^{- \xi_{l_1}^\mathrm{op} x_1}
\PR
\\&
+
\P{\cS_{l_2}^{[1,0]}(x_2 Q^2/\nu)}\,
(1 - e^{- \xi_{l_2}^\mathrm{op} x_2})
\\&\quad\times
\P{\cS_{l_1}^{[1,0]}(x_1 Q^2/\nu)}\,
(1 - e^{- \xi_{l_1}^\mathrm{op} x_1})
\Big\}
\;.
\end{split}
\end{equation}
In both terms we have a factor $(1 - e^{- \xi_{l_2} x_2})$  so there is an effective integration range $1 \lesssim x_2 < x_1 < x$. This implies that $\cY^{[2]}(x Q^2/\nu;\nu) \to 0$ for $x \ll 1$. In the first term, there is a factor $e^{- \xi_{l_1} x_1}$, so that the integrand is small for $1 \ll x_1$. However the second term contains no such factor. The operators $\cS_{l_1}^{[1,0]}(x_1 Q^2/\nu)$ and $\cS_{l_2}^{[1,0]}(x_2 Q^2/\nu)$ can give us logarithms of their arguments. For this reason, $\cY^{[2]}(x Q^2/\nu;\nu)$ can grow slowly, like a power of $\log(x)$, for $1 \ll x$.

If we take $x = 1$ in Eq.~(\ref{eq:cY2}), the effective integration range for $x_1$ and $x_2$ is $1 \lesssim x_2 < x_1 < 1$. Thus $x_2 \sim x_1 \sim 1$. Then there are no factors of $\log(\nu)$ produced by the integrations over $x_1$ and $x_2$. Each factor of $\cS_{l}^{[1,0]}(Q^2/\nu)$ contains one factor of $\log(\nu)$. Thus $\cY^{[2]}(Q^2/\nu;\nu)$ contains at most 2 factors of $\log(\nu)$.

We can generalize these observations to suggest induction hypotheses for $\cS_\cY^{[k]}$ and $\cY^{[k]}$ for $k \ge 2$:

\begin{enumerate}

\item The operator $\cS_\cY^{[k]}(x Q^2/\nu;\nu)$ is suppressed by a factor $x$ times logarithms for $x\to 0$ and by an exponential $e^{-cx}$ times logarithms for $x \to \infty$. Its only unsuppressed region is for $x \sim 1$.

\item The operator $\cY^{[k]}(x Q^2/\nu;\nu)$ is suppressed by a factor $x$ times logarithms for $x\to 0$ and grows at most logarithmically for $x \to \infty$. 

\item The operators $\cS_\cY^{[k]}(Q^2/\nu;\nu)$ and $\cY^{[k]}(Q^2/\nu;\nu)$ each contain at most $k$ factors of $\log(\nu)$ at order $\as^k(Q^2/\nu)$. 

\end{enumerate}

In property 3, we note that the operators $\cS_\cY^{[k]}(Q^2/\nu;\nu)$ and $\cY^{[k]}(Q^2/\nu;\nu)$ contain higher powers of $\as(Q^2/\nu)$ that arise from expanding the running couplings in their definitions in powers of $\as(Q^2/\nu)$. This expansion can yield one more power of $\log(\nu)$ per power of $\as(Q^2/\nu)$. Thus there are at most $n$ powers of $\log(\nu)$ at order $\as^{n}(Q^2/\nu)$.

We have found that these properties hold at order $k=2$. We now establish that they hold for any larger order by assuming that they hold at order $k$ and showing that they hold at order $k+1$.

Begin with $\cS_\cY^{[k+1]}$. From Eq.~(\ref{eq:cScYk}) we have
\begin{align}
\cS_\cY^{[k+1]}(x Q^2/\nu;\nu) \hskip - 1.7 cm &  \notag
\\={}&  \notag
\PL\cY^{[k]}(x Q^2/\nu;\nu)
\\&\times \notag
\big\{
\cO(\nu)\cS(x Q^2/\nu)\cO(\nu)
-\cS_\cY^{[1]}(x Q^2/\nu;\nu)
\big\}\PR
\\ & \notag
- \sum_{j=2}^{k-1}
\P{\cY^{[k+1-j]}(x Q^2/\nu;\nu)\,
\cS_\cY^{[j]}(x Q^2/\nu;\nu)}
\\ &
- 
\P{\cY^{[1]}(x Q^2/\nu;\nu)\,
\cS_\cY^{[k]}(x Q^2/\nu;\nu)}
\;.
\end{align}
We use Eq.~(\ref{eq:cScY1soln}) to simplify the first term and Eqs.~(\ref{eq:cY1}) and (\ref{eq:cScY1soln}) to simplify the last term:
\begin{align}
\cS_\cY^{[k+1]}(x Q^2/\nu;\nu) \hskip - 1.7 cm & \notag 
\\={}&  \notag
\PL\cY^{[k]}(x Q^2/\nu;\nu)
\omP{\cO(\nu)\cS(x Q^2/\nu)\cO(\nu)}
\PR
\\ & \notag
- \sum_{j=2}^{k-1}
\P{\cY^{[k+1-j]}(x Q^2/\nu;\nu)\,
\cS_\cY^{[j]}(x Q^2/\nu;\nu)}
\\ & \notag
- 
{\color{red}\bigg[}
\int_0^x\!\frac{d\bar x}{\bar x}\,
\P{\cO(\nu)\cS(\bar x Q^2/\nu)\cO(\nu)}
\\ & \qquad \times
\cS_\cY^{[k]}(x Q^2/\nu;\nu){\color{red}\bigg]_\mathbb{P}}
\;.
\end{align}
Now we can use Eq.~(\ref{eq:cS1omP}) in the first term and Eq.~(\ref{eq:cS1P}) in the last term, giving us
\begin{align}
\label{eq:cScYnp1}
\cS_\cY^{[k+1]}(x Q^2/\nu;\nu) \hskip - 1.7 cm &
\notag
\\={}&  \notag
\sum_l
\PL\cY^{[k]}(x Q^2/\nu;\nu)
\omP{\cS_l^{[1,0]}(x Q^2/\nu)}\PR 
e^{- \xi_l^\mathrm{op} x}
\\ & \notag
- \sum_{j=2}^{k-1}
\P{\cY^{[k+1-j]}(x Q^2/\nu;\nu)\,
\cS_\cY^{[j]}(x Q^2/\nu;\nu)}
\\ & \notag
+\sum_l 
{\color{red}\bigg[}
\int_0^x\!\frac{d\bar x}{\bar x}\,
\P{\cS_l^{[1,0]}(\bar x Q^2/\nu)}
(1 - e^{- \xi_l^\mathrm{op} \bar x})
\\ & \qquad \times
\cS_\cY^{[k]}(x Q^2/\nu;\nu){\color{red}\bigg]_\mathbb{P}}
\;.
\end{align}
In the first term, property 2 for $\cY^{[k]}(x Q^2/\nu;\nu)$ implies that this term is unsuppressed only for $1 \lesssim x$, while the factor $\exp(- \xi_l^\mathrm{op} x)$ implies that this term is unsuppressed only for $x \lesssim 1$. Thus this term is unsuppressed only for $x \sim 1$. In the second term, property 1 for $\cS_\cY^{[j]}(x Q^2/\nu;\nu)$ implies that this term is unsuppressed only for $x \sim 1$. In the third term, property 1 for $\cS_\cY^{[k]}(x Q^2/\nu;\nu)$ implies that this term is unsuppressed only for $x \sim 1$. This gives us property 1 for $\cS_\cY^{[k+1]}(x Q^2/\nu;\nu)$.

Now set $x = 1$ in Eq.~(\ref{eq:cScYnp1}). There is an integration over $\bar x$ in the third term, but, accounting for the factor $[1 - \exp(- \xi_l^\mathrm{op} \bar x)]$, the integration region is $1 \lesssim \bar x < 1$. That is, $\bar x \sim 1$. We can then use property 3 for the operators that appear in order to count the maximum possible number of factors of $\log(\nu)$ in each term. At order $\as^{k+1}$, this gives the maximum number of factors of $\log(\nu)$ as $k+1$, thus verifying property 3 for $\cS_\cY^{[k+1]}(Q^2/\nu;\nu)$.

Now we examine $\cY(x Q^2/\nu;\nu)$. We use Eq.~(\ref{eq:cYk}) to write for $k \ge 1$,
\begin{align}
\cY^{[k+1]}&(x Q^2/\nu;\nu) \notag
\\ ={}& \notag
\int_{0}^{x}\!\frac{d\bar x}{\bar x}\
\cY^{[k]}(\bar x Q^2/\nu;\nu)\,
\cS_\cY^{[1]}(\bar x Q^2/\nu;\nu)
\\ & \notag
+ \sum_{j=2}^{k}
\int_{0}^{x}\!\frac{d\bar x}{\bar x}
\cY^{[k+1-j]}(\bar x Q^2/\nu;\nu)\, \cS_\cY^{[j]}(\bar x Q^2/\nu;\nu)
\\ &
+ \int_{0}^{x}\!\frac{d\bar x}{\bar x}\
\cS_\cY^{[k+1]}(\bar x Q^2/\nu;\nu)
\;.
\end{align}
We use Eq.~(\ref{eq:cScY1soln}) and (\ref{eq:cS1P}) to simplify the first term:
\begin{align}
\label{eq:cYnp1}
\cY^{[k+1]}(x Q^2/\nu;\nu) \hskip - 1.8 cm &
\notag
\\ ={}& \notag
-\sum_l
\int_{0}^{x}\!\frac{dx_1}{x_1}\
\cY^{[k]}(x_1 Q^2/\nu;\nu)\,
\P{\cS_l^{[1,0]}(x_1 Q^2/\nu)}
\\&\qquad\times \notag
(1 - e^{- \xi_l^\mathrm{op} x_1})
\\ & \notag
+ \sum_{j=2}^{k}
\int_{0}^{x}\!\frac{dx_1}{x_1}\
\cY^{[k+1-j]}(x_1 Q^2/\nu;\nu)\,
\cS_\cY^{[j]}(x_1 Q^2/\nu;\nu)
\\ & 
+
\int_{0}^{x}\!\frac{d x_1}{x_1}\
\cS_\cY^{[k+1]}(x_1 Q^2/\nu;\nu)
\;.
\end{align}
In each term, condition 2 for $\cY^{[k]}(x_1 Q^2/\nu;\nu)$ or $\cY^{[k+1-j]}(x_1 Q^2/\nu;\nu)$ or condition 1 for $\cS_\cY^{[k+1]}(x_1 Q^2/\nu;\nu)$ implies that the integrand of the $x_1$ integration is unsuppressed only for $1 \lesssim x_1$. Since $x_1 < x$, $\cY^{[k+1]}(x Q^2/\nu;\nu)$ is unsuppressed only for $1 \lesssim x$. This establishes property 2 for $\cY^{[k+1]}(x Q^2/\nu;\nu)$.

Now set $x = 1$ in Eq.~(\ref{eq:cYnp1}). There is an integration over $x_1$ in each term, but the integration region is $1 \lesssim x_1 < 1$. We can then use property 3 for the operators that appear in order to count the maximum possible number of factors of $\log(\nu)$ in each term. At order $\as^{k+1}(Q^2/\nu)$, this gives the maximum number of factors of $\log(\nu)$ as $k+1$, thus verifying property 3 for $\cY^{[k+1]}(Q^2/\nu;\nu)$.

We call the properties 1,2, and 3 above the {\em LL exponentiation property} of $\cS_\cY(\mu^2;\nu)$, as discussed at the start of this section. In the following section we analyze the NLL contributions to $\cS_\cY(\mu^2;\nu)$.


\section{Parton shower at next-to-leading log}
\label{sec:showeratNLL}

We have seen that $\cS_\cY(\mu^2;\nu)$ has the proper perturbative structure to allow $\tilde g(\nu)$ as given by a leading order parton shower using the $\Lambda$-ordered \textsc{Deductor} algorithm to exponentiate correctly at the leading log level. 

First, the operator $\cS_\cY^{[1]}(\mu^2;\nu)$, constructed from one power of the shower splitting operator $\cS(\mu^2)$ has the right structure to reproduce the known QCD result \cite{thrustsum} at LL accuracy and even at NLL accuracy, provided that the argument the running coupling $\as$ in $\cS(\mu^2)$ is properly defined. For $\cI^{[1]}(\nu)$, we can state this in terms of an expansion in powers of $\as(Q^2/\nu)$. We consider the integral $\cI^{[1]}(\nu)$ of $\cS_\cY^{[1]}(\mu^2;\nu)$ defined in Eq.~(\ref{eq:Inux}). When the running $\as$ in $\cI^{[1]}(\nu)$ is expanded in powers of $\as(Q^2/\nu)$, the coefficients of $\as^n(Q^2/\nu) \log^{n+1}(\nu)$, that is the LL coefficients, are correct and the coefficients of $\as^n(Q^2/\nu) \log^{n}(\nu)$, the NLL coefficients, are also correct.

Second, each of the operators $\cS_\cY^{[k]}(\mu^2;\nu)$ for $k \ge 2$ has the right structure so that in the integral $\cI^{[k]}(\nu)$, the coefficient of $\as^n(Q^2/\nu) \log^{n+1}(\nu)$, which contributes to the exponent in $\tilde g(\nu)$ at LL accuracy, vanish. That is, the coefficient $\cI^{[k]}_n(\nu)$ of $\as^n(Q^2/\nu)$ in $\cI^{[k]}(\nu)$ contains at most $n$ powers of $\log(\nu)$. 

This LL exponentiation property arises from two features of $\cS_\cY^{[k]}(\mu^2;\nu)$. First, $\cS_\cY^{[k]}(\mu^2;\nu)$ is suppressed for $\mu^2 \gg Q^2/\nu$ and for $\mu^2 \ll Q^2/\nu$, so that only the integration region $\mu^2 \sim Q^2/\nu$ contributes to $\cI^{[k]}(\nu)$ and no factor of $\log(\nu)$ arises from integrating over $\mu^2$ from $Q^2/\nu$ to $Q^2$. Second, $\cS_\cY^{[k]}(Q^2/\nu;\nu)$ at order $\as^n(Q^2/\nu)$ contains at most $n$ factors of $\log(\nu)$.

Now, if the coefficients of $\as^n(Q^2/\nu) \log^n(\nu)$ in $\cI^{[k]}(\nu)$ were to vanish for $k \ge 2$, then $\cI^{[k]}(\nu)$ would not contribute to $\tilde g(\nu)$ at NLL level. Then the only NLL contributions to $\tilde g(\nu)$ would come from the expansion of the running coupling in $\cI^{[1]}(\nu)$. Since these contributions match the known QCD result \cite{thrustsum}, we would conclude that the first order parton shower according to the \textsc{Deductor} algorithm generates the known QCD result at NLL accuracy.

Remarkably, this is the case: in $\cI^{[k]}(\nu)$ for $k \ge 2$ the coefficients $\cI^{[k]}_n(\nu)$ of $\as^n(Q^2/\nu)$ contain at most $n-1$ powers of $\log(\nu)$ for large $\nu$. The proof of this result is somewhat involved, so we present it in Appendix \ref{sec:NLLproof}. 

The proof in Appendix \ref{sec:NLLproof} requires that color be treated exactly. Although, in principle, the \textsc{Deductor} algorithm allows color to be treated with arbitrarily high accuracy \cite{NSNewColor, NSColoriPi}, high accuracy requires substantial computer resources. The use of a less exact version of color, the LC+ approximation \cite{NScolor}, is more practical and is adequate for most purposes. With the LC+ approximation, some of the NLL contributions to the exponent in $\tilde g(\nu)$ will be incorrect at order $1/N_\Lc^2$, where $N_\Lc = 3$ is the number of colors.


\section{Numerical behavior of $\cI^{[2]}(\nu)$}
\label{sec:numericalcScY2}

We have considered analytically the coefficient $\cI^{[k]}_n(\nu)$ of $[\as(Q^2/\nu)/(2\pi)]^n$ in $\cI^{[k]}(\nu)$, Eq.~(\ref{eq:Inu}). We have seen analytically in Secs.~\ref{sec:showeratLL} and \ref{sec:showeratNLL} and in Appendix \ref{sec:NLLproof} that $\cI^{[k]}_n(\nu)$ for $k\ge 2$ contains no more than $n-1$ powers of $\log(\nu)$ for large $\nu$.

The first nontrivial example of this is that $\cI^{[2]}_2(\nu)$, when calculated at large $\log(\nu)$, is proportional to $\log(\nu)$ plus a constant but has no $\log^2(\nu)$ contribution. Similarly, $\cI^{[2]}_3(\nu)$ has at most a $\log^2(\nu)$ contribution at large $\nu$. We can check these results numerically.

We define the second order term in the exponent in $\tilde g(\nu)$, Eq.~(\ref{eq:sigmanualt}):
\begin{equation}
\label{eq:cI2def}
\big\langle \cI^{[2]}(\nu) \big\rangle =
\int_0^{Q^2}\!\frac{d\mu^2}{\mu^2}\, 
\sbra{1} \cS_\cY^{[2]}(\mu^2;\nu)
\sket{\{p,f,c,c\}_{2}}
\;.
\end{equation}
We expand $\langle \cI^{[2]}(\nu) \rangle$ in powers of $\as(Q^2/\nu)/(2\pi)$ and calculate numerically the first two coefficients, $\langle \cI^{[2]}_2(\nu) \rangle$ and $\langle \cI^{[2]}_3(\nu) \rangle$,
\begin{equation}
\begin{split}
\label{eq:I22andI23}
\langle \cI^{[2]}(\nu) \rangle
={}& \langle \cI^{[2]}_2(\nu) \rangle \left(\frac{\as(Q^2/\nu)}{2\pi}\right)^2
\\&+
\langle \cI^{[2]}_3(\nu) \rangle \left(\frac{\as(Q^2/\nu)}{2\pi}\right)^3
+ \cdots
\;.
\end{split}
\end{equation}

The state $\isket{\{p,f,c,c\}_{2}}$ in Eq.~(\ref{eq:cI2def}) is a color singlet, flavor singlet, $q \bar q$ state with $p_1 + p_2 = Q$. The results are the same with any quark flavor choice and there is only one possible color state. The state is normalized to $\isbrax{1}\isket{\{p,f,c,c\}_{2}} = 1$. The operator $\cS_\cY^{[2]}(\mu^2;\nu)$ is calculated using the exact \textsc{Deductor} splitting functions according to Eq.~(\ref{eq:cScY2soln}). We use the exact definition of thrust to calculate $\tau$ in $\cO(\nu)$, Eq.~(\ref{cOnudef}). The calculation is performed with full color, not just leading color or the LC+ approximation. The integrals over scale in $\langle \cI^{[2]}_n(\nu) \rangle$ are infrared convergent so there is no need to impose a lower cutoff on the shower scale $\mu^2$. Then the coefficients $\langle \cI^{[2]}_n(\nu) \rangle$ are independent of $Q^2$.

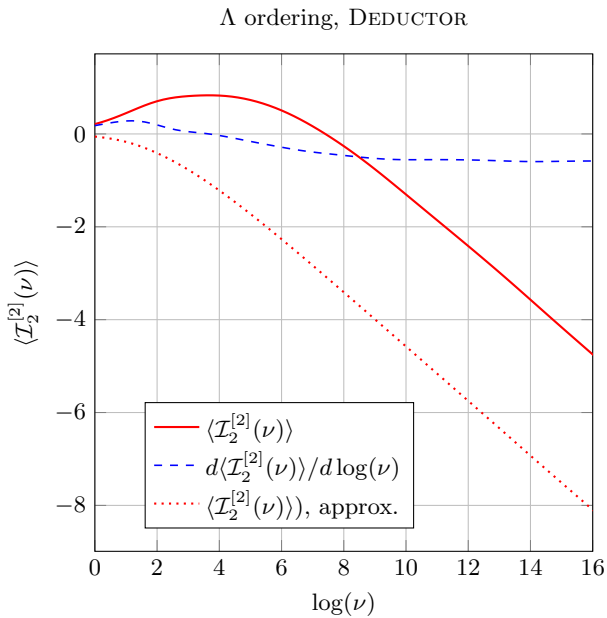
\begin{figure}
\begin{center}

\ifusefigs 

\begin{tikzpicture}

  \begin{axis}[title = {$\Lambda$ ordering, \textsc{Deductor}},
    xlabel={$\log(\nu)$}, ylabel={$\langle\cI^{[2]}_2(\nu)\rangle$},
    xmin=0, xmax=16,
    legend cell align=left,
    every axis legend/.append style = {
    at={(0.1,0.3)},
    anchor=north west}
    ]

     \def\normalization{1.0}

     \pgfplotstableread{Tables/I2-duct-lmd.dat}\datatable
    
     \addplot [red, thick, no markers] table 
     [x = {0}, y expr = \normalization*\thisrow{1}] 
     \datatable;
     \addlegendentry{$\langle \cI^{[2]}_2(\nu)\rangle$}
     
     \addplot [blue, dashed, semithick, no markers] table 
     [x = {0}, y expr = \normalization*\thisrow{3}] 
     \datatable;
     \addlegendentry{$d\langle \cI^{[2]}_2(\nu)\rangle/d\log(\nu)$}
     
     \pgfplotstableread{Tables/I2-duct-y.dat}\datatable
     
     \addplot [red, dotted, thick, no markers] table 
     [x = {0}, y expr = \normalization*\thisrow{1}] 
     \datatable;
     \addlegendentry{$\langle \cI^{[2]}_2(\nu)\rangle)$, approx.}

  \end{axis}
\end{tikzpicture}

\fi

\end{center}
\caption{
Plot of $\langle \cI^{[2]}_2(\nu)\rangle$, Eqs.~(\ref{eq:cI2def}) and (\ref{eq:I22andI23}), versus $\log(\nu)$ (solid red curve). For large $\log(\nu)$ the graph is approximately a straight line, corresponding to only one factor of $\log(\nu)$, indicating that the shower generates $\langle \cI^{[2]}_2(\nu)\rangle$ at NLL accuracy. The dashed blue curve is $d\langle \cI^{[2]}_2(\nu)\rangle/d\log(\nu)$. The dotted red curve shows an approximate version of  $\langle \cI^{[2]}_2(\nu)\rangle$ described in the text.
}
\label{fig:cI2}
\end{figure}

We plot $\langle \cI^{[2]}_2(\nu)\rangle$ versus $\log(\nu)$ as the solid red curve in  Fig.~\ref{fig:cI2}. We first note that $\langle \cI^{[2]}_2(\nu)\rangle$ is small. For instance, $\log(\nu) = 8$ corresponds roughly to $\tau = e^{-8} \approx 3\times 10^{-4}$ in the thrust distribution. For $\log(\nu) < 8$, we find $|\langle \cI^{[2]}_2(\nu)\rangle| \lesssim 1$. Then if we take $\as \approx 0.1$, we have $[\as/(2\pi)]^2 |\langle \cI^{[2]}_2(\nu)\rangle| \lesssim 0.0003$. The function $\cI^{[2]}(\nu)$ appears in the exponent of the Laplace transform of the thrust distribution, but for such a small value of $\cI^{[2]}(\nu)$, one would not have needed to exponentiate it.

Our primary concern is the behavior of $\langle \cI^{[2]}_2(\nu)\rangle$ for very large $\log(\nu)$.\footnote{The function $\tau(\{p\}_m)$ is a complicated function of the parton momenta. Evaluation of this function becomes numerically unstable for parton states $\{p\}_m$ that give very small $\tau$. For this reason, in this and later figures, we limit $\log(\nu)$ to $\log(\nu) < 16$, although in some cases the numerical results appear to be reliable for larger values of $\log(\nu)$.} Our analytical results indicate that $\langle \cI^{[2]}_2(\nu) \rangle$ should be a straight line for large $\log(\nu)$. The numerical result supports this conclusion. We also evaluate the integrand for $d\langle \cI^{[2]}_2(\nu)\rangle/d\log(\nu)$ analytically and then integrate this expression numerically and display the result as the dashed blue curve in Fig.~\ref{fig:cI2}. The analytical result implies that $d\langle \cI^{[2]}_2(\nu) \rangle/d\log(\nu)$ should  approach a constant for large $\log(\nu)$ and the numerical result supports this conclusion.

In our analysis, we argued that $\hat\tau - \tau = y$ should be a good approximation  in the second splitting for the purpose of determining how many powers of $\log(\nu)$ can appear in $\langle \cI^{[2]}_2(\nu) \rangle$. We tried calculating $\langle \cI^{[2]}_2(\nu)\rangle$ with this approximation. The result is shown as the dotted red line in Fig.~\ref{fig:cI2}. This curve is, as expected, a straight line for large $\log(\nu)$ and has the same slope as the curve for the exact $\langle \cI^{[2]}_2(\nu) \rangle$. We were a bit surprised to find that $\langle \cI^{[2]}_2(\nu) \rangle$ with the exact $\hat\tau - \tau$ differs by a noticeable amount from the result with the approximate thrust value. The difference is in the direction of making $|\langle \cI^{[2]}_2(\nu) \rangle|$ smaller. We do not have an analytical explanation for this behavior.

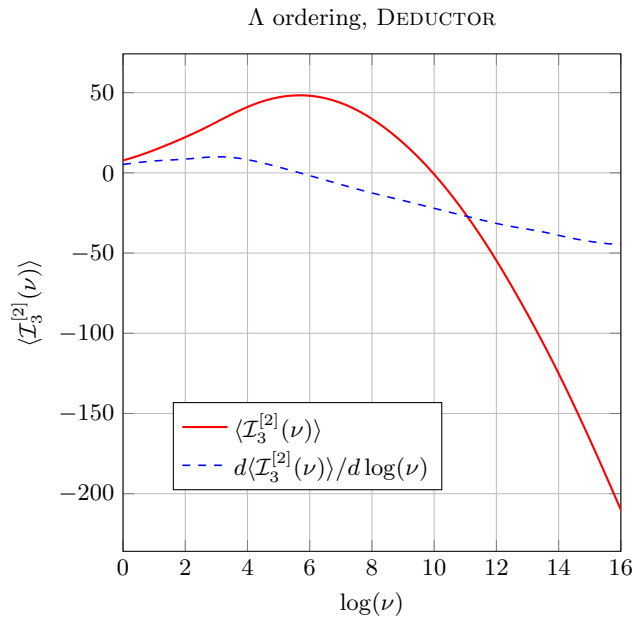
\begin{figure}
\begin{center}

\ifusefigs 

\begin{tikzpicture}

  \begin{axis}[title = {$\Lambda$ ordering, \textsc{Deductor}},
    xlabel={$\log(\nu)$}, ylabel={$\langle\cI^{[2]}_3(\nu)\rangle$},
    xmin=0, xmax=16,
    legend cell align=left,
    every axis legend/.append style = {
    at={(0.1,0.3)},
    anchor=north west}
    ]

     \def\normalization{3.83333} 

     \pgfplotstableread{Tables/I23-lambda.dat}\datatable
    
     \addplot [red, thick, no markers] table 
     [x = {0}, y expr = \normalization*\thisrow{1}] 
     \datatable;
     \addlegendentry{$\langle \cI^{[2]}_3(\nu)\rangle$}
     
     \addplot [blue, dashed, semithick, no markers] table 
     [x = {0}, y expr = \normalization*\thisrow{3}] 
     \datatable;
     \addlegendentry{$d\langle \cI^{[2]}_3(\nu)\rangle/d\log(\nu)$}

  \end{axis}
\end{tikzpicture}

\fi

\end{center}
\caption{
Plot of $\langle \cI^{[2]}_3(\nu)\rangle$, Eqs.~(\ref{eq:cI2def}) and (\ref{eq:I22andI23}), versus $\log(\nu)$ (solid red curve). The dashed blue curve is $d\langle \cI^{[2]}_3(\nu)\rangle/d\log(\nu)$. For large $\log(\nu)$ the graph of $d\langle \cI^{[2]}_3(\nu)\rangle/d\log(\nu)$ is approximately a straight line, indicating that the shower generates $\langle \cI^{[2]}_3(\nu)\rangle$ at NLL accuracy.
}
\label{fig:cI23}
\end{figure}

We also calculated $\langle \cI^{[2]}_3(\nu)\rangle$ as a numerical integral. We plot $\langle \cI^{[2]}_3(\nu)\rangle$ versus $\log(\nu)$ as the solid red curve in  Fig.~\ref{fig:cI23}.  We note first that $[\as/(2\pi)]^3 |\langle \cI^{[2]}_3(\nu)\rangle|$ is small for $\log(\nu) < 8$ if we take $\as \approx 0.1$. Our analytical results indicate that for large $\nu$ the highest power of $\log(\nu)$ in $\langle \cI^{[2]}_3(\nu) \rangle$ should be $\log^2(\nu)$. This implies that for large $\nu$ the highest power of $\log(\nu)$ in $d\langle \cI^{[2]}_3(\nu) \rangle/d\log(\nu)$ should be $\log^1(\nu)$. The numerical result, graphed as the dashed blue line in Fig.~\ref{fig:cI23}, supports this conclusion.

\section{Numerical behavior of the thrust distribution}
\label{sec:numericalg}

We have seen that the operator $\cS_\cY(\mu^2;\nu)$ directly generates the Laplace transform $\tilde g(\nu)$ of the thrust distribution according to  Eq.~(\ref{eq:sigmanualt}). The first order term $\cS_\cY^{[1]}(\mu^2;\nu)$ in this operator is obtained from the shower splitting function for a first order $\Lambda$-ordered parton shower. We have further seen that this term generates the known \cite{thrustsum} summation of logarithms of $\tau$ at the NLL level as long as the shower splitting function is suitably defined. Furthermore, the higher order terms $\cS_\cY^{[k]}(\mu^2;\nu)$ obtained from this first order shower splitting function generate only contributions beyond the NLL level.

According to Eq.~(\ref{eq:sigmanuencore}), same result for $\tilde g(\nu)$ as in Eq.~(\ref{eq:sigmanualt}) is obtained by running the $\Lambda$-ordered shower and measuring the Laplace transform of the thrust distribution. However, we do not need to take the Laplace transform. We can simply run the $\Lambda$-ordered shower and measure the thrust distribution $g(\tau)$, as in Eq.~(\ref{eq:sigmatau}). Will this give the same result as the NLL analytical result listed in Eqs.~(\ref{eq:gfromf}) and (\ref{eq:Laplacef})?

In this section, we try this experiment. It is not useful to set $Q^2 = M_\LZ^2$, which would be relevant for LEP (Large Electron Positron) experiments because a parton shower needs an infrared cutoff. We can take the cutoffs on allowed shower splittings to be $\Lambda > 1 \GeV$ and $k_\LT > 1 \GeV$, but then there is not much range between $(1 \GeV)^2$ and the starting scale $Q^2$ of the shower. The result is that there is not a wide range in $\tau$ in which we can examine the dependence of $g(\tau)$ on $\log(1/\tau)$ free of the effects of the infrared cutoffs. Instead, we retain $(1 \GeV)^2$ cutoffs but set $Q^2 = (10 \TeV)^2$. We then run the $\Lambda$-ordered \textsc{Deductor} shower with the LC+ approximation for color \cite{NScolor}. We turn off the top quark, so that the shower is based on 5-flavor QCD.

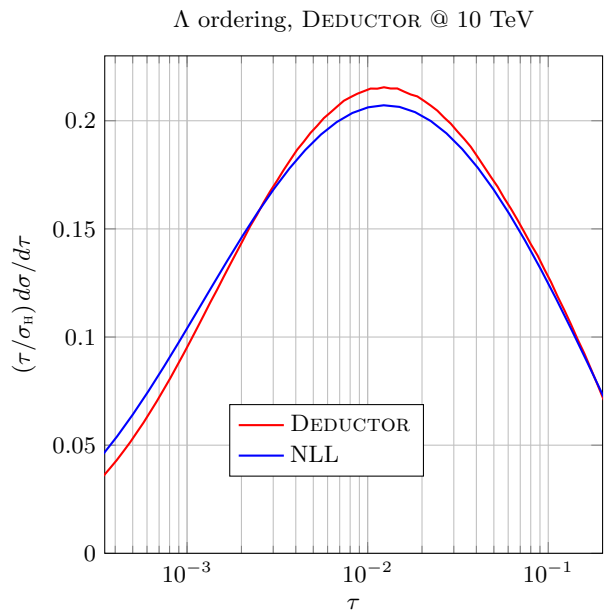
\begin{figure}
\begin{center}

\ifusefigs 

\begin{tikzpicture}

  \begin{semilogxaxis}
  [title = {$\Lambda$ ordering, \textsc{Deductor} @ 10 TeV},
    xlabel={$\tau$}, 
    ylabel={$(\tau/\sigma_\scH)\, d\sigma/d\tau$},
    xmin = 0.00035,
    xmax = 0.2,
    ymin = 0.0,
    ymax = 0.23,
    legend cell align=left,
    every axis legend/.append style = {
    at={(0.25,0.3)},
    anchor=north west
    }]


     \pgfplotstableread{Tables/thrust10TeVkT1.dat}\datatable
     \addplot [red, thick, no markers] table 
     [x = {1}, y = {3}] 
     \datatable; 
     \addlegendentry{\textsc{Deductor}}
     
     
     \pgfplotstableread{Tables/g10TeVtable.dat}\datatable 
     \addplot [blue, thick, no markers] table 
     [x = {0}, y = {1}] 
     \datatable;
     \addlegendentry{NLL}
     
  \end{semilogxaxis}
\end{tikzpicture}

\fi

\end{center}
\caption{
Plot of $(\tau/\sigma_\scH)\, d\sigma/d\tau$ according to  \textsc{Deductor} with $\Lambda$ ordering at $Q^2 = (10 \TeV)^2$ compared to the NLL expectation, Eqs.~(\ref{eq:gfromf}) and (\ref{eq:Laplacef}). In \textsc{Deductor}, we use a cutoff for splittings: $k_\LT > 1 \GeV$ and $\Lambda > 1 \GeV$. The \textsc{Deductor} curve is higher than the NLL curve at $\tau \approx 0.01$ and lower for small $\tau$.
}
\label{fig:thrustLambdatest}
\end{figure}

We compare $\tau g(\tau)$ according to \textsc{Deductor} with $\tau g(\tau)$ according to the NLL formula, Eqs.~(\ref{eq:gfromf}) and (\ref{eq:Laplacef}), in Fig.~\ref{fig:thrustLambdatest}.  We see that the \textsc{Deductor} curve is a bit higher than the NLL curve around $\tau = 0.01$ and a bit lower at the smallest values of $\tau$. Generally, the results agree to within about 0.01.

Do these results agree within the expected errors?  
\begin{itemize}[leftmargin = 0.3 cm]

\item The \textsc{Deductor} shower produces contributions beyond the NLL level. If we look at $\tau = 0.01$ so that $\log(1/\tau) = 4.6$, NNLL terms lack a factor 4.6 compared to NLL terms. A simple calculation shows that the NLL terms contribute approximately $-0.03$ to $\tau g(\tau)$ at $\tau = 0.01$. Thus we might expect that the NNLL terms in \textsc{Deductor} would contribute $\pm 0.03/4.6 \approx \pm 0.007$ to $\tau g(\tau)$. This gives us an error estimate from terms in \textsc{Deductor} beyond NLL of $\pm 0.007$.

\item There are typically about 20 parton splittings between the 10 TeV scale at which the shower starts and the 1 GeV scale at which it ends. We cannot be confident that there are not 0.1\% errors for each splitting resulting from approximations within the \textsc{Deductor} code, so we cannot rule out a 2\% systematic error in $g(\tau)$ resulting from these approximations. A 2\% error on the value $\tau g(\tau) \approx 0.2$ at $\tau = 0.01$ amounts to an error of $\pm 0.004$ in $\tau g(\tau)$.

\item The infrared cutoffs have some effect. The most important effect comes from the limit on the transverse momentum in a splitting, which we set to $k_\LT > 1 \GeV$. To test for sensitivity to this cutoff, we change the cut to $k_\LT > 3 \GeV$. In the range $0.0005 < \tau < 0.2$, we find that this change in cutoff produces a change in $\tau g(\tau)$ that is generally smaller than 0.003. Thus we estimate an error of $\pm 0.003$ in $\tau g(\tau)$ due to the influence of the infrared cutoff.

\item The \textsc{Deductor} splitting kernel omits the $\beta_1$ term in Eq.~(\ref{eq:asrunning}) for evaluating the dependence of $\as((1-z)y Q^2)$ on $(1-z)$. This changes the \textsc{Deductor} result at the NLL level. We examine this effect below.

\item The LC+ approximation used by default in \textsc{Deductor} is not the same as exact color. This introduces spurious terms of order $1/N_\Lc^2$ times logarithms of $1/\tau$ into the LC+ \textsc{Deductor} result, where $N_\Lc = 3$ is the number of colors. We examine this effect below.

\end{itemize}

\begin{figure}
\begin{center}

\ifusefigs 

\begin{tikzpicture}

  \begin{semilogxaxis}
  [title = {$\Lambda$ ordering, \textsc{Deductor} @ 10 TeV},
    xlabel={$\tau$}, 
    ylabel={$(\tau/\sigma_\scH)\, d\sigma/d\tau$},
    xmin = 0.00035,
    xmax = 0.2,
    ymin = 0.0,
    ymax = 0.23,
    legend cell align=left,
    every axis legend/.append style = {
    at={(0.2,0.3)},
    anchor=north west
    }]


     \pgfplotstableread{Tables/thrust10TeVkT1.dat}\datatable
     \addplot [black, thin, dashed, no markers] table 
     [x = {1}, y = {3}] 
     \datatable;
     \addlegendentry{\textsc{Deductor}}
     
     \pgfplotstableread{Tables/duct10TeVMod.dat}\datatable
     \addplot [red, thick, no markers] table 
     [x = {1}, y = {3}] 
     \datatable;
     \addlegendentry{\textsc{Duct}-corr}
     
     \pgfplotstableread{Tables/color10TeVMod.dat}\datatable
     \addplot [purple, no markers] table 
     [x = {1}, y = {3}] 
     \datatable;
     \addlegendentry{\textsc{Duct}-corr-color}
     
     \pgfplotstableread{Tables/g10TeVtable.dat}\datatable
     \addplot [blue, thick, no markers] table 
     [x = {0}, y = {1}] 
     \datatable;
     \addlegendentry{NLL}
     
  \end{semilogxaxis}
\end{tikzpicture}

\fi

\end{center}
\caption{
Plots of $(\tau/\sigma_\scH)\, d\sigma/d\tau$ using \textsc{Deductor} with $\Lambda$ ordering at $Q^2 = (10 \TeV)^2$. The black dashed curve is the \textsc{Deductor} curve from Fig.~\ref{fig:thrustLambdatest}. The blue solid curve is the NLL formula from Fig.~\ref{fig:thrustLambdatest}. The red solid curve is the \textsc{Deductor} result corrected with the factor $r(\tau)$ to include more exact $\as$ evolution. The purple solid curve, with noticeable statistical fluctuations, is the corrected \textsc{Deductor} result with two units of extra color beyond the LC+ approximation.
}
\label{fig:thrustLambdatestmore}
\end{figure}
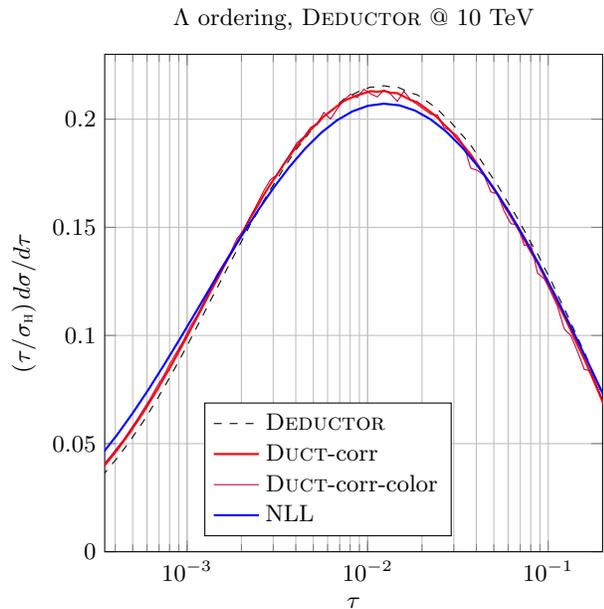

We examine the effects of missing NLL terms and of color in  Fig.~\ref{fig:thrustLambdatestmore}. Here the NLL curve is copied from Fig.~\ref{fig:thrustLambdatest} and the \textsc{Deductor} curve from Fig.~\ref{fig:thrustLambdatest} is displayed as a dashed (black) line. The remaining two curves are modified versions of the curves in Fig.~\ref{fig:thrustLambdatest}. 

We first address the fact that \textsc{Deductor} omits the $\beta_1$ term for evaluating the dependence of $\as((1-z)y Q^2)$ on $(1-z)$. This means that in the summation of logarithms of $\log(1/\tau)$, \textsc{Deductor} is missing the term $-(\beta_1/\beta_0)\log((1-\lambda^2)/(1-2\lambda))$ in $f_2(\lambda)$ in Eq.~(\ref{eq:f2lambda}). In order to see the effect of this term, we calculate the ratio
\begin{equation}
\label{eq:rtau}
r(\tau) = \frac{g_\mathrm{NLL}(\tau)}{g_\mathrm{NLL}^\mathrm{mod}(\tau)}
\;,
\end{equation}
where $g^\mathrm{mod}_\mathrm{NLL}(\tau)$ is obtained by omitting the term $-(\beta_1/\beta_0)\log((1-\lambda^2)/(1-2\lambda))$ in the calculation of $g(\tau)$. Then we correct the $\textsc{Deductor}$ result for $g(\tau)$ by multiplying it by $r(\tau)$. We plot the corrected \textsc{Deductor} curve in  Fig.~\ref{fig:thrustLambdatestmore}. We see that the corrected \textsc{Deductor} curve is quite close to the uncorrected curve. However the difference is visible in Fig.~\ref{fig:thrustLambdatestmore} and acts in the direction of reducing the discrepancy between the analytical summation of logarithms and the numerical \textsc{Deductor} result.\footnote{In a future version of \textsc{Deductor}, we may add this contribution to the splitting kernel, although its practical effect is quite small.}

We next address the fact that in Fig.~\ref{fig:thrustLambdatest} we used the default color approximation in \textsc{Deductor}, the LC+ approximation \cite{NScolor}. This approximation is an improvement over the leading color approximation, but it is far from being exact. In the LC+ approximation, we replace the exact first order splitting function $\cS(\mu^2) = \cS^{[1,0]}(\mu^2) + \cS^{[0,1]}(\mu^2)$ by an approximate version $\cS_\mathrm{LC+}(\mu^2) = \cS^{[1,0]}_\mathrm{LC+}(\mu^2) + \cS^{[0,1]}_\mathrm{LC+}(\mu^2)$. \textsc{Deductor} has the option of expanding in powers of $\cS(\mu^2) - \cS_\mathrm{LC+}(\mu^2)$ and keeping terms up to and including $[\cS(\mu^2) - \cS_\mathrm{LC+}(\mu^2)]^n$, where $n$ can be chosen by the user \cite{NSColorTheory}. In order to assess what difference a more exact treatment of color could make, we plot in Fig.~\ref{fig:thrustLambdatest} the result of calculating the thrust distribution at 10 TeV with $n=2$. We have corrected this result using the factor $r(\tau)$ from Eq.~(\ref{eq:rtau}). Of course, using $n=2$ slows the calculation down, increasing the statistical errors. Within the statistical errors, we find that improving the color treatment makes no difference.

In summary, we have made a numerical comparison of the expected NLL result for the thrust distribution and a direct calculation using a $\Lambda$-ordered parton shower with a global momentum mapping, setting $Q^2$ to $(10 \TeV)^2$ so as to allow $\log(1/\tau)$ to be adequately large to provide a real test. We have found good agreement within the estimated errors.

\section{$k_\LT$ ordering}
\label{sec:kTordering}

The default ordering variable in \textsc{Deductor} is $\Lambda$, Eq.~(\ref{eq:Lambdadef}). However, there is an option to use $k_\LT$ ordering.\footnote{For $k_\LT$ ordering, $k_\LT^2 = - k_\perp^2$ where the vector $k_\perp$ is orthogonal to the momentum $p_l$ of the emitting parton and to $Q$, rather than being orthogonal to $p_l$ and the momentum $p_k$ of the dipole partner parton.} We can define $\cI^{[2]}(\nu)$ with $k_\LT$ ordering using Eqs.~(\ref{eq:cScY2soln}) and (\ref{eq:cI2def}). We simply set the scale parameters to $\mug = k_\LT^2$ for the first splitting and $\bar \mu ^2 = \bar k_\LT^2$ for the second splitting. Then $k_\LT$ ordering means that $\bar k_\LT^2 < k_\LT^2$ in Eq.~(\ref{eq:cScY2soln}).

With $k_\LT$ ordering, the reasoning supporting NLL accuracy of the $\Lambda$-ordered shower from Sec.~\ref{sec:showeratLL} and Appendix \ref{sec:NLLproof} is lost. However, it appears that we can still get cancellation of $\log(\nu)$ factors in $\cI^{[2]}_2(\nu)$ at the NLL level. That is, the integral has contributions proportional to $\log^4(\nu)$ at large $\log(\nu)$, but after these contributions are summed, only terms proportional to $\log^1(\nu)$ and $\log^0(\nu)$ remain. The mechanism is that the contributions from the two terms specified by the $\iomP{\cdots}$ operation in the last line of Eq.~(\ref{eq:cScY2soln}), representing real emissions and virtual emissions, cancel each other. A complete proof is beyond the scope of this paper, but we present an argument that makes this conclusion plausible in Appendix \ref{sec:kTorderingcancellation}.

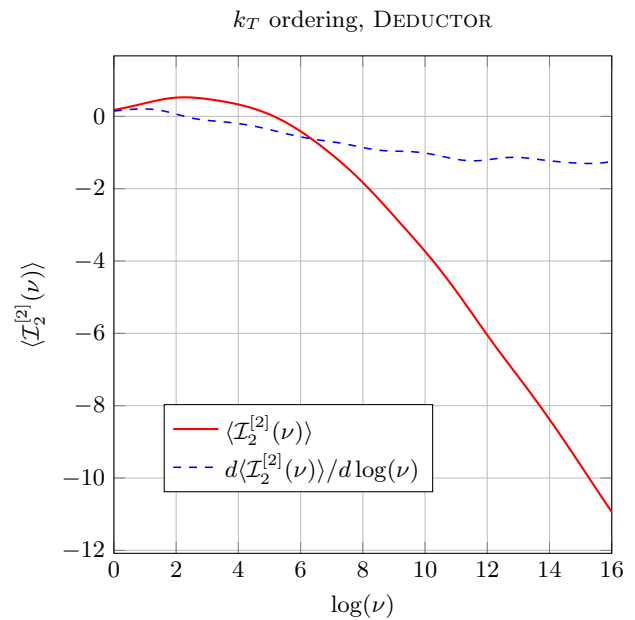
\begin{figure}
\begin{center}

\ifusefigs 
\begin{tikzpicture}
  \begin{axis}[title = {$k_T$ ordering, \textsc{Deductor}},
    xlabel={$\log(\nu)$}, ylabel={$\langle\cI^{[2]}_2(\nu)\rangle$},
    xmin=0, xmax=16,
    legend cell align=left,
    every axis legend/.append style = {
    at={(0.1,0.3)},
    anchor=north west}
    ]

     \def\normalization{1.0}
    
     \pgfplotstableread{Tables/I2-duct-kT.dat}\datatable
     \addplot [red, thick, no markers] table 
     [x = {0}, y expr = \normalization*\thisrow{1}] 
     \datatable;
     \addlegendentry{$\langle \cI^{[2]}_2(\nu)\rangle$}
     
     \addplot [blue, dashed, semithick, no markers] table 
     [x = {0}, y expr = \normalization*\thisrow{3}] 
     \datatable;
     \addlegendentry{$d\langle \cI^{[2]}_2(\nu)\rangle/d\log(\nu)$}
     

  \end{axis}
\end{tikzpicture}

\fi

\end{center}
\caption{
Plot of $\langle \cI^{[2]}_2(\nu) \rangle$ versus $\log(\nu)$, as in Fig.~\ref{fig:cI2}, for the \textsc{Deductor} shower algorithm with $k_\LT$ ordering. The blue dashed curve is $d\langle \cI^{[2]}_2(\nu) \rangle/ d\log(\nu)$.
}
\label{fig:cI2-kT}
\end{figure}

\begin{figure}
\begin{center}

\ifusefigs 
\begin{tikzpicture}
  \begin{axis}[title = {$k_T$ ordering, \textsc{Deductor}},
    xlabel={$\log(\nu)$}, ylabel={$\langle\cI^{[2]}_3(\nu)\rangle$},
    xmin=0, xmax=16,
    legend cell align=left,
    every axis legend/.append style = {
    at={(0.1,0.3)},
    anchor=north west}
    ]

     \def\normalization{3.83333} 
    
     \pgfplotstableread{Tables/I23-kT.dat}\datatable
     \addplot [red, thick, no markers] table 
     [x = {0}, y expr = \normalization*\thisrow{1}] 
     \datatable;
     \addlegendentry{$\langle \cI^{[2]}_3(\nu)\rangle$}
     
     \addplot [blue, dashed, semithick, no markers] table 
     [x = {0}, y expr = \normalization*\thisrow{3}] 
     \datatable;
     \addlegendentry{$d\langle \cI^{[2]}_3(\nu)\rangle/d\log(\nu)$}
     

  \end{axis}
\end{tikzpicture}

\fi

\end{center}
\caption{
Plot of $\langle \cI^{[2]}_3(\nu) \rangle$ versus $\log(\nu)$, as in Fig.~\ref{fig:cI23},  for the \textsc{Deductor} shower algorithm with $k_\LT$ ordering. The blue dashed curve is $d\langle \cI^{[2]}_3(\nu) \rangle/ d\log(\nu)$.
}
\label{fig:cI23-kT}
\end{figure}
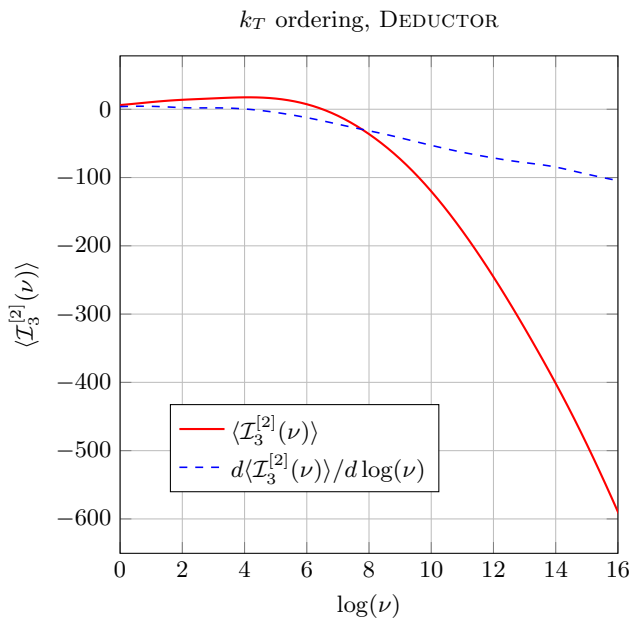

We can check the effect of the choice of ordering variable on the summation of $\log(\nu)$ factors in the thrust distribution by calculating $\langle \cI^{[2]}_2(\nu) \rangle$ numerically using the \textsc{Deductor} shower algorithm with $k_\LT$ ordering. The result is shown as the solid red curve in Fig.~\ref{fig:cI2-kT}. We see that $\langle \cI^{[2]}_2(\nu) \rangle$ is quite small, $|\langle \cI^{[2]}_2(\nu) \rangle| < 2$ for $\log(\nu) < 8$. For NLL accuracy, this curve should be linear for large $\log(\nu)$. To quite good, but not perfect, accuracy, it is. 

We have also checked the behavior of $\langle \cI^{[2]}_3(\nu) \rangle$ as a function of $\log(\nu)$. The results are shown in Fig.~\ref{fig:cI23-kT}. For large $\nu$ the highest power of $\log(\nu)$ in $\langle \cI^{[2]}_3(\nu) \rangle$ should be $\log^2(\nu)$. This implies that for large $\nu$ the highest power of $\log(\nu)$ in $d\langle \cI^{[2]}_3(\nu) \rangle/d\log(\nu)$ should be $\log^1(\nu)$. The numerical result, graphed as the dashed blue line in Fig.~\ref{fig:cI23-kT}, supports this conclusion.

We have investigated only $\langle \cI^{[2]}_2(\nu) \rangle$ and $\langle \cI^{[2]}_3(\nu) \rangle$. We have found results consistent with NLL accuracy for the \textsc{Deductor} shower with $k_\LT$ ordering, but there could still be inconsistencies with NLL accuracy for $\langle \cI^{[k]}_n(\nu) \rangle$ for other values of $k$ and $n$. A promising approach to investigating this issue would be to automate the calculation of $\langle \cI^{[k]}_n(\nu) \rangle$ so that these functions could be calculated numerically for any not-too-large values of $k$ and $n$. We leave this approach to future work. 

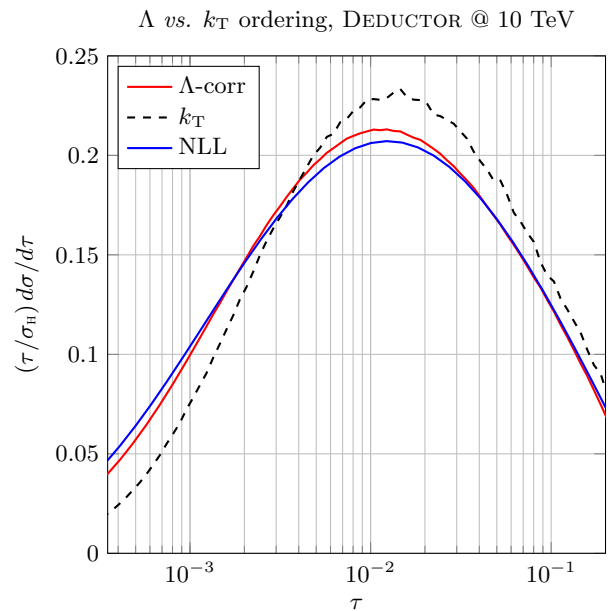
\begin{figure}
\begin{center}

\ifusefigs 

\begin{tikzpicture}

  \begin{semilogxaxis}
  [title = {$\Lambda$ {\em vs.} $k_\LT$ ordering, \textsc{Deductor}  @ 10 TeV},
    xlabel={$\tau$}, 
    ylabel={$(\tau/\sigma_\scH)\, d\sigma/d\tau$},
    xmin = 0.00035,
    xmax = 0.2,
    ymin = 0.0,
    ymax = 0.25,
    legend cell align=left,
    every axis legend/.append style = {
    at={(0.02,0.98)},
    anchor=north west
    }]


     \pgfplotstableread{Tables/duct10TeVMod.dat}\datatable
     \addplot [red, thick, no markers] table 
     [x = {1}, y = {3}] 
     \datatable;
     \addlegendentry{$\Lambda$-corr}
     
     
     \pgfplotstableread{Tables/thrustkT10TeVkT1.dat}\datatable
     \addplot [black, dashed, thick, no markers] table 
     [x = {1}, y = {3}] 
     \datatable;
     \addlegendentry{$k_\LT$}
     
           
     \pgfplotstableread{Tables/g10TeVtable.dat}\datatable
     \addplot [blue, thick, no markers] table 
     [x = {0}, y = {1}] 
     \datatable;
     \addlegendentry{NLL}

  \end{semilogxaxis}
\end{tikzpicture}

\fi

\end{center}
\caption{
Plot of $(\tau/\sigma_\scH)\, d\sigma/d\tau$ with $\Lambda$ ordering and $k_\LT$ ordering at $Q^2 = (10 \TeV)^2$. Both are compared to the NLL expectation, Eqs.~(\ref{eq:gfromf}) and (\ref{eq:Laplacef}). We use a cutoff on the transverse momentum in splittings: $k_\LT > 1 \GeV$.
}
\label{fig:thrustLambdaKTtest}
\end{figure}

We can also look directly at $(\tau/\sigma_\scH)\, d\sigma/d\tau$ with $Q^2 = (10 \TeV)^2$. We use either \textsc{Deductor} with its default $\Lambda$ ordering or \textsc{Deductor} with $k_\LT$ ordering. The result with $\Lambda$ ordering, from Fig.~\ref{fig:thrustLambdatestmore}, includes the correction factor $r(\tau)$ from Eq.~(\ref{eq:rtau}). The result with $k_\LT$ ordering needs no correction factor because $k_\LT^2$ in $\as(\lambda_\LR k_\LT^2)$ in the \textsc{Deductor} splitting function is the same as the ordering variable. We do not include hadronization. Thus we examine only perturbative effects and the effects of the shower cutoff. With $\Lambda$ ordering, the shower stops at $\Lambda = 1 \GeV$ and there is also a cut that prevents the $k_\LT$ in any splitting from being smaller than $1 \GeV$. With $k_\LT$ ordering, the shower stops at $k_\LT = 1 \GeV$. The result is shown in Fig.~\ref{fig:thrustLambdaKTtest}. We see that the shower ordering does make a difference. Although $(\tau/\sigma_\scH)\, d\sigma/d\tau$ calculated with $k_\LT$ ordering is similar to the NLL expectation $\tau g(\tau)$  from Eqs.~(\ref{eq:gfromf}) and (\ref{eq:Laplacef}), the difference between these two results is greater than the expected uncertainties discussed for $\Lambda$ ordering in Sec.~\ref{sec:numericalg}.

\begin{figure}
\begin{center}

\ifusefigs 

\begin{tikzpicture}

  \begin{semilogxaxis}
  [title = {$k_\LT$ ordering, \textsc{Deductor}, ratio to NLL},
    xlabel={$\tau$}, 
    ylabel={$(\tau/\sigma_\scH)\, d\sigma/d\tau\ /(\tau g(\tau))$},
    xmin = 0.001,
    xmax = 0.1,
    ymin = 0.75,
    ymax = 1.25,
    legend cell align=left,
    every axis legend/.append style = {
    at={(0.64,0.24)},
    anchor=north west
    }]


     \pgfplotstableread{Tables/NLLratio1TeV.dat}\datatable
     \addplot [black, thick, no markers] table 
     [x = {0}, y = {1}] 
     \datatable;
     \addlegendentry{1 TeV}
     
     \pgfplotstableread{Tables/NLLratio10TeV.dat}\datatable
     \addplot [red, thick, no markers] table 
     [x = {0}, y = {1}] 
     \datatable;
     \addlegendentry{10 TeV}
     
     \pgfplotstableread{Tables/NLLratio100TeV.dat}\datatable
     \addplot [blue, thick, no markers] table 
     [x = {0}, y = {1}] 
     \datatable;
     \addlegendentry{100 TeV}
     
     \addplot [red, thick, dashed, domain=0.013:0.1] {1.112};
     \addplot [black, thick, dashed, domain=0.03:0.1] {1.190};
     \addplot [blue, thick, dashed, domain=0.004:0.1] {1.070};

  \end{semilogxaxis}
\end{tikzpicture}

\fi

\end{center}
\caption{
Ratio of $(\tau/\sigma_\scH)\, d\sigma/d\tau$ with  $k_\LT$ ordering to the NLL expectation, $\tau g(\tau)$, Eqs.~(\ref{eq:gfromf}) and (\ref{eq:Laplacef}). The ratio is calculated at $Q^2 = (1 \TeV)^2$, $Q^2 = (10 \TeV)^2$, and $Q^2 = (100 \TeV)^2$.  We use a cutoff on the transverse momentum in splittings: $k_\LT > 1 \GeV$ in each case.
}
\label{fig:thrusthigherQ}
\end{figure}
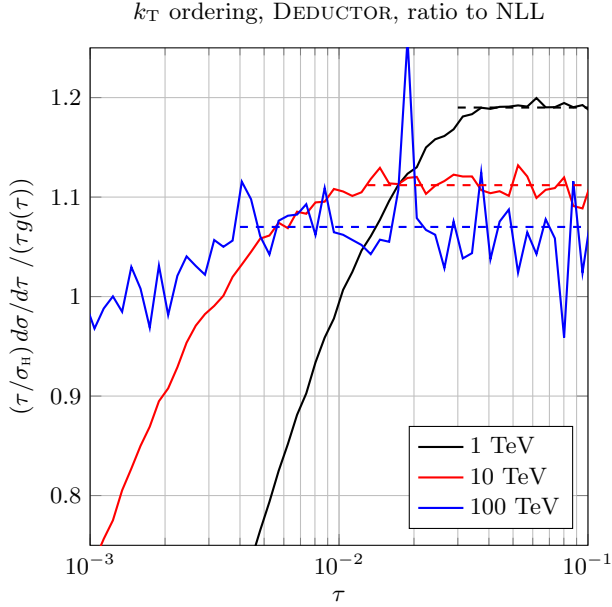

As an alternative, we can follow the method of Ref.~\cite{DasguptaShowerSum} and calculate $(\tau/\sigma_\scH)\, d\sigma/d\tau$ for various values of $Q^2$, and thus for various values of $\as(Q^2)$. We choose $Q^2 = (1 \TeV)^2$, $(10 \TeV)^2$, and $(100 \TeV)^2$, corresponding to $\as(Q^2) = 0.087$, $0.069$, and $0.058$.\footnote{Ref.~\cite{DasguptaShowerSum} considers $\as(Q^2)$ as small as 0.005, corresponding to $Q^2 \approx (10^{70} \GeV)^2$ but \textsc{Deductor} is not capable of working with values of $Q^2$ as large as this.} For each value of $Q^2$, we calculate the expected NLL function $\tau g(\tau)$, Eqs.~(\ref{eq:gfromf}) and (\ref{eq:Laplacef}). Then we plot the ratio 
\begin{equation}
R(\tau,Q^2) = \frac{(\tau/\sigma_\scH)\, d\sigma/d\tau}{\tau g(\tau)}
\;.
\end{equation}
The results are displayed in Fig.~\ref{fig:thrusthigherQ}. In the case $Q^2 = (100 \TeV)^2$, there are typically around 100 partons produced in each event. This causes \textsc{Deductor} to operate very slowly, which leads to substantial statistical fluctuations that are visible in the plot.  

If the log summation is working at the NLL level, the ratio plotted should be close to 1 and should get closer to 1 as $Q^2$ increases. We note two features of the results. First, for any fixed value of $Q^2$, $\tau g(\tau)$ fails to match the parton shower result for sufficiently small $\tau$. The value of $\tau$ at which this failure sets in decreases as $Q^2$ grows. For larger values of $\tau$, but still with $\tau < 0.1$, $R(\tau,Q^2)$ is approximately constant:
\begin{equation}
R(\tau,Q^2) \approx R_0(Q^2)
\;.
\end{equation}
These values ($R_0 = 1.190, 1.112, 1.070$) are shown as dashed lines in Fig.~\ref{fig:thrusthigherQ}. Second, we note that $R(\tau,Q^2)$ is fairly close to 1 and gets closer to 1 as $Q^2$ increases. In fact, to within about 10\%,
\begin{equation}
R_0(Q^2) -1 \approx 23\,\as^2(Q^2)
\;.
\end{equation}
This is consistent with the expectation that $R_0(Q^2) \to 0$ as $\as(Q^2) \to 0$. We tentatively conclude from these results that the $k_\LT$-ordered \textsc{Deductor} shower is correctly summing thrust logarithms at the NLL level, even though the difference between the shower result and the NLL analytical result is larger for $k_\LT$ ordering than for $\Lambda$ ordering.

\section{Effect of the momentum mapping for $\Lambda$ ordering}
\label{sec:momentummapping}

Recall from Sec.~\ref{sec:changeintau} that in a splitting $p_l \to \hat p_l + \hat p_\mpone$, we always have $p_l \ne \hat p_l + \hat p_\mpone$. In order to conserve momentum, we need to map the momenta $p_i$ into new momenta $\hat p_i$ such that 
\begin{equation}
\sum_{i=1}^{m+1} \hat p_i = \sum_{i=1}^{m}  p_i
\;.
\end{equation}
In the \textsc{Deductor} algorithm, this is accomplished by using a Lorentz transformation \cite{NSI}
\begin{equation}
\label{eq:deductorboost}
\hat p_i^\mu = \Lambda^\mu_\nu\, p_i^\nu \;,
\hskip 1 cm 
i \notin \{l,\mpone\}
\;.
\end{equation}
The Lorentz transformation is defined to be a boost in the plane of $p_l$ and $Q$. We have found in Sec.~\ref{sec:changeintau} that the boost angle $\omega$ is small, of order $y$, and that the effect of this small Lorentz transformation on the thrust is small compared to the order $y$ effect produced by the splitting itself. 

For any parton shower, one will need a momentum mapping that preserves the total momentum. The {\em global} mapping produced by a Lorentz transformation is not the only possibility. A more widely used {\em local} choice is provided by the Catani-Seymour dipole splitting formalism \cite{CataniSeymour} or the local mapping in \textsc{Pythia} \cite{SjostrandSkands}. For the Catani-Seymour choice, we start with the parton $l$ that splits and its dipole partner $k$, with momenta $p_l$ and $p_k$. After the splitting, we have a new parton $m+1$ and new momenta $\hat p_i$, $\hat p_\mpone$ and $\hat p_k$. The definition is
\begin{equation}
\begin{split}
\label{eq:CataniSeymourMapping}
\hat p_\mpone ={}& (1-z)\, p_l + zy\, p_k + k_\perp
\;,
\\
\hat p_l ={}& z p_l + (1-z)\,y\, p_k - k_\perp
\;,
\\
\hat p_k ={}& \left(1 -  y\right) p_k
\;,
\end{split}
\end{equation}
with $k_\perp\cdot p_j = k_\perp\cdot p_k = 0$. Here $z$, $y$, and $k_\perp$ are different from $z$, $y$ and $k_\perp$ defined for \textsc{Deductor} kinematics. The momenta of the other partons is unchanged: 
\begin{equation}
\label{eq:pi}
\hat p_i = p_i\, \qquad i \notin \{l,k,\mpone\}
\;.
\end{equation}
With this definition,
\begin{equation}
\label{eq:plkmp1}
p_l + p_k = \hat p_l + \hat p_\mpone + \hat p_k
\;.
\end{equation}
Thus the total momentum is conserved. We have $\hat p_\mpone + \hat p_l = p_l + y\,  p_k$ so
\begin{equation}
y = \frac{\hat p_l\cdot \hat p_\mpone}{p_l\cdot p_k}
\;.
\end{equation}
From $\hat p_\mpone^2 = 0$ we derive
\begin{equation}
-k_\perp^2 = z(1-z)y\,2p_l\cdot p_k
\;.
\end{equation}

Note that if we start with a two parton state, $m=2$, and let one of the two partons, $l$, split to produce parton $\mpone$, then there is precisely one parton $i$ with $i\notin \{l,\mpone\}$ in Eq.~(\ref{eq:deductorboost}) and this is the same as parton $k$ in Eq.~(\ref{eq:CataniSeymourMapping}). That is, the global and local mappings are the same for $\cS_\cY^{[1]}(\mu^2;\nu)$ for $m=2$. The operators $\cS_\cY^{[k]}(\mu^2;\nu)$, with $k$ real or virtual splittings, do depend on the choice of momentum mapping for $k \ge 2$ . 

The local momentum mapping has a feature for thrust that one might regard as peculiar. Suppose that parton $l$ is in the right thrust hemisphere, $l \in \LR$. Then for a small angle splitting, the daughter partons $l$ and $\mpone$ will also be in the right hemisphere. In the case that $k \in \LR$, we split a dipole that is entirely in $R$. Then  Eqs.~(\ref{eq:pi}) and (\ref{eq:plkmp1}) imply that both $\tau_\LR$ and $\tau_\LL$ in Eq.~(\ref{eq:tauLR}) are unchanged by the splitting, so that $\tau = \tau_\LR + \tau_\LL$ is unchanged. Since, in this class of choices for the dipole that splits, the thrust is not changed, the real-virtual cancelation between $\cS^{[1,0]}(\mu^2)$ and $\cS^{[0,1]}(\mu^2)$ simply removes contributions of these dipoles from the calculation of the thrust distribution. 

With $\Lambda$ ordering and a local momentum mapping, the argument in Sec.~\ref{sec:showeratLL} that the shower sums logarithms of thrust at LL level still works, but the argument in Appendix \ref{sec:NLLproof} for cancellations at the NLL level fails. Thus we cannot expect a $\Lambda$-ordered parton shower that uses a local momentum mapping following Eqs.~(\ref{eq:pi}) and (\ref{eq:plkmp1}) to properly sum the logarithms of $\nu$ at NLL accuracy.

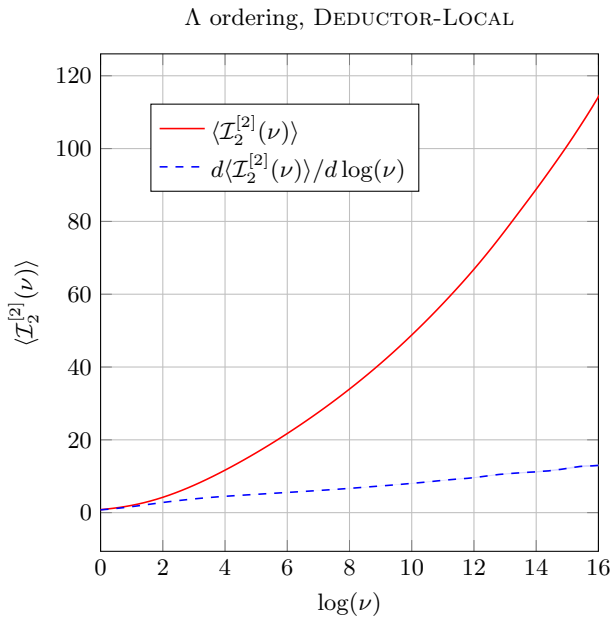
\begin{figure}
\begin{center}

\ifusefigs 

\begin{tikzpicture}
  \begin{axis}[title = {$\Lambda$ ordering, \textsc{Deductor-Local}},
    xlabel={$\log(\nu)$}, ylabel={$\langle\cI^{[2]}_2(\nu)\rangle$},
    xmin=0, xmax=16,
    legend cell align=left,
    every axis legend/.append style = {
    at={(0.1,0.9)},
    anchor=north west}
    ]


     \errorband[red,semithick]{fill=red!30!white, opacity=0.5}
       {Tables/I2-ductCS-lmd.dat}{0}{1}{2}
       \addlegendentry{$\langle \cI^{[2]}_2(\nu)\rangle$}
       
     \errorband[blue,dashed,semithick]{fill=blue!30!white, opacity=0.5}
       {Tables/I2-ductCS-lmd.dat}{0}{3}{4}
       \addlegendentry{$d\langle \cI^{[2]}_2(\nu)\rangle/d\log(\nu)$}

  \end{axis}
\end{tikzpicture}

\fi

\end{center}
\caption{
Plot of $\langle \cI^{[2]}_2(\nu)\rangle$, as in Fig.~\ref{fig:cI2}, for the \textsc{Deductor} splitting functions with the Catani-Seymour local momentum mapping \cite{CataniSeymour}. $\langle \cI^{[2]}_2(\nu) \rangle$ is approximately quadratic in $\log(\nu)$, indicating that $\cI^{[2]}_2(\nu)$ that changes the NLL result.
}
\label{fig:I2-ductCS}
\end{figure}

We can check what happens numerically by calculating $\langle \cI^{[2]}_2(\nu) \rangle$, Eq.~(\ref{eq:cI2def}), using the $\Lambda$-ordered \textsc{Deductor} parton shower algorithm but with the Catani-Seymour momentum mapping substituted for the global momentum mapping. The result is shown as the solid red curve in Fig.~\ref{fig:I2-ductCS}. We note immediately that this result is completely different from the result in Fig.~\ref{fig:cI2}: in the range $\log(\nu) < 8$, $|\langle \cI^{[2]}_2(\nu)\rangle|$ with the global momentum mapping is less than 1  while with the local mapping it reaches values greater than 30. Leaving aside the magnitude of $\langle \cI^{[2]}_2(\nu) \rangle$, if the parton shower algorithm with a local momentum mapping produced NLL accuracy for summing $\log(\nu)$ factors, the graph of $\langle \cI^{[2]}_2(\nu)\rangle$ would be a straight line, but it is not.  The dashed blue curve is $d\langle \cI^{[2]}_2(\nu)\rangle/d\log(\nu)$. This curve is not a constant but rather a straight line. This implies that at large $\log(\nu)$, $\langle \cI^{[2]}_2(\nu) \rangle$ is has contributions up to $\log^{2}(\nu)$. 

We conclude from the combination of the analytical argument and the numerical results that using a local momentum mapping destroys the NLL accuracy of the result from a $\Lambda$-ordered parton shower, although LL accuracy is maintained.


\section{Local momentum mapping with other orderings}
\label{sec:kTorderingCataniSeymour}

As we have seen in Sec.~\ref{sec:momentummapping}, a parton shower algorithm needs to conserve momentum while accommodating the approximation that a parton that splits to two partons was on shell before the splitting. \textsc{Deductor} uses a global recoil strategy that spreads the needed momentum over all of the other partons in the event. With a local momentum mapping in the style of Catani-Seymour, Eq.~(\ref{eq:CataniSeymourMapping}), the recoil momentum is taken up by a single parton, possibly a very soft parton. For this reason the global recoil strategy seems less likely to lead to problems than the local recoil strategy. 

Nevertheless, a local momentum mapping can certainly work. Indeed, we present an argument in Appendix \ref{sec:kTorderingcancellation} that $\cI^{[2]}_2(\nu)$ in \textsc{Deductor} with $k_\LT$ ordering is well behaved. In this construction, the local and global momentum mappings were equivalent in the limits considered. Thus $\cI^{[2]}_2(\nu)$ with $k_\LT$ ordering and a local momentum mapping should be well behaved.

\begin{figure}
\begin{center}

\ifusefigs 
\begin{tikzpicture}
  \begin{axis}[title = {$\beta = 0.0$ ($k_\LT$) ordering, \textsc{PanLocal}},
    xlabel={$\log(\nu)$}, ylabel={$\langle\cI^{[2]}_2(\nu)\rangle$},
    xmin=0, xmax=16,
    legend cell align=left,
    every axis legend/.append style = {
    at={(0.1,0.9)},
    anchor=north west}
    ]

         
     \errorband[red,semithick]{fill=red!30!white, opacity=0.5}
       {Tables/I2-panlocal-beta0.dat}{0}{1}{2}
       \addlegendentry{$\langle \cI^{[2]}_2(\nu)\rangle$}
       
     \errorband[blue,dashed,semithick]{fill=blue!30!white, opacity=0.5}
       {Tables/I2-panlocal-beta0.dat}{0}{3}{4}
       \addlegendentry{$d\langle \cI^{[2]}_2(\nu)\rangle/d\log(\nu)$}
       
  \end{axis}
\end{tikzpicture}

\fi

\end{center}
\caption{
Plot of $\langle \cI^{[2]}_2(\nu)\rangle$, as in Fig.~\ref{fig:cI2}, for a shower with $k_\LT$ $(\beta = 0.0)$ ordering and the Catani-Seymour local momentum mapping \cite{CataniSeymour} according to an algorithm based on the \textsc{PanLocal} dipole shower of Ref.~\cite{DasguptaShowerSum}. For large $\log(\nu)$, $\langle \cI^{[2]}_2(\nu)\rangle$ is approximately linear in $\log(\nu)$, indicating that $\cI^{[2]}_2(\nu)$ leaves the NLL result intact.
}
\label{fig:I2-panlocalbeta0}
\end{figure}

\begin{figure}
\begin{center}

\ifusefigs 

\begin{tikzpicture}
  \begin{axis}[title = {$\beta = 0.5$ ordering, \textsc{PanLocal}},
    xlabel={$\log(\nu)$}, ylabel={$\langle\cI^{[2]}_2(\nu)\rangle$},
    xmin=0, xmax=16,
    legend cell align=left,
    every axis legend/.append style = {
    at={(0.1,0.9)},
    anchor=north west}
    ]

    
     \errorband[red,semithick]{fill=red!30!white, opacity=0.5}
       {Tables/I2-panlocal-betahalf.dat}{0}{1}{2}
       \addlegendentry{$\langle \cI^{[2]}_2(\nu)\rangle$}
       
     \errorband[blue,dashed,semithick]{fill=blue!30!white, opacity=0.5}
       {Tables/I2-panlocal-betahalf.dat}{0}{3}{4}
       \addlegendentry{$d\langle \cI^{[2]}_2(\nu)\rangle/d\log(\nu)$}

  \end{axis}
\end{tikzpicture}

\fi

\end{center}
\caption{
Plot of $\langle \cI^{[2]}_2(\nu)\rangle$, as in Fig.~\ref{fig:cI2}, versus $\log(\nu)$, for a shower with $\beta = 0.5$ ordering and the Catani-Seymour local momentum mapping \cite{CataniSeymour} according to an algorithm based on the \textsc{PanLocal} dipole shower of Ref.~\cite{DasguptaShowerSum}. For large $\log(\nu)$, $\langle \cI^{[2]}_2(\nu)\rangle$ is approximately linear in $\log(\nu)$, indicating that $\cI^{[2]}_2(\nu)$ leaves the NLL result intact.
}
\label{fig:I2-panlocalbetahalf}
\end{figure}

\begin{figure}
\begin{center}

\ifusefigs 
\begin{tikzpicture}
  \begin{axis}[title = {$\beta = 0.0$ ($k_\LT$) ordering, \textsc{PanLocal}},
    xlabel={$\log(\nu)$}, ylabel={$\langle\cI^{[2]}_3(\nu)\rangle$},
    xmin=0, xmax=16,
    legend cell align=left,
    every axis legend/.append style = {
    at={(0.1,0.9)},
    anchor=north west}
    ]

         
     \errorbandMOD[red,semithick]{fill=red!30!white, opacity=0.5}
       {Tables/I23-panlocal-0.0.dat}{0}{1}{2}
       \addlegendentry{$\langle \cI^{[2]}_3(\nu)\rangle$}
       
     \errorbandMOD[blue,dashed,semithick]{fill=blue!30!white, opacity=0.5}
       {Tables/I23-panlocal-0.0.dat}{0}{3}{4}
       \addlegendentry{$d\langle \cI^{[2]}_3(\nu)\rangle/d\log(\nu)$}
       
  \end{axis}
\end{tikzpicture}

\fi

\end{center}
\caption{
Plot of $\langle \cI^{[2]}_3(\nu)\rangle$, as in Fig.~\ref{fig:cI23}, for a shower with $k_\LT$ $(\beta = 0.0)$ ordering and the Catani-Seymour local momentum mapping \cite{CataniSeymour} according to an algorithm based on the \textsc{PanLocal} dipole shower of Ref.~\cite{DasguptaShowerSum}.
}
\label{fig:I23-panlocalbeta0}
\end{figure}

\begin{figure}
\begin{center}

\ifusefigs 

\begin{tikzpicture}
  \begin{axis}[title = {$\beta = 0.5$ ordering, \textsc{PanLocal}},
    xlabel={$\log(\nu)$}, ylabel={$\langle\cI^{[2]}_3(\nu)\rangle$},
    xmin=0, xmax=16,
    legend cell align=left,
    every axis legend/.append style = {
    at={(0.1,0.9)},
    anchor=north west}
    ]

    
     \errorbandMOD[red,semithick]{fill=red!30!white, opacity=0.5}
       {Tables/I23-panlocal-0.5.dat}{0}{1}{2}
       \addlegendentry{$\langle \cI^{[2]}_3(\nu)\rangle$}
       
     \errorbandMOD[blue,dashed,semithick]{fill=blue!30!white, opacity=0.5}
       {Tables/I23-panlocal-0.5.dat}{0}{3}{4}
       \addlegendentry{$d\langle \cI^{[2]}_3(\nu)\rangle/d\log(\nu)$}

  \end{axis}
\end{tikzpicture}

\fi

\end{center}
\caption{
Plot of $\langle \cI^{[2]}_3(\nu)\rangle$, as in Fig.~\ref{fig:cI23}, versus $\log(\nu)$ for a shower with $\beta = 0.5$ ordering and the Catani-Seymour local momentum mapping \cite{CataniSeymour} according to an algorithm based on the \textsc{PanLocal} dipole shower of Ref.~\cite{DasguptaShowerSum}.
}
\label{fig:I23-panlocalbetahalf}
\end{figure}
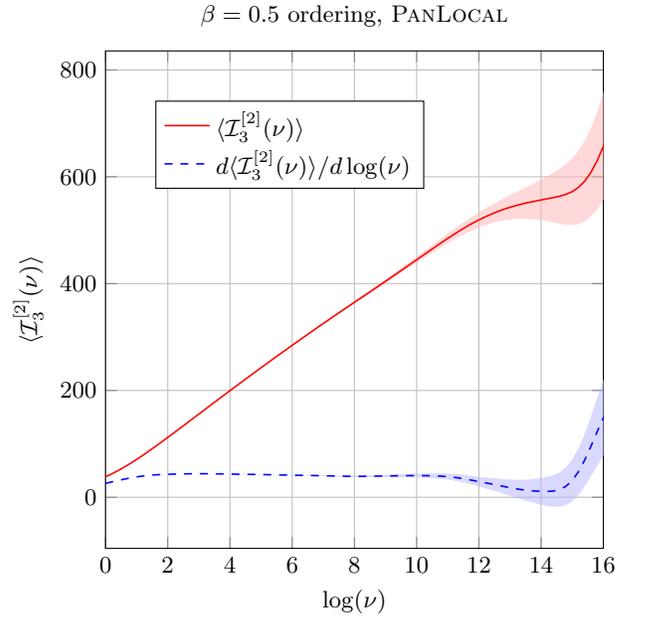

We can investigate this issue by calculating $\langle \cI^{[2]}_2(\nu)\rangle$ using two shower algorithms with a local momentum mapping following Eq.~(\ref{eq:CataniSeymourMapping}). The algorithms we use follow closely the PanLocal shower of Ref.~\cite{DasguptaShowerSum}. 
In the first algorithm that we use, the parameter $\beta$ that defines the ordering variable in the PanLocal algorithm is set to $\beta = 0$. That corresponds to $k_\LT$ ordering. In the second algorithm, we choose $\beta = 0.5$. Roughly, that is half way between $k_\LT$ ordering and $\Lambda$ ordering. Ref.~\cite{DasguptaShowerSum} claims that these \textsc{PanLocal} showers sum the trust distribution at NLL accuracy at leading color.

The results are shown in Figs.~\ref{fig:I2-panlocalbeta0} and \ref{fig:I2-panlocalbetahalf}. In each case, in the range $\log(\nu) < 8$, $|\langle \cI^{[2]}_2(\nu)\rangle|$ reaches values greater than 10, while for $\textsc{Deductor}$ with $\Lambda$ ordering this same quantity is less than 1. Nevertheless, in each case, we see that $\langle \cI^{[2]}_2(\nu)\rangle$ is, to a good approximation, a linear function of $\log(\nu)$ for large $\log(\nu)$. This is consistent with NLL accuracy for summing logarithms of $\nu$. 

In Figs.~\ref{fig:I23-panlocalbeta0} and \ref{fig:I23-panlocalbetahalf}, we plot $\langle \cI^{[2]}_3(\nu)\rangle$ for the two \textsc{PanLocal} shower algorithms. To be consistent with NLL accuracy $\langle \cI^{[2]}_3(\nu)\rangle$ at large $\log(\nu)$ should not contain terms $\log^j(\nu)$ for $j = 3$ or higher. The numerical results are consistent with this NLL expectation. In fact, in each case the highest power of $\log(\nu)$ numerically is $\log^1(\nu)$. The coefficient of $\log^2(\nu)$ vanishes to a good approximation. This tells us that the average value of the scale of the coupling inside the integrations is about $Q^2/\nu$.


\section{Conclusions}
\label{sec:conclusions}

We have explored how to gain direct access to how a parton shower sums large logarithms, following the general program outlined in Ref.~\cite{NSlogsum}. In this paper, we have limited ourselves to electron-positron annihilation and to just one observable, the thrust distribution. We have, however, looked at results for more than one shower algorithm.

The method that we propose works with the appropriate integral transform of the distribution of interest. In this case, we need the Laplace transform $\tilde g(\nu)$, Eq.~(\ref{eq:LaplaceTransform}), of the thrust distribution. We seek to find how $\tilde g(\nu)$ behaves for large $\nu$.

We rearrange the cross section calculation so as to write $\tilde g(\nu)$ in the form from Eq.~(\ref{eq:sigmanualt}),
\begin{equation}
\begin{split}
\label{eq:sigmanualtencore}
\tilde g(\nu) ={}& 
\exp\!\Big(
\sbra{1}\cI(\nu)\sket{\{p,f,c,c\}_{2}}
\Big)
\;.
\end{split}
\end{equation}
Here $\isket{\{p,f,c,c\}_{2}}$ is a color and flavor singlet $q \bar q$ basis state with $p_1 + p_2 = Q$ and the operator $\cI(\nu)$ is an integral,
\begin{equation}
\label{eq:Inudefencore}
\cI(\nu) = \int_{\mu_\Lf^2}^{Q^{2}}\!\frac{d\bar \mu^2}{\bar\mu^2}\,
\cS_\cY(\bar \mu^2;\nu)
\;.
\end{equation}
We expand $\cI(\nu)$ in powers of the shower evolution operator $\cS(\mu^2)$. Then the coefficients $\cI^{[k]}(\nu)$, proportional to $k$ powers of $\cS(\mu^2)$, can be further expanded as 
\begin{equation}
\label{eq:Iknbis}
\cI^{[k]}(\nu) = 
\sum_{n=k}^\infty \left[\frac{\as(Q^2/\nu)}{2\pi}\right]^n
\cI^{[k]}_n(\nu)
\;,
\end{equation}
in which the strong coupling is evaluated at a fixed scale $Q^2/\nu$. Thus the shower result is quite directly expressed in exponentiated form in terms of an operator $\cI(\nu)$ with a known perturbative expansion.

For the \textsc{Deductor} shower algorithm with either $\Lambda$ or $k_\LT$ ordering, $\cI^{[1]}(\nu)$ provides the standard NLL summation of $\log(\nu)$ factors.\footnote{The current \textsc{Deductor} code with $\Lambda$ ordering, as distinct from the algorithm that it is based on, lacks the term with coefficient $\beta_1$ needed to evaluate the dependence of $\as((1-z) y Q^2)$ on $(1-z)$. This changes the \textsc{Deductor} result at the NLL level.} In order for the contributions $\cI^{[k]}_n(\nu)$ for $k \ge 2$ to not spoil the NLL summation, $\cI^{[k]}_n(\nu)$ should not contain more than $n-1$ powers of $\log(\nu)$.

For the \textsc{Deductor} shower algorithm with its default $\Lambda$ ordering, we find analytically that $\cI^{[k]}_n(\nu)$ does not contain more than $n-1$ powers of $\log(\nu)$.

We have no such result for \textsc{Deductor} with $k_T$ ordering, but we outline an argument in Appendix \ref{sec:kTorderingcancellation} that real-virtual cancellations in $\cI^{[2]}_2(\nu)$ reduce its large $\nu$ behavior from $\log^4(\nu)$ to $\log^1(\nu)$.

We evaluate $\cI^{[2]}_2(\nu)$ numerically. In order not to spoil NLL summation, its large $\nu$ behavior should be no more than $\log^1(\nu)$. For the \textsc{Deductor} algorithm with $\Lambda$ ordering but with a local momentum mapping instead of the global momentum mapping used in \textsc{Deductor}, we find $\log^2(\nu)$ behavior, implying a failure of NLL accuracy (Fig.~\ref{fig:I2-ductCS}). In other cases, we find $\log^1(\nu)$ behavior, consistently with NLL accuracy. These cases include \textsc{Deductor}-$\Lambda$ (Fig.~\ref{fig:cI2}), \textsc{Deductor}-$k_\LT$ (Fig.~\ref{fig:cI2-kT}), \textsc{PanLocal}-($\beta$=0) (Fig.~\ref{fig:I2-panlocalbeta0}), and \textsc{PanLocal}-($\beta$=0.5) (Fig.~\ref{fig:I2-panlocalbetahalf}).

We also evaluate $\cI^{[2]}_3(\nu)$ numerically for the shower algorithms \textsc{Deductor}-$\Lambda$ (Fig.~\ref{fig:cI23}), \textsc{Deductor}-$k_\LT$ (Fig.~\ref{fig:cI23-kT}), \textsc{PanLocal}-($\beta$=0) (Fig.~\ref{fig:I23-panlocalbeta0}), and \textsc{PanLocal}-($\beta$=0.5) (Fig.~\ref{fig:I23-panlocalbetahalf}). In each case, we find large $\log(\nu)$ behavior with no more than 2 powers of $\log(\nu)$, consistently with NLL accuracy.

We emphasize in this paper writing the appropriate integral transform of the distribution of interest, such as the thrust distribution, as an exponential and examining the exponent $\cI(\nu)$. However, it is also possible to simply look directly at the distribution of interest as it is generated by a given parton shower. For this, one needs to simulate collisions at large values of $Q^2$. We have not pushed this method to nearly as large a value of $Q^2$ as in Ref.~\cite{DasguptaShowerSum}. However, we find that, at least for electron-positron annihilation, this direct method can be useful.

Specifically, we examine directly the thrust distribution $\tau g(\tau)$ for \textsc{Deductor} with $\Lambda$ and $k_\LT$ ordering, using $Q^2 = (10 \TeV)^2$. With $\Lambda$ ordering, this works well (Figs.~\ref{fig:thrustLambdatest} and \ref{fig:thrustLambdatestmore}). With $k_\LT$ ordering (Fig.~\ref{fig:thrustLambdaKTtest}), the agreement with the analytic NLL expectation is not as good. However, when we compare $\tau g(\tau)$ to the NLL expectation at a sequence of values of $Q^2$, we find what appears to be convergence to the NLL result as $Q^2$ increases (Fig.~\ref{fig:thrusthigherQ}).

There are several avenues available for future research that extends the results of this paper. 

First, the method of this paper applies to several observables in electron-positron annihilation. We have tried variations on the shower algorithm examined, but have looked at only one observable, the thrust distribution. It is certainly worthwhile to see what patterns emerge if we look at other observables.

Second, the method developed in Ref.~\cite{NSlogsum} applies to observables in hadron-hadron collisions as well as in electron-positron collisions. It is of interest to see how this method works in practice for some hadron-hadron observables, starting with the $k_\LT$ distribution in the Drell-Yan process.

Third, we construct numerical implementations of $\cI^{[2]}_2(\nu)$ and $\cI^{[2]}_3(\nu)$ for the particular observable examined and for several shower algorithms. This allows one to test numerically if the large $\nu$ behaviors of $\cI^{[2]}_2(\nu)$ and $\cI^{[2]}_3(\nu)$ are consistent with NLL summation. When we find for a certain shower algorithm that NLL summation fails at the level of $\cI^{[2]}_2(\nu)$ or $\cI^{[2]}_3(\nu)$, then NLL summation fails for that shower algorithm and observable. However, if NLL summation is not spoiled by $\cI^{[2]}_2(\nu)$ or $\cI^{[2]}_3(\nu)$, it could still fail in $\cI^{[k]}_n(\nu)$ for some larger values of $k$ and $n$. Thus it would be valuable to have numerical implementations of $\cI^{[k]}_n(\nu)$ for some larger values of $k$ and $n$. Then one would have more stringent numerical tests of NLL summation for a given shower algorithm and a given observable.

Fourth, it would be helpful to have analytical insight into the behavior of the operators $\cI^{[k]}(\nu)$ for $k \ge 3$ in cases that are similar to the thrust distribution using a $k_\LT$-ordered shower.

We close with the observation that it is expecting a lot to expect that a first order shower algorithm will sum logarithms at the LL or NLL level. If we had a parton shower based on splitting functions at order $\as^N$ \cite{NSAllOrder, NSlogsum}, then we could expect to correctly produce contributions to $\cI(\nu)$ of order $\as^n \log^{j}(\nu)$ with $n \le N$, $j \le n+1$. We might not correctly produce contributions of order $\as^n \log^{j}(\nu)$ with $n > N$, $j \le n+1$ because we lack the order $\as^n$ contributions to the shower splitting functions. However, contributions of order $\as^n \log^{j}(\nu)$ with $j > n+1$ should vanish because these contributions can never be provided by $\as^n$ contributions to the shower splitting functions. Currently, all that we have (in several variations) is a first order shower, $N=1$. Thus we can expect to correctly produce contributions of order $\as^1 \log^{2}(\nu)$ and $\as^1 \log^{1}(\nu)$. We can also expect to obtain exponentiation of logarithms of $\nu$: contributions of order $\as^n \log^{j}(\nu)$ with $j > n+1$ should vanish.  With care, we can hope to have LL or NLL summation of $\log(\nu)$ factors, but this relies on incorporating the most important parts of higher order splitting operators into the first order operator $\cS$.

\acknowledgments{ 
This work was supported in part by the United States Department of Energy under grant DE-SC0011640. The work benefited from access to the University of Oregon high performance computer cluster, Talapas.  
}

\appendix


\section{Structure of $\cS_\cY$ at NLL accuracy}
\label{sec:NLLproof}

We examine $\cS_\cY^{[k]}(xQ^2/\mu)$ for $x$ of order 1 and $k \ge 2$. We prove that this operator has at most $n-1$ factors of $\log(\nu)$ at order $\as^{n}(Q^2/\nu)$. 

Recall from Sec.~\ref{sec:showeratLL} that $\cY^{[k]}(xQ^2/\mu)$ for $x$ of order 1 and $k \ge 2$ has at most $n$ factors of $\log(\nu)$ at order $\as^{n}(Q^2/\nu)$. 

We also note that $\cS_l^{[1,0]}(x Q^2/\nu)$ for $x$ of order 1 has one power of $\log(\nu)$ at order $\as(Q^2/\nu)$, where the $\log(\nu)$ factor arises from an integration $d(1-z)/(1-z)$ down to a lower limit proportional to $1/\nu$, as in Eq.~(\ref{eq:lambday3}). Thus $\cS_l^{[1,0]}(x Q^2/\nu)$ for $x$ of order 1 has at most $n$ powers of $\log(\nu)$ at order $\as^{n}(Q^2/\nu)$.

To proceed, we prove that $\cS_\cY^{[k]}(xQ^2/\mu)$ with $k=2$ contains at most $n-1$ factors of $\log(\nu)$ at order $\as^{n}(Q^2/\nu)$ and we prove that if this property holds for $k = 2,3,\dots,N$, then it holds for $k=N+1$.

Consider Eq.~(\ref{eq:cScYnp1}) for $\cS_\cY^{[k+1]}(x Q^2/\nu;\nu)$ for $k \ge 2$. In the first term, at order $\as^{k+1}(Q^2/\nu)$, there are $k$ powers of $\log(\nu)$ from $\cY^{[k]}$ and one power from $\cS_l^{[1,0]}$. In the second term (if $k \ge 3$) at order $\as^{k+1}(Q^2/\nu)$ there are there are $k+1-j$ powers of $\log(\nu)$ from $\cY^{[k+1-j]}$ and $j-1$ powers from $\cS_\cY^{[j]}$, for a total of just $k$ powers of $\log(\nu)$. That is, this contribution is NNLL. In the third term, at order $\as^{k+1}(Q^2/\nu)$ there is one power of $\log(\nu)$ from $\cS_l^{[1,0]}$ and $k-1$ powers of $\log(\nu)$ from $\cS_\cY^{[k]}$, for a total of $k$ powers of $\log(\nu)$. That is, this contribution is NNLL. If we expand the NNLL contributions to higher order in $\as(Q^2/\nu)$, we add just one power of $\log(\nu)$ per $\as$, so the contributions remain NNLL. This gives us
\begin{align}
\label{eq:cScYnp1NLL}
\cS_\cY^{[k+1]}(x Q^2/\nu;\nu) \hskip - 1.7 cm &
\notag
\\={}&  \notag
\sum_l
\PL\cY^{[k]}(x Q^2/\nu;\nu)
\omP{\cS_l^{[1,0]}(x Q^2/\nu)}\PR 
e^{- \xi_l^\mathrm{op} x}
\\ & 
+ \mathrm{NNLL}
\;.
\end{align}
This leaves us with an NLL contribution if the NLL contribution does not cancel. This result does not include $\cS_\cY^{[2]}$. For $\cS_\cY^{[2]}$, Eq.~(\ref{eq:cScY2result1}) gives us 2 powers of $\log(\nu)$ at order $\as^2(Q^2)$. This is an NLL contribution if the NLL contribution does not cancel.

If we use Eq.~(\ref{eq:cScYnp1NLL}), then we need information on $\cY^{[k]}$. We can use Eq.~(\ref{eq:cYnp1}) for $\cY^{[k+1]}(x Q^2/\nu;\nu)$ for $k\ge 1$. In the first term at order $\as^{k+1}(Q^2/\nu)$ there are $k$ powers of $\log(\nu)$ from $\cY^{[k]}$ and one power of $\log(\nu)$ from $\cS_l^{[1,0]}$, giving us a total of $k+1$ powers of $\log(\nu)$. This is an NLL contribution. In the second term (for $k \ge 2$) at order $\as^{k+1}(Q^2/\nu)$ there are $k+1-j$ powers of $\log(\nu)$ from $\cY^{[k+1-j]}$ and $j-1$ powers of $\log(\nu)$ from $\cS_\cY^{[j]}$, giving us a total of $k$ powers of $\log(\nu)$. This is an NNLL contribution.  In the third term at order $\as^{k+1}(Q^2/\nu)$ there are $k$ powers of $\log(\nu)$ from $\cS_\cY^{[k+1]}$. This is an NNLL contribution. Again, if we expand the NNLL contributions to higher order in $\as(Q^2/\nu)$, we add just one power of $\log(\nu)$ per $\as$, so the contributions remain NNLL. This leaves us with
\begin{align}
\label{eq:cYnp1NLL}
\cY^{[k+1]}(x Q^2/\nu;\nu) \hskip - 1.8 cm &
\notag
\\ ={}& \notag
-\sum_l
\int_{0}^{x}\!\frac{dx_1}{x_1}\
\cY^{[k]}(x_1 Q^2/\nu;\nu)\,
\P{\cS_l^{[1,0]}(x_1 Q^2/\nu)}
\\&\qquad\times \notag
(1 - e^{- \xi_l^\mathrm{op} x_1})
\\ &  + \mathrm{NNLL}
\;.
\end{align}
This derivation does not include $\cY^{[1]}$. For $\cY^{[1]}$  we can use Eq.~(\ref{eq:cY1result}), which gives us just Eq.~(\ref{eq:cYnp1NLL}) with $\cY^{[0]}$ replaced by 1 and no NNLL additional contribution.

Eq.~(\ref{eq:cYnp1NLL}) gives us a recursion relation that we can solve to NLL accuracy in the form
\begin{align}
\label{eq:cYNLLsoln}
\cY^{[k]}(x Q^2&/\nu;\nu) 
\notag
\\ ={}& \notag
(-1)^k \sum_{l_1 \dots l_k}
\int_{0}^{x}\!\frac{dx_1}{x_1}
\int_{0}^{x_1}\!\frac{dx_2}{x_2}\cdots
\int_{0}^{x_{k-1}}\!\frac{dx_k}{x_k}
\\&\times \notag
\P{\cS_{l_k}^{[1,0]}(x_k Q^2/\nu)}(1 - e^{- \xi_{l_k}^\mathrm{op} x_k})
\\ \notag& \times\cdots
\\&\times \notag
\P{\cS_{l_2}^{[1,0]}(x_2 Q^2/\nu)}(1 - e^{- \xi_{l_2}^\mathrm{op} x_2})
\\&\times \notag
\P{\cS_{l_1}^{[1,0]}(x_1 Q^2/\nu)}(1 - e^{- \xi_{l_1}^\mathrm{op} x_1})
\\ &  + \mathrm{NNLL}
\;.
\end{align}
We can substitute this solution for $\cY^{[k]}$ into Eq.~(\ref{eq:cScYnp1NLL}) to give us
\begin{align}
\label{eq:cScYNLLsoln0}
\cS_\cY^{[k+1]}(&x_0 Q^2/\nu;\nu) 
\notag
\\ ={}& \notag
(-1)^k \sum_{l_0 \dots l_k}
\int_{0}^{x_0}\!\frac{dx_1}{x_1}
\int_{0}^{x_1}\!\frac{dx_2}{x_2}\cdots
\int_{0}^{x_{k-1}}\!\frac{dx_k}{x_k}
\\&\times \notag
\PL\P{\cS_{l_k}^{[1,0]}(x_k Q^2/\nu)}(1 - e^{- \xi_{l_n}^\mathrm{op} x_k})
\\ \notag& \times\cdots
\\&\times \notag
\P{\cS_{l_2}^{[1,0]}(x_2 Q^2/\nu)}(1 - e^{- \xi_{l_2}^\mathrm{op} x_2})
\\&\times \notag
\P{\cS_{l_1}^{[1,0]}(x_1 Q^2/\nu)}(1 - e^{- \xi_{l_1}^\mathrm{op} x_1})
\\&\times \notag
\omP{\cS_{l_0}^{[1,0]}(x_0 Q^2/\nu)}\PR\, 
e^{- \xi_{l_0}^\mathrm{op} x_0}
\\ &  + \mathrm{NNLL}
\;.
\end{align}
The explicit exponential exponential factors restrict the $x_i$ integrations to $x_i$ of order 1 (as we have already seen). We now want to find how many factors of $\log(\nu)$ are contained in the operators $\cS_{l}^{[1,0]}(x Q^2/\nu)$. Since $\log(x/\nu)$ is equivalent for this purpose to $\log(1/\nu)$ when $x$ is of order 1, we can replace all of the $x_i$ factors in the arguments of $\cS_{l}^{[1,0]}(x Q^2/\nu)$ by 1.

In Eq.~(\ref{eq:cScYNLLsoln0}), we have factors $\exp(-\hat\xi_l^\mathrm{op} x_i)$. The parameters $\xi_l$, are defined in Eq.~(\ref{eq:cldef}). They are close to 1: $\xi_l - 1$ is proportional to $[1 - \cos(\theta(l,\vec n_\LT))]$. It is a good approximation to take the thrust axis $\vec n_\LT$ to be the direction of either the quark or the antiquark in the $q$-$\bar q$ state at the start of the shower. Then the angle between $\vec p_l$ at a later stage of the shower and $\vec n$ is determined by the emission angles at the intervening stages. But in order to accumulate the maximal number of $\log(\nu)$ factors in these splittings, all of these emission angles must be small. That is, if we expand $\exp(-\xi_l x_i)$ in powers of $[1 - \cos(\theta)]$, where $\theta$ is one of the splitting angles, then a factor $[1 - \cos(\theta)]$ will eliminate a $\log(\nu)$ factor in an integration $d\cos(\theta)/[1 - \cos(\theta)]$ with limits analogous to the limits in Eq.~(\ref{eq:anglerange}). We conclude that for the purpose of our present NLL calculation we can set all of the $\xi_l^\mathrm{op}$ factors in Eqs.~(\ref{eq:cScYNLLsoln0}) to 1.

These changes gives us
\begin{align}
\label{eq:cScYNLLsoln}
\cS_\cY^{[k+1]}(&x_0 Q^2/\nu;\nu) \sket{\{p,f,c,c'\}_m}
\notag
\\ ={}& \notag
(-1)^k \sum_{l_0 \dots l_k}
\int_{0}^{x_0}\!\frac{dx_1}{x_1}
\int_{0}^{x_1}\!\frac{dx_2}{x_2}\cdots
\int_{0}^{x_{k-1}}\!\frac{dx_k}{x_k}
\\&\times \notag
\PL\P{\cS_{l_k}^{[1,0]}(Q^2/\nu)}(1 - e^{- x_k})
\\ \notag& \times\cdots
\\&\times \notag
\P{\cS_{l_2}^{[1,0]}(Q^2/\nu)}(1 - e^{- x_2})
\\&\times \notag
\P{\cS_{l_1}^{[1,0]}(Q^2/\nu)}(1 - e^{- x_1})
\\&\times \notag
\omP{\cS_{l_0}^{[1,0]}(Q^2/\nu)}\PR\, 
e^{- x_0}
\\ & \notag \times \sket{\{p,f,c,c'\}_m}
\\ &  + \mathrm{NNLL}
\;.
\end{align}
The first $\cS_{l}^{[1,0]}(Q^2/\nu)$ factor in Eq.~(\ref{eq:cScYNLLsoln}) is $$\omP{\cS_{l_0}^{[1,0]}(Q^2/\nu)} = \cS_{l_0}^{[1,0]}(Q^2/\nu) - \iP{\cS_{l_0}^{[1,0]}(Q^2/\nu)}\;.$$ The contribution from $\iP{\cS_{l_0}^{[1,0]}(Q^2/\nu)}$ is rather simple and we will consider it later. 

We begin by considering the contribution from ${\cS_{l_0}^{[1,0]}(Q^2/\nu)}$. This operator, acting on the state $\isket{\{p,f,c,c'\}_m}$, produces a linear combination of states with $m+1$ partons, $\isket{\{\hat p,\hat f,\hat c,\hat c'\}_{m+1}}$,
\begin{equation}
  \label{eq:cS10onm}
  \begin{split}
    \sum_{l_0 = 1}^m&\cS_{l_0}^{[1,0]}(Q^2/\nu)
    e^{- x_0}
    \sket{\{p, f, c, c'\}_m} 
    \\\approx{}& 
    - 
    \sum_{l_0 = 1}^m \sum_{\substack{k_0 = 1 \\ k_0 \ne l_0}}^m 
    \cC_0(l_0,k_0)\sket{\{c,c'\}_{m}}
    \\&\times
    \int\!\frac{d\phi_0}{2\pi}
    \int\!\frac{dz_0}{1-z_0}\
    \frac{\as\big(\lambda_\LR (1-z_0) Q^2/(\nu a_{l_0})\big)}{2\pi}
    \\&\times
    \Theta\!\left(\frac{a_{l_0}}{\nu\,\vartheta(l_0,k_0)} 
      < 1-z_0 < 1\right)
    \\&\times
    e^{- x_0}
    \sket{\{\hat p, \hat f\}_{m+1}}
    \;.
  \end{split}
\end{equation}
Here we use the approximate form of $\cS^{[1,0]}(Q^2/\nu)$ given in Eq.~(\ref{eq:S10}). We split parton $l_0$ with dipole partner parton $k_0$, creating a new parton $m+1$, which we consider to be a gluon. The color operator is
\begin{equation}
\label{eq:cl0k0def}
\cC_0(l_0,k_0) = T_{l_0} \otimes  T_{k_0}^\dagger
+ T_{k_0} \otimes  T_{l_0}^\dagger
\;,
\end{equation}
as defined below Eq.~(\ref{eq:S10}). We have specified a scale argument based on the transverse momentum for the splitting for $\as$. The new momentum $\hat p_{m+1}$ and the new momentum $\hat p_l$ are given by the splitting variables $y = 1/\nu$, $z_0$ and $\phi_0$. The new momenta $\hat p_i$ for $i \ne l_0, m+1$ are slightly different from the starting momenta, as specified by the momentum mapping. 

Let us consider what the one of the operators, $\iP{\cS_{l_i}^{[1,0]}(Q^2/\nu)}$, in Eq.~(\ref{eq:cScYNLLsoln}) does to this state. We consider the quantity
\begin{equation}
\label{A1def}
\sket{A_i} = \sum_{l_i = 1}^{m+1} \P{\cS_{l_i}^{[1,0]}(Q^2/\nu)}
(1 - e^{- x_i})
\sket{\{\hat p, \hat f, \hat c, \hat c'\}_{m+1}}
\;.
\end{equation}
Again, we use the approximate form of $\cS^{[1,0]}(Q^2/\nu)$ given in Eq.~(\ref{eq:S10}), so that
\begin{equation}
  \begin{split}
    \label{eq:splitting1a}
    \sket{A_i}
    \approx{}& 
    -
    \sum_{l = 1}^{m+1} \sum_{\substack{k = 1 \\ k \ne l}}^{m+1} 
    \cC(l,k)
    \\&\times
    \int\!\frac{d\phi}{2\pi}
    \int\!\frac{dz}{1-z}\
    \frac{\as\big(\lambda_\LR (1-z)Q^2/(\nu \hat a_{l})\big)}{2\pi}
    \\&\times
    \Theta\!\left(\frac{1}{\nu\,\hat\vartheta(l,k)} < \frac{1-z}{\hat a_l} < \frac{1}{\hat a_l}\right)
    \\&\times
    (1 - e^{- x_i})
    \sket{\{\hat p, \hat f, \hat c, \hat c'\}_{m+1}}
    \;.
  \end{split}
\end{equation}
Here the hats in $\hat\vartheta(l,k)$ and $\hat a_{l}$ indicate that these quantities are based on the momenta in $\isket{\{\hat p, \hat f\}_{m+1}}$. In Eq.~(\ref{eq:splitting1a}), we split parton $l$ with dipole partner parton $k$, creating a new parton $m+2$, which we consider to be a gluon.\footnote{We omit splittings $\Lg \to q\bar q$ since these splittings lack a soft singularity. For a $q \to q\Lg$ or $\bar q \to \bar q \Lg$ splitting from an $m+1$ parton state, the daughter gluon is labelled $m+2$.} However, the $\iP{\cdots}$ operation, Eqs.~(\ref{eq:Pcolorform1}) and (\ref{eq:Pcolorform2}), returns us to the starting momentum and flavor state $\isket{\{\hat p, \hat f\}_{m+1}}$. With the $\iP{\cdots}$ operation, the color operator is 
\begin{equation}
\begin{split}
\label{eq:cl1k1def}
\cC(l,k) ={}& 
\P{T_{l} \otimes  T_{k}^\dagger
+ T_{k} \otimes  T_{l}^\dagger}
\\={}&
\bm T_{l}\!\cdot\!\bm T_{k} \otimes  1 
+ 1 \otimes \bm T_{l}\!\cdot\!\bm T_{k}
\;.
\end{split}
\end{equation}
In the first term in the second line, the operator $\bm T_{l}\!\cdot\!\bm T_{k}$ operates on the ket color state and leaves the number of partons in the color state unchanged. The operator inserts a color matrix $T^a$ with gluon color index $a$ on line $l$ and another $T^a$ on line $k$. The dot in $\bm T_{k}\!\cdot\!\bm T_{l}$ indicates a sum over $a$. In the second term, the same operator is applied to the bra state.

There is an integration over the splitting variables $\phi$ and $z$. It will prove helpful to define a function $L(w,u)$ given by performing this integration,
\begin{equation}
\label{eq:Lwudef}
  L(w,u) = \int_0^{2\pi}\!\frac{d\phi}{2\pi}
  \int_{1/w}^{1/u}\!\frac{dx}{x}\
  \frac{\as(\lambda_\LR x Q^2/\nu)}{2\pi}
  \;.
\end{equation}
This function is to be expanded in powers of $\as(Q^2/\nu)$. At lowest order, this integration gives simply $[\as/(2\pi)]\log(w/u)$. At higher orders in an expansion in powers of $\as(Q^2/\nu)$ the result is more complicated. With this notation,
\begin{equation}
  \begin{split}
    \label{eq:splitting1b}
    \sket{A_i}
    \approx{}& 
    -
    \sum_{l = 1}^{m+1} \sum_{\substack{k = 1 \\ k \ne l}}^{m+1} 
    \cC(l,k)\,
    L\big(\nu\,\hat\vartheta(l,k), \hat a_{l}\big)
    \\&\times
    (1 - e^{- x_i})
    \sket{\{\hat p, \hat f, \hat c, \hat c'\}_{m+1}}
    \;.
  \end{split}
\end{equation}
We break up the sums in the form
\begin{align}
\label{eq:splitting1c}
    \sket{A_i}
    \approx{}& 
    -\bigg\{
    \sum_{\substack{l = 1\\ l \ne l_0}}^{m} \sum_{\substack{k = 1 \\ k \ne l}}^{m+1} 
    \cC(l,k)
    L\big(\nu\,\hat\vartheta(l,k), \hat a_{l}\big)
    \notag
    \\&+ 
    \sum_{\substack{k = 1 \\ k \ne l_0}}^{m} 
    \cC(l_0,k)
    L\big(\nu\,\hat\vartheta^2(l_0,k), \hat a_{l_0}\big)
    \\&+ 
    \sum_{\substack{k = 1 \\ k \ne l_0}}^{m} 
    \cC(m+1,k)
    L\big(\nu\,\hat\vartheta^2(m+1,k), \hat a_{\mpone}\big)
     \notag
    \\&+ 
    \cC(l_0,m+1)
    L\big(\nu\,\hat\vartheta^2(l_0,m+1), \hat a_{l_0}\big)
     \notag
    \\&+ 
    \cC(m+1,l_0)
    L\big(\nu\,\hat\vartheta(m+1,l_0), \hat a_{m+1}\big)
    \bigg\}
    \notag
    \\&\times
    (1 - e^{- x_i})
    \sket{\{\hat p, \hat f, \hat c, \hat c'\}_{m+1}}
    \;. \notag
\end{align}
Now, as long as neither $l$ nor $k$ equals $m+1$, the angle variable $\hat\vartheta(l,k)$ is very close to the
corresponding angle variable $\vartheta(l,k)$ in the state $\isket{\{p,f,c,c'\}_{m}}$ before the first splitting. The
angle variable $\hat\vartheta(m+1,k)$ for $k \ne l_0$ is very close to $\vartheta(l_0,k)$ in the state before the first
splitting, since partons $l_0$ and $m + 1$ are nearly collinear in the integration region that can lead to a $\log(\nu)$
factor in the first splitting. Thus we regard these angles as fixed when calculating $\cS_\cY^{[k+1]}(x_0 Q^2/\nu;\nu)
\isket{\{p,f,c,c'\}_m}$. On the other hand, $\hat\vartheta(l_0,m+1)$ is the angle variable for the first splitting and
is thus an integration variable in this calculation. Integrating over this variable can produce a $\log(\nu)$
factor. Thus we treat $\hat\vartheta(l_0,m+1)$ as potentially small in Eq.~(\ref{eq:splitting1c}), but we treat the
other angle variables as being finite. For the purpose of finding $\log(\nu)$ factors, we simply replace these finite
angle variables by 1. These substitutions give us 
\begin{equation}
  \begin{split}
    \label{eq:splitting1d}
    \sket{A_i}
    \approx{}& 
    -\bigg\{
    \sum_{\substack{l = 1\\ l \ne l_0}}^{m} \sum_{\substack{k = 1 \\ k \ne l}}^{m+1} 
    \cC(l,k) L\big(\nu, \hat a_{l}\big)
    \\&+ 
    \sum_{\substack{k = 1 \\ k \ne l_0}}^{m} 
    \cC(l_0,k) L\big(\nu,\hat a_{l_0}\big)
    \\&+ 
    \sum_{\substack{k = 1 \\ k \ne l_0}}^{m} 
    \cC(m+1,k) L\big(\nu, \hat a_{\mpone}\big)
    \\&+ 
    \cC(l_0,m+1) L\big(\nu\,\hat\vartheta(l_0,m+1), \hat a_{l_0}\big)
    \\&+ 
    \cC(l_0,m+1) L\big(\nu\,\hat\vartheta(l_0,m+1), \hat a_{m+1}\big)
    \bigg\}
    \\&\times
    (1 - e^{- x_i})
    \sket{\{\hat p, \hat f, \hat c, \hat c'\}_{m+1}}
    \;.
  \end{split}
\end{equation}

In two of the terms in Eq.~(\ref{eq:splitting1d}), the parameter $\hat a_{m+1}$ appears. This parameter is large when the momentum fraction $1-z_0$ of parton $m+1$ in the first splitting is small: 
\begin{equation}
\hat a_\mpone \approx  \frac{a_{l_0}}{1-z_0}
\;.
\end{equation}
We also note that the angle variable $\hat\vartheta(l_0,\mpone)$ is proportional to $1/(1-z_0)$ according to Eq.~(\ref{eq:splittinganglelimit}). We have 
\begin{equation}
  \hat\vartheta(l_0,\mpone)
  \approx
  \frac{a_{l_0}}{\nu (1-z_0)}
  \;.
\end{equation}
Combining these equations gives us
\begin{equation}
\hat a_\mpone \approx  \nu\hat\vartheta(l_0,\mpone)
\;.
\end{equation}
With this replacement, the function $L$, Eq.~(\ref{eq:Lwudef}), in the last term in Eq.~(\ref{eq:splitting1d}) is approximately
\begin{equation}
L\big(\nu\,\hat\vartheta(l_0,m+1), \hat a_{m+1}\big)
\approx L(\hat a_{m+1},\hat a_{m+1}) = 0
\;.
\end{equation}
In the fourth term in Eq.~(\ref{eq:splitting1d}), we use this replacement to eliminate $\hat\vartheta(l_0,m+1)$ in favor of $\hat a_\mpone$. With these substitutions, we have
\begin{equation}
  \begin{split}
    \label{eq:splitting1e}
    \sket{A_i}
    \approx{}& 
    -\bigg\{
    \sum_{\substack{l = 1\\ l \ne l_0}}^{m} \sum_{\substack{k = 1 \\ k \ne l}}^{m+1} 
    \cC(l,k) L\big(\nu, \hat a_{l}\big)
    \\&+ 
    \sum_{\substack{k = 1 \\ k \ne l_0}}^{m} 
    \cC(l_0,k) L\big(\nu, \hat a_{l_0}\big)
    \\&+ 
    \sum_{\substack{k = 1 \\ k \ne l_0}}^{m} 
    \cC(m+1,k) L\big(\nu,\hat a_{m+1}\big)
    \\&+ 
    \cC(l_0,m+1) L\big(\hat a_{m+1}, \hat a_{l_0}\big)
    \bigg\}
    \\&\times
    (1 - e^{- x_i})
    \sket{\{\hat p, \hat f, \hat c, \hat c'\}_{m+1}}
    \;.
  \end{split}
\end{equation}

Using the definition (\ref{eq:Lwudef}) of $L(w,u)$, this function in the last term can be written as
\begin{equation}
L\big(\hat a_{m+1}, \hat a_{l_0}\big) = -L\big(\nu, \hat a_{m+1}\big) + L\big(\nu, \hat a_{l_0}\big)\;.
\end{equation}
In the sum in the second term in Eq.~(\ref{eq:splitting1e}) we can add and subtract a contribution from $k = m+1$. After adding this contribution, the sum includes $k = m+1$, so that this sum can be combined with the sums in the first term. Then in the first term we can include $l = l_0$ in the sum over $l$. In the third term in Eq.~(\ref{eq:splitting1e}) we can add and subtract a contribution from $k = l_0$, so that after adding this contribution the sum includes $k = l_0$. With these changes, we have
\begin{equation}
  \begin{split}
    \label{eq:splitting1f}
    \sket{A_i}
    \approx{}& 
    -\bigg\{
    \sum_{l = 1}^{m} \sum_{\substack{k = 1 \\ k \ne l}}^{m+1} 
    \cC(l,k) L\big(\nu, \hat a_{l}\big)
    \\&+ 
    \sum_{k=1}^{m} 
    \cC(m+1,k) L\big(\nu, \hat a_{m+1}\big)
    \\&- 
    2 \cC(l_0,m+1) L\big(\nu, \hat a_{m+1}\big)
    \bigg\}
    \\&\times
    (1 - e^{- x_i})
    \sket{\{\hat p, \hat f, \hat c, \hat c'\}_{m+1}}
    \;.
  \end{split}
\end{equation}

In the first term in Eq.~(\ref{eq:splitting1f}), we can use color conservation to write
\begin{equation}
\begin{split}
\label{eq:cl1k1sum}
\sum_{\substack{k = 1 \\ k \ne l}}^{m+1} \cC(l,k) 
={}& 
\sum_{\substack{k = 1 \\ k \ne l}}^{m+1}
[ \bm T_{l}\!\cdot\! \bm T_{k} \otimes  1 
+ 1 \otimes \bm T_{l}\!\cdot\! \bm T_{k}]
\\={}& - [ \bm T_{l}\!\cdot\!\bm T_{l} \otimes  1 
+ 1 \otimes \bm T_{l}\!\cdot\!\bm T_{l}]
\\={}& - 2 C_{l}[1 \otimes 1]
\;,
\end{split}
\end{equation}
where $C_{l} = C_\LA$ if parton $l$ is a gluon and $C_{l} = C_\LF$ if parton $l$ is a quark or antiquark. The same applies to the second term:
\begin{equation}
\begin{split}
\label{eq:mp1k1sum}
\sum_{k=1}^{m} \cC(m+1,k) 
=  - 2 C_\LA[1 \otimes 1]
\;,
\end{split}
\end{equation}
where we have used $C_{\mpone} = C_\LA$ since parton $m+1$ must be a gluon in order to give a leading $\log(\nu)$ contribution. These substitutions give us
\begin{align}
    \label{eq:splitting1g}
    \sket{A_i}
    \approx{}& 
    \bigg\{
    \sum_{l = 1}^{m} 
    2 C_l [1 \otimes 1]\,
    L\big(\nu, \hat a_{l}\big)
    \\&+ \notag
    2\big[C_\LA [1 \otimes 1] +  \cC(l_0,m+1) \big]
    L\big(\nu, \hat a_{m+1}\big)
    \bigg\}
    \\&\times \notag
    (1 - e^{- x_i})
    \sket{\{\hat p, \hat f, \hat c, \hat c'\}_{m+1}}
    \;.
\end{align}

Consider now the term in Eq.~(\ref{eq:splitting1g}) that contains a color operator $\cC(l_0,\mpone)$, defined in Eq.~(\ref{eq:cl1k1def}). We apply this operator after the color operator for the initial splitting, $\cC_0(l_0,k_0)$, defined in Eq.~(\ref{eq:cl0k0def}). This gives us an operator with four terms,
\begin{equation}
\begin{split}
\label{eq:cCidef}
\cC_i ={}& 
(\bm T_{l_0}\!\cdot\!\bm T_{\mpone})\, T_{l_0}\!\otimes \! T_{k_0}^\dagger
+  T_{l_0}\!\otimes \! T_{k_0}^\dagger\,(\bm T_{l_0}\!\cdot\!\bm T_{\mpone})
\\&
+ (\bm T_{l_0}\!\cdot\!\bm T_{\mpone})\,  T_{k_0}\!\otimes \! T_{l_0}^\dagger
+  T_{k_0}\!\otimes \!  T_{l_0}^\dagger\, (\bm T_{l_0}\!\cdot\!\bm T_{\mpone})
\;.
\end{split}
\end{equation}
There can be several factors of $\iP{\cS_{l}^{[1,0]}(x Q^2/\nu)}$ in Eq.~(\ref{eq:cScYNLLsoln}) and in some of those factors we can select the $\cC(l_0,\mpone)$ term in Eq.~(\ref{eq:splitting1g}). Finally, there is a $\iP{\cdots}$ operation. This gives us a sum of color operators of the form
\begin{equation}
\begin{split}
\label{eq:cCdef}
\P{\cC} ={}& 
\PL (\bm T_{l_0}\!\cdot\!\bm T_{\mpone})^A\,T_{l_0}\!\otimes \!T_{k_0}^\dagger\,
(\bm T_{l_0}\!\cdot\!\bm T_{\mpone})^B
\\&
+ (\bm T_{l_0}\!\cdot\!\bm T_{\mpone})^A\,T_{k_0}\!\otimes \!T_{l_0}^\dagger\,
(\bm T_{l_0}\!\cdot\!\bm T_{\mpone})^B \PR
\;.
\end{split}
\end{equation}
Using Eq.~(\ref{eq:Pcolorform2}), this becomes
\begin{equation}
\begin{split}
\label{eq:PcCA}
\P{\cC} ={}& 
\PL (\bm T_{l_0}\!\cdot\!\bm T_{\mpone})^{A+B}\,
T_{l_0}\!\otimes \!T_{k_0}^\dagger
\\&
+ T_{k_0}\!\otimes \!T_{l_0}^\dagger\,
(\bm T_{l_0}\!\cdot\!\bm T_{\mpone})^{A + B} \PR
\;.
\end{split}
\end{equation}
Now consider the color operator $\bm T_{l_0}\!\cdot\!\bm T_{\mpone}\,T_{l_0}^{a}$. In diagrams, parton $l_0$ emits a gluon with label $\mpone$, leaving parton $l_0$ in a new color state. Then a gluon is exchanged between partons $l_0$ and $\mpone$. This gives us a color triangle diagram,
\begin{equation}
\bm T_{l_0}\!\cdot\!\bm T_{\mpone}\,T_{l_0}^{a}
= \mi f_{abc} T_{l_0}^b T_{l_0}^c
\;.
\end{equation}
Then we can use
\begin{equation}
\begin{split}
\mi f_{abc} T_{l_0}^b T_{l_0}^c
={}& \frac{1}{2} \mi f_{abc} [T_{l_0}^b, T_{l_0}^c]
= \frac{1}{2} \mi f_{abc} \mi f_{bcd} T_{l_0}^d
\\
={}& -\frac{C_\LA}{2}\,T_{l_0}^a
\;.
\end{split}
\end{equation}
Thus
\begin{equation}
\label{eq:triangle}
\bm T_{l_0}\!\cdot\!\bm T_{\mpone}\,T_{l_0}^{a}
= -\frac{C_\LA}{2}\,T_{l_0}^a
\;.
\end{equation}
This gives us
\begin{equation}
\label{eq:multitriangle0}
(\bm T_{l_0}\!\cdot\!\bm T_{\mpone})^{A+B}\,T_{l_0}\!\otimes \!T_{k_0}^\dagger
= \left[-\frac{C_\LA}{2}\right]^{A+B} T_{l_0}\!\otimes \!T_{k_0}^\dagger
\;.
\end{equation}
The second term in Eq.~(\ref{eq:PcCA}) gives the same result, so that the net color operator defined in Eq.~(\ref{eq:cCdef}) is
\begin{equation}
\begin{split}
\label{eq:PcC}
\P{\cC} ={}&
\left[-\frac{C_\LA}{2}\right]^{A+B} \!\!
\P{T_{l_0} \otimes T_{k_0}^\dagger
+ T_{k_0} \otimes T_{l_0}^\dagger}
\;.
\end{split}
\end{equation}

We conclude that when $\cC(l_0,\mpone)$ in Eq.~(\ref{eq:splitting1g}) is part of $\cS_\cY^{[k+1]}(x_0 Q^2/\nu;\nu)$ in Eq.~(\ref{eq:cScYNLLsoln}), we get the same result for $\cS_\cY^{[k+1]}(x_0 Q^2/\nu;\nu)$ by making the replacement
\begin{equation}
\cC(l_0,\mpone) \to - C_\LA [ 1\otimes 1]
\;.
\end{equation}
There is a factor 2 for each $C_\LA$ here because there are two $T_{l_0} \otimes T_{k_0}^\dagger$ terms and two $T_{k_0} \otimes T_{l_0}^\dagger$ terms in Eq.~(\ref{eq:cCidef}).

With this replacement, the terms in Eq.~(\ref{eq:splitting1g}) proportional to $L\big(\nu, \hat a_{m+1}\big)$ cancel. Thus we get the same result for $\cS_\cY^{[k+1]}(x_0 Q^2/\nu;\nu)$ by making the replacement 
\begin{equation}
\sket{A_i} \to \sket{A_i^\mathrm{eff}}
\;,
\end{equation}
where
\begin{equation}
  \begin{split}
    \label{eq:splitting1h}
    \sket{A_i^\mathrm{eff}}
    \approx{}& 
    \sum_{l = 1}^{m} 
    2 C_l [1 \otimes 1]\,
    L\big(\nu,\hat a_{l}\big)
    \\&\times
    (1 - e^{- x_i})
    \sket{\{\hat p, \hat f, \hat c, \hat c'\}_{m+1}}
    \;.
  \end{split}
\end{equation}
Note that $\sket{A_i^\mathrm{eff}}$ is a number, which we may call $\lambda_i$, times the starting state vector,
\begin{equation}
\label{eq:Aieff}
\sket{A_i^\mathrm{eff}} = \lambda_i\sket{\{\hat p, \hat f, \hat c, \hat c'\}_{m+1}}
\;.
\end{equation}

Return now to Eq.~(\ref{eq:cScYNLLsoln}) for  $\cS_\cY^{[k+1]}(x_0 Q^2/\nu;\nu)$ applied to the starting state $\sket{\{p,f,c,c'\}_m}$. In the last factor, we have dealt with the operator $\cS_{l_0}^{[1,0]}(Q^2/\nu)$, which creates a new parton with label $m+1$. Now we turn to the remaining operator, $-\P{\cS_{l_0}^{[1,0]}(Q^2/\nu)}$. This operator, acting on the state $\sket{\{p,f,c,c'\}_m}$, produces a linear combination of states with $m$ partons, $\sket{\{p, f,\hat c,\hat c'\}_{m}}$. Here the momentum and flavors are the same as in the initial state, but the colors change. More precisely,
\begin{equation}
  \label{eq:cS10onmP}
  \begin{split}
    \sum_{l_0 = 1}^m&\P{\cS_{l_0}^{[1,0]}(Q^2/\nu)}
    e^{- x_0}
    \sket{\{p, f, c, c'\}_m} 
    \\\approx{}& 
    - 
    \sum_{l_0 = 1}^m \sum_{\substack{k_0 = 1 \\ k_0 \ne l_0}}^m 
    \P{\cC(l_0,k_0)}\sket{\{p,f,c, c'\}_{m}}
    \\&\times
    \int\!\frac{d\phi_0}{2\pi}
    \int\!\frac{dz_0}{1-z_0}\
    \frac{\as\big(\lambda_\LR (1-z_0) Q^2/(\nu a_{l_0})\big)}{2\pi}
    \\&\times
    \Theta\!\left(\frac{a_{l_0}}{\nu\,\vartheta(l_0,k_0)} 
      < 1-z_0 < 1\right)
    \\&\times
    e^{- x_0}
    \;.
  \end{split}
\end{equation}

Let us consider what the one of the operators, $\P{\cS_{l_i}^{[1,0]}(Q^2/\nu)}$, in Eq.~(\ref{eq:cScYNLLsoln}) does to this state. We consider the quantity
\begin{equation}
\label{B1def}
\sket{B_i} = \sum_{l_i = 1}^{m+1} \P{\cS_{l_i}^{[1,0]}(Q^2/\nu)}
(1 - e^{- x_i})
\sket{\{p, f, \hat c, \hat c'\}_{m}}
\;.
\end{equation}
With an analysis similar to but simpler than our previous analysis, we obtain
\begin{equation}
  \begin{split}
    \label{eq:splitting2A}
    \sket{B_i}
    \approx{}& 
    \sum_{l = 1}^{m} 
    2 C_l [1 \otimes 1]\,
    L\big(\nu,\hat a_{l}\big)
    \\&\times
    (1 - e^{- x_i})
    \sket{\{p, f, \hat c, \hat c'\}_{m}}
    \;.
  \end{split}
\end{equation}
This gives us
\begin{equation}
\label{eq:Bi}
\sket{B_i} = \lambda_i\sket{\{p, f, \hat c, \hat c'\}_{m}}
\;,
\end{equation}
where the eigenvalue $\lambda_i$ is exactly the $\lambda_i$ in Eq.~(\ref{eq:Aieff}).
 
We can substitute Eqs.~(\ref{eq:Bi}) and (\ref{eq:Aieff}) into Eq.~(\ref{eq:cScYNLLsoln}) to obtain
\begin{equation}
\begin{split}
\cS_\cY^{[k+1]}(&x_0 Q^2/\nu;\nu) \sket{\{p,f,c,c'\}_m}
\\ ={}&
(-1)^n 
\int_{0}^{x_0}\!\frac{dx_1}{x_1}
\int_{0}^{x_1}\!\frac{dx_2}{x_2}\cdots
\int_{0}^{x_{k-1}}\!\frac{dx_k}{x_k}
\\&\times 
\lambda_k \cdots \lambda_2\,\lambda_1
\\&\times 
\sum_{l_0}
\P{\omP{\cS_{l_0}^{[1,0]}(Q^2/\nu)}}
\\ &  \times 
e^{- x_0}
\sket{\{p,f,c,c'\}_m}
\\ &  + \mathrm{NNLL}
\;.
\end{split}
\end{equation}
However
\begin{equation}
\begin{split}
\P{\omP{\cS_{l_0}^{[1,0]}(Q^2/\nu)}} \hskip - 1.5 cm &
\\={}& 
\P{\cS_{l_0}^{[1,0]}(Q^2/\nu)
-\P{\cS_{l_0}^{[1,0]}(Q^2/\nu)}}
\\
={}& \P{\cS_{l_0}^{[1,0]}(Q^2/\nu)}
-\P{\cS_{l_0}^{[1,0]}(Q^2/\nu)}
\\
={}& 0
\;.
\end{split}
\end{equation}
Thus the NLL contributions to $\cS_\cY^{[k+1]}(x_0 Q^2/\nu;\nu)$ vanish:
\begin{equation}
\cS_\cY^{[k+1]}(x_0 Q^2/\nu;\nu) \sket{\{p,f,c,c'\}_m}
= \mathrm{NNLL}
\;.
\end{equation}
%


\section{Cancellation with $k_\LT$ ordering}
\label{sec:kTorderingcancellation}

In this appendix, we explore the cancellation of large $\log(\nu)$ factors in $\cI^{[2]}_2(\nu)$ with $k_\LT$ ordering. We can write $\isbra{1}\cI^{[2]}_2(\nu)\sket{\{\tilde p,\tilde f,\tilde c,\tilde c'\}_2}$ in the form 
\begin{equation}
\begin{split}
\label{eq:I2kTordered}
\sbra{1}\cI^{[2]}&(\nu) \sket{\{\tilde p,\tilde f, \tilde c, \tilde c'\}_2}
\\ ={}& 
\int_0^{Q^2}\!\frac{d\tilde k_T}{\tilde k_T}
\int\!d\tilde\eta\int\!\frac{d\tilde\phi}{2\pi}
\int_0^{Q^2}\!\frac{dk_T}{k_T}\int\!d\eta\int\!\frac{d\phi}{2\pi}\
\\&\times
\Theta(k_\LT < \tilde k_\LT)
(1 - e^{\nu(\hat \tau - \tau)})
e^{-\nu\tau}
\\ &\times
\sbra{1}
\cS^{[1,0]}(k_\LT,\eta,\phi)
\\&\times
\big\{\cS^{[1,0]}(\tilde k_\LT,\tilde\eta,\tilde\phi)
-\P{\cS^{[1,0]}(\tilde k_\LT,\tilde\eta,\tilde\phi)}
\big\}
\\&\times
\sket{\{\tilde p,\tilde f, \tilde c, \tilde c'\}_2}
\;.
\end{split}
\end{equation}
We begin with a $q\bar q$ state with parton momenta $\tilde p_1$ and $\tilde p_2$ aligned along the $+$ and $-$ $z$ axis, respectively. Then one of these two partons splits, producing parton 3. We suppose that it is parton $1$ that splits. After the splitting, we have partons with momenta $p_1$, $p_2$, and $p_3$. The value of $1-T$ in this state is $\tau$ and we suppose that $\tau \ll 1$. Then there is a second splitting, producing partons with momenta $\hat p_1$, $\hat p_2$, $\hat p_3$, and $\hat p_4$ with a thrust variable $\hat \tau \ll 1$. We consider either the splitting of parton 3 with parton 2 as the dipole partner or the splitting of parton 2 with parton 3 as dipole partner. Other splitting possibilities are not as important and we omit consideration of them here. We limit our consideration to the leading color approximation.

We begin with the first splitting, which we describe with splitting variables $ \tilde k_\LT$, $\tilde \eta$, $\tilde \phi$ that relate $p_3$ to $\tilde p_1$ and $\tilde p_2$: 
\begin{equation}
\label{eq:p3kinematics}
p_3 =
e^{\tilde \eta}\, \frac{\tilde k_\LT}{|Q|}\, \tilde p_1 
+ e^{-\tilde \eta}\, \frac{\tilde k_\LT}{|Q|}\, \tilde p_2  
+ \tilde k_\perp
\;.
\end{equation}
Here $|Q| = [Q^2]^{1/2} = [2 \tilde p_1\cdot \tilde p_2]^{1/2}$  and $\tilde k_\perp$ is a vector that is orthogonal to $\tilde p_1$ and $\tilde p_2$:
\begin{equation}
\tilde k_\perp \cdot \tilde p_1 =  \tilde k_\perp \cdot \tilde p_2 = 0
\;.
\end{equation}
We have defined the scalar $\tilde k_\LT$ by
\begin{equation}
\tilde k_\LT = \left[- \tilde k_\perp^2\right]^{1/2}
\;.
\end{equation}
This definition gives $p_3^2 = 0$. The variable $\tilde \eta$ is the rapidity of $p_3$. We need one more splitting variable, the azimuthal angle $\tilde \phi$ of $\tilde k_\perp$. 

For emission from parton 1, the splitting function is small for $\tilde \eta < 0$. There is a maximum value of $\tilde \eta$ for fixed $\tilde k_\LT$, set by the condition for a maximally collinear emission
\begin{equation}
e^{\tilde \eta}\, \frac{\tilde k_\LT}{|Q|} = 1
\;.
\end{equation}
When $\tilde \eta$ is close to this upper bound, the splitting function tends to zero. Thus we integrate over the splitting variables with measure $d\tilde \eta \,d\log(\tilde k_\LT/|Q|)$ over the range $0 \lesssim \tilde \eta \lesssim -\log(\tilde k_\LT/|Q|)$. In this range, as long as $\tilde \eta$ is not near either endpoint, the splitting function is approximately constant. For small $\tilde k_\LT$, this is a large range. The integration gives us a large logarithm, which comes from integrating over the interior of the range, omitting the regions near the endpoints:
\begin{equation}
\label{eq:etaintegrationrange}
0 \ll \tilde \eta \ll -\log(\tilde k_\LT/|Q|)
\;.
\end{equation}
We will assume that $\tilde \eta$ lies in this range in the analysis that follows.

For an emission from parton 1, we define the momentum of parton 1 after the emission to be
\begin{equation}
\label{eq:p1kinematics}
p_1 =
\left[ 1 - e^{\tilde \eta}\, \frac{\tilde k_\LT}{|Q|}\right] \tilde p_1  
+  \frac{\tilde k_\LT^2/|Q|^2}{1 - e^{\tilde \eta} \tilde k_\LT/|Q|}\,
\tilde p_2
- \tilde k_\perp
\;.
\end{equation}
With this definition, $p_1^2=0$ and $p_1 - \tilde p_1 + p_3$ lies entirely in the direction of $\tilde p_2$:
\begin{equation}
\label{eq:p1p3kinematics}
p_1 - \tilde p_1 + p_3 = 
\frac{e^{-\tilde \eta} \tilde k_\LT/|Q|}{1 - e^{\tilde \eta} \tilde k_\LT/|Q|}\,
\tilde p_2
\;.
\end{equation}

Finally, we need to define the momentum $p_2$ of parton 2 after the splitting so that momentum is conserved: $p_1 + p_2 + p_3 = \tilde p_1 + \tilde p_2$.
Using Eq.~(\ref{eq:p1p3kinematics}) we obtain $p_2$ by applying a small boost in the $z$ direction to $\tilde p_2$:
\begin{equation}
p_2 = \left[1 - \frac{e^{-\tilde \eta} \tilde k_\LT/|Q|}
{1 - e^{\tilde \eta} \tilde k_\LT/|Q|}\right] 
\tilde p_2
\;.
\end{equation}
This is the exact relation. In the integration range (\ref{eq:etaintegrationrange}), this relation becomes
\begin{equation}
\label{eq:Deltap2}
\tilde p_2 - p_2 \approx e^{-\tilde \eta}\,\frac{\tilde k_\LT}{|Q|}\,
\tilde p_2
\;.
\end{equation}

We use Eqs.~(\ref{eq:tauparts}) and (\ref{eq:tauLR}) to calculate the thrust for the state after the first splitting:
\begin{equation}
\begin{split}
\tau ={}& \frac{1}{Q^-}\left(
p_1^- + p_3^- + p_2^+\right)
\\
={}& \frac{1}{Q^-}\left(
\tilde p_2^- - p_2^-
+ \tilde p_1^- + p_2^+\right)
\;.
\end{split}
\end{equation}
We can use $\tilde p_1^- = p_2^+ = 0$. Then we can use $\tilde p_2^- = Q^-$ and  Eq.~(\ref{eq:Deltap2}) for $\tilde p_2^- - p_2^-$. This gives $\tau \approx  e^{-\tilde \eta}\, {\tilde k_\LT}/{|Q|}$ or
\begin{equation}
\label{eq:thrust1}
\nu\tau \approx \nu e^{-\tilde \eta}\,\frac{\tilde k_\LT}{|Q|}
\;.
\end{equation}

This relation is significant because this emission is accompanied by a measurement function $\exp(-\nu\tau)$. The measurement function is approximately 1 for $\nu\tau \ll 1$ but approximately zero for $1 \ll \nu\tau$. Thus we effectively integrate over the range
\begin{equation}
\label{eq:taurange}
\nu\tau < 1
\;.
\end{equation}

In the analysis that follows, we will need a relation between $2 p_3\cdot Q$ and the values of $\tilde \eta$ and $\tilde k_\LT$ for the splitting. We can use Eq.~(\ref{eq:p3kinematics}) with $\eta \gg 0$ together with $2\tilde p_1 \cdot Q = Q^2$ to give
\begin{equation}
\label{eq:p3dotQ1}
\frac{2p_3\cdot Q}{Q^2} = e^{\tilde \eta}\, \frac{\tilde k_\LT}{|Q|}
\;.
\end{equation}
%

\begin{figure}
\begin{center}
\includegraphics[width = 7.5 cm]{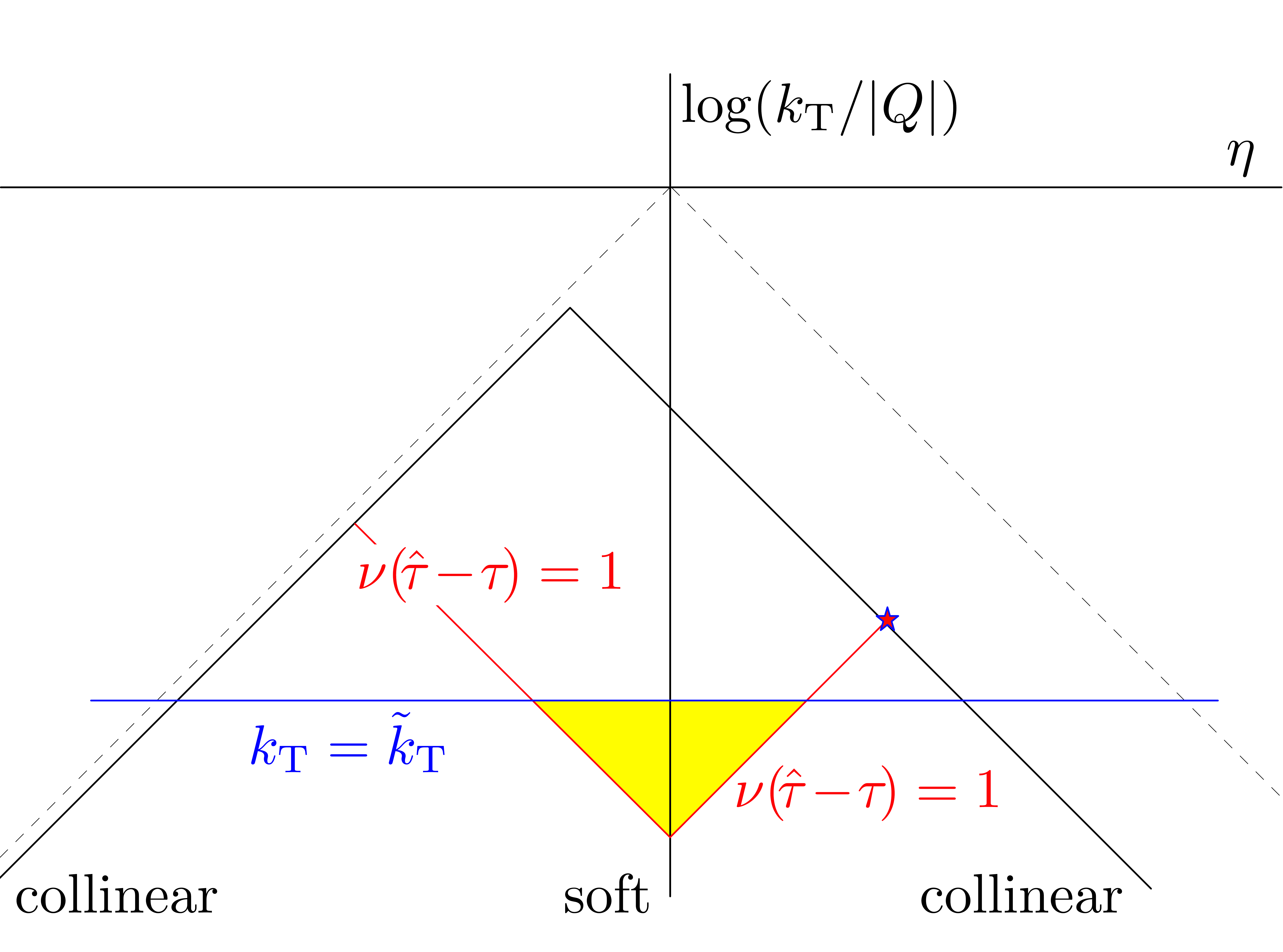}
\end{center}
\caption{
Integration regions for second splitting.
}
\label{fig:PanLocal}
\end{figure}

We now turn to the second splitting. We will describe the splitting using variables and a momentum mapping that are slightly different from what is used in \textsc{Deductor} with $k_\LT$ ordering. In fact, we will use a local momentum mapping. However, in the kinematic limit of interest, the description used here reduces to the description used in \textsc{Deductor}. The splitting kinematics are illustrated in Fig.~\ref{fig:PanLocal}. We describe the second splitting with splitting variables $k_\LT$, $\eta$, $\phi$ that relate $\hat p_4$ to $p_2$ and $p_3$: 
\begin{equation}
\begin{split}
\label{eq:p4kinematics}
\hat p_4 ={}& 
A_{32}\,e^{\eta}\frac{k_\LT}{|Q|} \, p_3 
+ A_{23}\,e^{-\eta}\frac{k_\LT}{|Q|}\, p_2 
+ k_\perp
\;,
\end{split}
\end{equation}
where
\begin{equation}
\begin{split}
A_{32} ={}& \left[\frac{Q^2}{2p_2\cdot p_3}\,
\frac{p_2\cdot Q}{p_3\cdot Q}\right]^{1/2}
\;,
\\
A_{23} ={}& \left[\frac{Q^2}{2p_2\cdot p_3}\,
\frac{p_3\cdot Q}{p_2\cdot Q}\right]^{1/2}
\;.
\end{split}
\end{equation}
Here $k_\perp$ is a vector that is orthogonal to $p_2$ and $p_3$:
\begin{equation}
k_\perp \cdot p_3 = k_\perp \cdot p_2 = 0
\;.
\end{equation}
As for the first splitting, we have defined the scalar $k_\LT = \left[-k_\perp^2\right]^{1/2}$. This definition gives $\hat p_4^2 = 0$. The variable $\eta$ describes the rapidity of $\hat p_4$ with respect to the emitting dipole, with a constant $\log(p_2\cdot Q/p_3\cdot Q)/2$ added \cite{DasguptaShowerSum}. We need one more splitting variable, the azimuthal angle $\phi$ of $k_\perp$ in the dipole c.m.\ frame.

There is a limit to how large $\eta$ can be: $\eta_\mathrm{min} < \eta < \eta_\mathrm{max}$. The limits are fixed by the requirements that the components of $\hat p_4$ along $p_3$ and $p_2$ cannot be larger than 1:
\begin{equation}
\begin{split}
\label{eq:etamaxmin}
\eta_\mathrm{max} ={}&
-\log\left(\frac{k_\LT}{|Q|}\right)
-\log\left(A_{32}\right)
\;,
\\
\eta_\mathrm{min} ={}&
\log\left(\frac{k_\LT}{|Q|}\right)
+\log\left(A_{23}\right)
\;.
\end{split}
\end{equation}
The lines $\eta = \eta_\mathrm{max}$ and $\eta = \eta_\mathrm{min}$ are indicated in Fig.~\ref{fig:PanLocal} as the lines labelled collinear. This is a large integration range. We will assume in what follows that $\eta$ is not near to the endpoints of the integration range:
\begin{equation}
\eta_\mathrm{min} \ll \eta \ll \eta_\mathrm{max}
\;.
\end{equation}

For emission from parton 3, we let the momentum of parton 3 after the emission be
\begin{equation}
\begin{split}
\hat p_3 \approx{}& 
\left[1 - A_{32}\,e^{\eta}\frac{k_\LT}{|Q|}\right]  p_3
\\&
+ \frac{k_\LT^2}{Q^2} \frac{Q^2}{2p_3\cdot p_2}
\left[1 - A_{32}\,e^{\eta}\frac{k_\LT}{|Q|}\right]^{-1}
p_2
- k_\perp
\;.
\end{split}
\end{equation}
With this definition, $\hat p_3^2 = 0$ and $\hat p_3 - p_3 + \hat p_4$ lies entirely in the direction of $p_2$. Then we can maintain momentum conservation, $\hat p_1 + \hat p_2 + \hat p_3 + \hat p_4 = p_1 + p_2 + p_3$ by setting $\hat p_1 = p_1$ and obtaining $\hat p_2$ by performing a small boost on $p_2$:
\begin{equation}
\label{eq:p4boost}
\hat p_2 = 
e^{-\omega}
p_2
\;.
\end{equation}
With a few algebraic steps, we find
\begin{equation}
\begin{split}
\label{eq:eminusomega}
e^{-\omega} ={}& 1 
- A_{23}\,e^{-\eta}\frac{k_\LT}{|Q|}
\left[1 - A_{32}\,e^{\eta}\frac{k_\LT}{|Q|}\right]^{-1}
\;.
\end{split}
\end{equation}

These definitions have been exact for the kinematic variables and momentum mapping chosen. We can now make some approximations. Given our kinematic conditions (\ref{eq:etaintegrationrange}) for the first emission, the momentum $p_3$ has large rapidity. That is, it makes a small angle with the $z$ axis. The transverse momentum vector defined in Eq.~(\ref{eq:p4kinematics}) is orthogonal to $p_3$ and $p_2$ whereas the transverse momentum vector in \textsc{Deductor} is orthogonal to $p_3$ and $Q$. However, since $p_3$ makes a small angle with the $z$ axis, this is almost the same thing. In \textsc{Deductor}, momentum is conserved by applying a boost in the plane of $p_3$ and $Q$. Since $p_3$ makes a small angle with the $z$ axis, this boost is almost exactly along the $z$ axis. The boost is applied to both $p_2$ and $p_1$, but this difference has only a tiny effect on the resulting thrust. Thus in the limit considered, the \textsc{Deductor} kinematics and the kinematics used here are equivalent.

We now examine the change in thrust produced by the emission of parton 4 from parton 3. We assume that $p_4$ is in the right thrust hemisphere. This is always the case when $\eta \gg 0$. There is a region near $\eta \approx 0$ in which this assumption fails. With the kinematics that we are using, the thrust axis is along $-\vec p_2$. That is, it is the $z$ axis. Then we have
\begin{equation}
\begin{split}
\hat \tau - \tau 
={}& \frac{1}{Q^-}\left[\hat p_4^- + \hat p_3^- - p_3^-
+\hat p_2^+ - p_2^+\right]
\\
={}& \frac{1}{Q^-}\left[p_2^- - \hat p_2^- 
+\hat p_2^+ - p_2^+\right]
\;.
\end{split}
\end{equation}
We have $p_2^+ = 0$,  $\hat p_2 = e^{-\omega} p_2$ from Eq.~(\ref{eq:p4boost}), and $p_2^- /Q^- = 2 p_2\cdot Q/Q^2$. This gives us
\begin{equation}
\begin{split}
\hat \tau - \tau 
={}& \frac{2p_2\cdot Q}{Q^2}[1 - e^{-\omega}]
\;.
\end{split}
\end{equation}
Now the condition $\eta \ll \eta_\mathrm{max}$ that we assume implies that $A_{32}\,e^{\eta}k_\LT/|Q| \ll 1$. Thus in Eq.~(\ref{eq:eminusomega}), we can replace the factor $1 - A_{32}\, e^{\eta}  k_\LT/|Q|$ in $e^{-\omega}$ by just 1. Then
\begin{equation}
\begin{split}
\hat \tau - \tau 
={}& \frac{2p_2\cdot Q}{Q^2}\,A_{23}\, e^{-\eta} \frac{k_\LT}{|Q|}
\;.
\end{split}
\end{equation}

Since $p_3$ makes a small angle with the $z$ axis, we obtain the approximations
\begin{equation}
\begin{split}
\label{eq:split2approximations}
2 p_2\cdot p_3 \approx{}& \frac{2 p_2\cdot Q\,2p_3\cdot Q}{Q^2}
\;,
\\
A_{32} \approx{}& \frac{Q^2}{2 p_3\cdot Q}
\;,
\\
A_{23} \approx{}& \frac{Q^2}{2 p_2\cdot Q}
\;.
\end{split}
\end{equation}
With these approximations, we have
\begin{equation}
\begin{split}
\label{eq:thrust2}
\nu(\hat \tau - \tau) 
\approx{}& \nu e^{-\eta} \frac{k_\LT}{|Q|}
\;.
\end{split}
\end{equation}
With the same approximations, we obtain for the change in thrust produced by an emission from parton 2 with the dipole partner being parton 3,
\begin{equation}
\begin{split}
\label{eq:nudeltatau}
\nu(\hat \tau - \tau) 
\approx{}& \nu e^{\eta} \frac{k_\LT}{|Q|}
\;.
\end{split}
\end{equation}
Again, this is for $|\eta| \gg 0$. For the soft emission region near $\eta = 0$, there is the possibility that $p_4$ is in the opposite thrust hemisphere from the parton that emitted it, so that the thrust calculation changes.

These relations are significant because the second emission is accompanied by a measurement function $1 - \exp(-\nu(\hat \tau - \tau))$. The measurement function is approximately 1 for $1 \ll \nu(\hat \tau - \tau)$ but approximately zero for $\nu(\hat \tau - \tau) \ll 1$. Thus we effectively integrate over the range
\begin{equation}
\nu(\hat \tau - \tau) > 1
\;.
\end{equation}
The boundary of this integration region is indicated in Fig.~\ref{fig:PanLocal} as straight lines with the labels $\nu(\hat \tau - \tau) = 1$.

There is one more restriction on the integration range for the second splitting. We are analyzing  a $k_\LT$ ordered shower, so
\begin{equation}
\label{eq:kTordering}
k_\LT < \tilde k_\LT
\;.
\end{equation}
The line $k_\LT = \tilde k_\LT$ is indicated in Fig.~\ref{fig:PanLocal}.

To analyze Eq.~(\ref{eq:kTordering}), we will need to know the value $k_{\LT,\star}$ of $k_\LT$ at the point labelled with a star in Fig.~\ref{fig:PanLocal}. We first note that the line for $\eta > 0$ labelled collinear in Fig.~\ref{fig:PanLocal} is given by $\eta = \eta_\mathrm{max}$ in Eq.~(\ref{eq:etamaxmin}), $e^\eta k_\LT/|Q| = 1/A_{32}$. We can use Eqs.~(\ref{eq:split2approximations}) and (\ref{eq:p3dotQ1}) for $A_{32}$, giving
\begin{equation}
\label{eq:collinear}
e^{\eta}\, \frac{k_\LT}{|Q|} \approx e^{\tilde\eta}\, \frac{\tilde k_\LT}{|Q|}
\;,
\hskip 1 cm \mathrm{collinear}
\;.
\end{equation}

Then using Eq.~(\ref{eq:thrust1}) to eliminate $\tilde\eta$ and Eq.~(\ref{eq:thrust2}) to eliminate $\eta$ we have
\begin{equation}
\frac{k_\LT^2}{Q^2} \approx   
\frac{\nu(\hat \tau - \tau)}{\nu\tau}\,
\frac{\tilde k_\LT^2}{Q^2}
\;,
\hskip 1 cm \mathrm{collinear}
\;.
\end{equation}
The point labelled with a star in Fig.~\ref{fig:PanLocal} is the intersection of the collinear line and the line $\nu(\hat \tau - \tau) = 1$. Thus,
\begin{equation}
\label{eq:kTstar}
\frac{k_{\LT,\star}^2}{Q^2} \approx   
\frac{1}{\nu\tau}\,
\frac{\tilde k_\LT^2}{Q^2}
\;.
\end{equation}
Since in the dominant integration region $\nu\tau < 1$, we conclude that $k_{\LT,\star} >  \tilde k_\LT$. Thus the line $k_\LT = \tilde k_\LT$ lies below the point $(\eta_\star,  k_{\LT,\star})$ in Fig.~\ref{fig:PanLocal}. This implies that the effective integration region for the second splitting is the region shaded in yellow in Fig.~\ref{fig:PanLocal}. Inside this region, the integrand is approximately 1.

Now consider the case in which the first splitting is virtual. The corresponding contribution comes from the term $\iP{\cS^{[1,0]}(\tilde k_\LT, \tilde\eta,\tilde\phi)\, e^{-\nu\tau}}$ in the last line of Eq.~(\ref{eq:I2kTordered}). We integrate over the splitting variables for the first splitting, including the measurement function $e^{-\nu\tau}$, but we start the second splitting from the $q\bar q$ state with just partons with momenta $\tilde p_1$ and $\tilde p_2$, but with the $k_\LT$ ordering requirement $k_\LT < \tilde k_\LT$. Now the limits on $\eta$ in Fig.~\ref{fig:PanLocal}, indicated by the lines labelled collinear, are expanded to the dotted lines in the figure. However, the effective integration region for the second splitting is the region shaded in yellow in Fig.~\ref{fig:PanLocal}. When we subtract the virtual contribution from the real contribution, we get zero within the approximations that we have used.

In Eq.~(\ref{eq:kTstar}), we have equality, $\tilde k_{\LT} = k_{\LT,\star}$, when the value of $\tau$ for the first splitting is given by $\nu\tau = 1$. The value of $\tilde k_{\LT}$ in the first splitting can be less than $k_{\LT,\star}$, but if $\tilde k_{\LT}$ is too small then the integration region in Fig.~\ref{fig:PanLocal} disappears. From Eq.~(\ref{eq:thrust2}) at $\eta = 0$, $\nu(\hat \tau - \tau) = 1$ and $\tilde k_\LT = k_\LT$, we see that this limits $\tilde k_\LT$ to
\begin{equation}
\label{eq:tildekTmin}
\frac{\tilde k_\LT}{|Q|} > \frac{1}{\nu}
\;.
\end{equation}

Our analysis above has assumed that the first emission is at large rapidity, $\tilde \eta \gg 0$. What happens when $\tilde \eta \approx 0$? The approximations that we have used are not adequate in this situation, so it might seem that there is nothing that we can say. However, we can examine what happens when $\tilde \eta$ is large enough that the approximations are still valid, but $\tilde \eta$ becomes smaller and smaller. Start with Eq.~(\ref{eq:collinear}) for the collinear line in Fig.~\ref{fig:PanLocal} and use Eq.~(\ref{eq:thrust1}) to eliminate $\tilde k_\LT$ and Eq.~(\ref{eq:thrust2}) to eliminate $k_\LT$, giving
\begin{equation}
e^{2\eta} \approx   
\frac{\nu\tau}{\nu(\hat \tau - \tau)}\,
e^{2\tilde\eta}
\;,
\hskip 1 cm \mathrm{collinear}
\;.
\end{equation}
The point labelled with a star in Fig.~\ref{fig:PanLocal} is the intersection of the collinear line and the line $\nu(\hat \tau - \tau) = 1$. Thus,
\begin{equation}
e^{2\eta_\star} \approx   
\nu\tau\,
e^{2\tilde\eta}
\;.
\end{equation}
In the effective integration range for the first splitting, we have $\nu\tau < 1$. Thus
\begin{equation}
\eta_\star < \tilde\eta
\;.
\end{equation}
This tells us that when the rapidity of the first splitting becomes small, $\tilde \eta \to 0$, we have $\eta_\star \to 0$. In this limit, the real-virtual cancellation in this region deteriorates, but this deterioration does not matter because the allowed integration region for the second splitting in Fig.~\ref{fig:PanLocal} shrinks to zero.

The cancellation will fail on a certain surface in the integration region. On this surface, the splitting variables for the second emission are given by  
\begin{equation}
(k_{\LT},\eta) \approx (k_{\LT,\star},\eta_\star)
\;.
\end{equation}
In this region, the second emission is collinear rather than both soft and collinear, so that the emission probability does not match the constant that appears  in the region in which the second emission is both soft and collinear. However in the virtual subtraction the second emission is both soft and collinear so that the emission probability is this constant. Thus the emission probabilities do not match between the real emission and the subtraction. 

The surface of non-matching probabilities is specified as follows. If $k_{\LT} = k_{\LT,\star}$, then the line $k_{\LT} = \tilde k_{\LT}$ in Fig.~\ref{fig:PanLocal} must pass through $(k_{\LT,\star},\eta_\star)$, so that $\tilde k_{\LT} = k_{\LT,\star}$. Then Eq.~(\ref{eq:kTstar}) implies that the value of $\tau$ for the fist emission is given by $\nu\tau = 1$. Then Eq.~(\ref{eq:thrust1}) gives
\begin{equation}
\tilde \eta \approx \log(\nu) + \log\!\left(\frac{\tilde k_{\LT}}{Q}\right)
\;.
\end{equation}
The transverse momentum for the first emission varies in the range
\begin{equation}
- \log(\nu) \ll \log\!\left(\frac{\tilde k_{\LT}}{Q}\right)
\ll - \frac{1}{2} \log(\nu) 
\;.
\end{equation}
Here the lower limit is from Eq.~(\ref{eq:tildekTmin}) and the upper limit is from Eqs.~(\ref{eq:etaintegrationrange}), (\ref{eq:thrust1}), and (\ref{eq:taurange}). For the second emission, $(k_{\LT},\eta) \approx (k_{\LT,\star},\eta_\star)$:
\begin{equation}
\begin{split}
\eta \approx {}& \tilde \eta
\;,
\\
\log\!\left(\frac{k_{\LT}}{Q}\right) \approx{}& 
\log\!\left(\frac{\tilde k_{\LT}}{Q}\right)
\;.
\end{split}
\end{equation}
Thus the integration region inside which cancellation fails is one dimensional, so we are left with a contribution to $\cI^{[2]}_2$ proportional to $\log^1(\nu)$.





\begin{thebibliography}{99}


\bibitem{NSdglap}
  Z.~Nagy and D.~E.~Soper,
  {\em Final state dipole showers and the DGLAP equation},
  \href{http://dx.doi.org/10.1088/1126-6708/2009/05/088}
  {JHEP \textbf{05}, 088 (2009)}
  [\href{http://inspirehep.net/search?p=find+doi+10.1088/1126-6708/2009/05/088}
  {\textsc{inSPIRE}}].

\bibitem{NSpT}
  Z.~Nagy and D.~E.~Soper,
  {\em On the transverse momentum in Z-boson production in a
  virtuality ordered parton shower},
  \href{http://dx.doi.org/10.1007/JHEP03(2010)097}
  {JHEP {\bf 1003} (2010) 097}
  [\href{http://inspirehep.net/search?p=find+doi+10.1007/JHEP03(2010)097}
  {\textsc{inSPIRE}}].
  

\bibitem{DasguptaShowerSum}
  M.~Dasgupta, F.~A.~Dreyer, K.~Hamilton, P.~F.~Monni, G.~P.~Salam and G.~Soyez,
  {\em Parton showers beyond leading logarithmic accuracy},
  \href{http://dx.doi.org/10.1103/PhysRevLett.125.052002}
  {Phys. Rev. Lett. \textbf{125}, 052002 (2020)}
  [\href{http://inspirehep.net/search?p=find+doi:10.1103/PhysRevLett.125.052002}
  {\textsc{inSPIRE}}].

  
\bibitem{NSlogsum} 
  Z.~Nagy and D.~E.~Soper,
  {\em Summations of large logarithms by parton showers},
  \href{https://arxiv.org/abs/2011.04773}
  {[arXiv:2011.04773 [hep-ph]]}
  [\href{https://inspirehep.net/literature/1829280}
  {\textsc{inSPIRE}}].


\bibitem{NSI}
  Z.~Nagy and D.~E.~Soper,
  {\em Parton showers with quantum interference},
  \href{http://dx.doi.org/10.1088/1126-6708/2007/09/114}
  {JHEP {\bf 0709} (2007) 114}
  [\href{http://inspirehep.net/search?p=find+doi+10.1088/1126-6708/2007/09/114}
  {\textsc{inSPIRE}}].
  
\bibitem{NSII}
  Z.~Nagy and D.~E.~Soper,
  {\em Parton showers with quantum interference: Leading color, spin averaged},
  \href{http://dx.doi.org/10.1088/1126-6708/2008/03/030}
  {JHEP {\bf 0803} (2008) 030}
  [\href{http://inspirehep.net/search?p=find+doi+10.1088/1126-6708/2008/03/030}
  {\textsc{inSPIRE}}].

\bibitem{NSspin}
  Z.~Nagy and D.~E.~Soper,
  {\em Parton showers with quantum interference: Leading color, with spin},
  \href{http://dx.doi.org/10.1088/1126-6708/2008/07/025}
  {JHEP {\bf 0807} (2008) 025}
  [\href{http://inspirehep.net/search?p=find+doi+10.1088/1126-6708/2008/07/025}
  {\textsc{inSPIRE}}].
 
\bibitem{NScolor}
  Z.~Nagy and D.~E.~Soper,
  {\em Parton shower evolution with subleading color},
  \href{http://dx.doi.org/10.1007/JHEP06(2012)044}
  {JHEP {\bf 1206} (2012) 044}
  [\href{http://inspirehep.net/search?p=find+doi+10.1007/JHEP06(2012)044}
  {\textsc{inSPIRE}}].

\bibitem{Deductor}
  Z.~Nagy and D.~E.~Soper,
  {\em A parton shower based on factorization of the quantum density matrix},
  \href{http://dx.doi.org/10.1007/JHEP06(2014)097}
  {JHEP {\bf 1406} (2014) 097}
  [\href{http://inspirehep.net/search?p=find+doi+10.1007/JHEP06(2014)097}
  {\textsc{inSPIRE}}].

\bibitem{ShowerTime}
  Z.~Nagy and D.~E.~Soper,
  {\em Ordering variable for parton showers},
  \href{http://dx.doi.org/10.1007/JHEP06(2014)178}
  {JHEP {\bf 1406} (2014) 178}
  [\href{http://inspirehep.net/search?p=find+doi+10.1007/JHEP06(2014)178}
  {\textsc{inSPIRE}}].
  
\bibitem{PartonDistFctns}
  Z.~Nagy and D.~E.~Soper,
  {\em Parton distribution functions in the context of parton showers},
  \href{http://dx.doi.org/10.1007/JHEP06(2014)179}
  {JHEP {\bf 1406} (2014) 179}
  [\href{http://inspirehep.net/search?p=find+doi+10.1007/JHEP06(2014)179}
  {\textsc{inSPIRE}}].

\bibitem{ColorEffects}
  Z.~Nagy and D.~E.~Soper,
  {\em Effects of subleading color in a parton shower},
  \href{http://dx.doi.org/10.1007/JHEP07(2015)119}
  {JHEP {\bf 1507} (2015) 119}
  [\href{http://inspirehep.net/search?p=find+doi+10.1007/JHEP07(2015)119}
  {\textsc{inSPIRE}}].

\bibitem{NSThreshold}
  Z.~Nagy and D.~E.~Soper,
  {\em Summing threshold logs in a parton shower},
  \href{http://dx.doi.org/10.1007/JHEP10(2016)019}
  {JHEP {\bf 1610} (2016) 019}
  [\href{http://inspirehep.net/search?p=find+doi+10.1007/JHEP10(2016)019}
  {\textsc{inSPIRE}}].
   
\bibitem{NSAllOrder} 
  Z.~Nagy and D.~E.~Soper,
  {\em What is a parton shower?},
  \href{http://dx.doi.org/10.1103/PhysRevD.98.014034}
  {Phys.\ Rev.\ D {\bf 98}, 014034 (2018)}
  [\href{http://inspirehep.net/search?p=find+doi+10.1103/PhysRevD.98.014034}
  {\textsc{inSPIRE}}].
 
 

\bibitem{CMW}
  S.~Catani, B.~R.~Webber and G.~Marchesini,
  {\em QCD coherent branching and semiinclusive processes at large x},
  \href{http://dx.doi.org/10.1016/0550-3213(91)90390-J}
  {Nucl.\ Phys.\ B {\bf 349}, 635  (1991)}
  [\href{http://inspirehep.net/search?p=find+doi+10.1016/0550-3213(91)90390-J}
  {\textsc{inSPIRE}}].

 
\bibitem{thrustdef1}
  S.~Brandt, C.~Peyrou, R.~Sosnowski and A.~Wroblewski,
  {\em The Principal axis of jets. An Attempt to analyze high-energy collisions 
  as two-body processes},
  \href{http://dx.doi.org/10.1016/0031-9163(64)91176-X}
  {Phys. Lett. \textbf{12}, 57-61 (1964)}
  [\href{http://inspirehep.net/search?p=find+doi+10.1016/0031-9163(64)91176-X}
  {\textsc{inSPIRE}}].
  
\bibitem{thrustdef2}
  E.~Farhi,
  {\em Quantum Chromodynamics Test for Jets},
  \href{http://dx.doi.org/10.1103/PhysRevLett.39.1587}
  {Phys. Rev. Lett. \textbf{39}, 1587-1588 (1977)}
  [\href{http://inspirehep.net/search?p=find+doi+10.1103/PhysRevLett.39.1587}
  {\textsc{inSPIRE}}].

  

\bibitem{thrustsum}
  S.~Catani, L.~Trentadue, G.~Turnock and B.~Webber,
  {\em Resummation of large logarithms in e+ e- event shape distributions},
  \href{http://dx.doi.org/10.1016/0550-3213(93)90271-P}
  {Nucl. Phys. B \textbf{407}, 3 (1993)}
  [\href{http://inspirehep.net/search?p=find+doi+10.1016/0550-3213(93)90271-P}
  {\textsc{inSPIRE}}].



\bibitem{SurguladzeSoper}
  D.~E.~Soper and L.~R.~Surguladze,
  {\em On the QCD perturbative expansion for $e^{+} e^{-} \to$ hadrons},
  \href{http://dx.doi.org/10.1103/PhysRevD.54.4566}
  {Phys. Rev. D \textbf{54}, 4566 (1996)}
  [\href{http://inspirehep.net/search?p=find+doi+10.1103/PhysRevD.54.4566}
  {\textsc{inSPIRE}}].


\bibitem{NSNewColor}
Z.~Nagy and D.~E.~Soper,
{\em Parton showers with more exact color evolution},
\href{http://dx.doi.org/10.1103/PhysRevD.99.054009}
{Phys. Rev. D \textbf{99},  054009 (2019)}
[\href{http://inspirehep.net/search?p=find+doi+10.1103/PhysRevD.99.054009}
{\textsc{inSPIRE}}].

\bibitem{NSColoriPi}
Z.~Nagy and D.~E.~Soper,
{\em Exponentiating virtual imaginary contributions in a parton shower},
\href{http://dx.doi.org/10.1103/PhysRevD.100.074005}
{Phys. Rev. D \textbf{100}, 074005 (2019)}
[\href{http://inspirehep.net/search?p=find+doi+10.1103/PhysRevD.100.074005}
{\textsc{inSPIRE}}].

\bibitem{NSColorTheory} 
  Z.~Nagy and D.~E.~Soper,
  {\em Parton showers with more exact color evolution},
  \href{http://dx.doi.org/10.1103/PhysRevD.99.054009}
  {Phys.\ Rev.\ D {\bf 99}, 054009 (2019)}
  [\href{http://inspirehep.net/search?p=find+doi+10.1103/PhysRevD.99.054009}
  {\textsc{inSPIRE}}].



\bibitem{CataniSeymour}
  S.~Catani and M.~H.~Seymour,
  {\em A General algorithm for calculating jet cross-sections in NLO QCD},
  \href{http://dx.doi.org/10.1016/S0550-3213(96)00589-5}
  {Nucl.\ Phys.\ B {\bf 485} (1997) 291}
  \href{http://dx.doi.org/10.1016/S0550-3213(98)81022-5}
  {[Erratum-ibid.\ B {\bf 510} (1998) 503]}
  [\href{http://inspirehep.net/search?p=find+doi:10.1016/S0550-3213(96)00589-5}
  {\textsc{inSPIRE}}].


\bibitem{SjostrandSkands}
  T.~Sjostrand and P.~Z.~Skands,
  {\em Transverse-momentum-ordered showers and interleaved multiple interactions},
  \href{http://dx.doi.org/10.1140/epjc/s2004-02084-y}
  {Eur. Phys. J. C \textbf{39}, 129 (2005)}
  [\href{http://inspirehep.net/search?p=find+doi:10.1140/epjc/s2004-02084-y}
  {\textsc{inSPIRE}}].
  
\end{thebibliography}
\end{document}